\documentclass[twocolumn, numberedappendix]{aastex62}
\usepackage{graphicx}
\usepackage{epstopdf}
\usepackage{epsfig}
\usepackage{subfigure}
\usepackage{hyperref}
\usepackage{bm}
\usepackage{natbib}
\usepackage{color}
\usepackage[mediumspace,Gray,squaren]{SIunits}
\usepackage{multirow}
\usepackage{comment}
\usepackage{amsmath}
\usepackage[pass]{geometry}
\usepackage{aas_macros}
\usepackage[T1]{fontenc}
\usepackage{aecompl}
\usepackage{comment}
\usepackage{longtable}

\begin{document}
\newcommand{\ngalactic}{321}
\newcommand{\galNcutOne}{169}
\newcommand{\galNcutOneEx}{23}
\newcommand{\galNcutTwo}{195}
\newcommand{\galNcutTwoEx}{3}
\newcommand{\galNcutTwoNotOne}{76}
\newcommand{\galNcutThree}{262}
\newcommand{\galNcutThreeEx}{49}
\newcommand{\galNcutFour}{9}
\newcommand{\galNcutFive}{18}
\newcommand{\galNmmf}{82}
\newcommand{\galNTwo}{226}
\newcommand{\galNOne}{13}
\newcommand{\galNCO}{178}
\newcommand{\galNearby}{12}

\newcommand{\nmaster}{797}
\newcommand{\masNdust}{268}
\newcommand{\masNagn}{376}
\newcommand{\masNduststar}{19}
\newcommand{\masNdustTot}{287}
\newcommand{\masNagnstar}{134}
\newcommand{\masNagnTot}{510}
\newcommand{\masNnoclass}{0}
\newcommand{\masNone}{504}
\newcommand{\masNtwo}{76}
\newcommand{\masNmmf}{217}
\newcommand{\masNauxfields}{112}
\newcommand{\masNearby}{68}

\newcommand{\fracNeighborFourAllData}{0.19}
\newcommand{\fracNeighborFourMaster}{0.01}
\newcommand{\fracNeighborFourSim}{0.03}
\newcommand{\fracExtendedNinetyFiveAllData}{0.25}
\newcommand{\fracExtendedNinetyFiveMaster}{0.05}

\newcommand{\medianMMFToTwoSNRatio}{1.30}
\newcommand{\corrMMFToTwoSNRatio}{1.22}

\newcommand{\nAGNone}{370}
\newcommand{\nAGNtwo}{0}
\newcommand{\nAGNmmf}{6}
\newcommand{\nDSFGone}{0}
\newcommand{\nDSFGtwo}{68}
\newcommand{\nDSFGmmf}{200}

\shorttitle{ACT Equatorial Sources}
\title{Atacama Cosmology Telescope: Dusty star-forming galaxies and active galactic nuclei in the equatorial survey}

\author[0000-0001-9032-1585]{Megan~B.~Gralla}
\affiliation{Department of Astronomy/Steward Observatory, University of Arizona, 933 N. Cherry Ave., Tucson, AZ 85721, USA}
\author[0000-0003-4496-6520]{Tobias~A.~Marriage}
\affiliation{Dept. of Physics and Astronomy, Johns Hopkins University, 3400 N. Charles St., Baltimore, MD 21218, USA}
\author[0000-0002-2147-2248]{Graeme~Addison}
\affiliation{Dept. of Physics and Astronomy, Johns Hopkins University, 3400 N. Charles St., Baltimore, MD 21218, USA}
\author{Andrew~J.~Baker}
\affiliation{Department of Physics and Astronomy, Rutgers University, 136 Frelinghuysen Road, Piscataway, NJ 08854-8019, USA}
\author{J.~Richard~Bond}
\affiliation{Canadian Institute for Theoretical Astrophysics, University of Toronto, 60 St. George St., Toronto, ON M5S 3H8, Canada}
\author[0000-0003-1204-3035]{Devin~Crichton}
\affiliation{Astrophysics \& Cosmology Research Unit, School of Mathematics, Statistics \& Computer Science, University of KwaZulu-Natal, Westville Campus, Durban 4041, South Africa}
\author{Rahul~Datta}
\affiliation{Dept. of Physics and Astronomy, Johns Hopkins University, 3400 N. Charles St., Baltimore, MD 21218, USA}
\author{Mark~J.~Devlin}
\affiliation{Department of Physics and Astronomy, University of Pennsylvania, 209 South 33rd Street, Philadelphia, PA 19104, USA}
\author{Joanna~Dunkley}
\affiliation{Joseph Henry Laboratories of Physics, Jadwin Hall, Princeton University, Princeton, NJ 08544, USA}
\affiliation{Department of Astrophysical Sciences, Peyton Hall, Princeton University, Princeton, NJ 08544, USA}
\author{Rolando~D{\"u}nner}
\affiliation{Instituto de Astrof{\'i}sica and Centro de Astro-Ingenier{\'i}a, Facultad de F{\'i}sica, Pontificia Universidad Cat{\'o}lica de Chile, Av. Vicu{\~n}a Mackenna 4860, 7820436 Macul, Santiago, Chile}
\author{Joseph~Fowler}
\affiliation{Joseph Henry Laboratories of Physics, Jadwin Hall, Princeton University, Princeton, NJ 08544, USA}
\affiliation{NIST Quantum Sensors Group, Boulder, CO 80305, USA}
\author{Patricio A. Gallardo}
\affiliation{Department of Physics, Cornell University, Ithaca, NY 14853, USA}
\author{Kirsten~Hall}
\affiliation{Dept. of Physics and Astronomy, Johns Hopkins University, 3400 N. Charles St., Baltimore, MD 21218, USA}
\author{Mark~Halpern}
\affiliation{Department of Physics and Astronomy, University of British
Columbia, Vancouver, BC V6T 1Z4, Canada}
\author{Matthew~Hasselfield}
\affiliation{Department of Astronomy and Astrophysics, The Pennsylvania State University, University Park, PA 16802}
\author{Matt~Hilton}
\affiliation{Astrophysics \& Cosmology Research Unit, School of Mathematics, Statistics \& Computer Science, University of KwaZulu-Natal, Westville Campus, Durban 4041, South Africa}
\author{Adam~D.~Hincks}
\affiliation{Canadian Institute for Theoretical Astrophysics, University of Toronto, 60 St. George St., Toronto, ON M5S 3H8, Canada}
\author[0000-0001-7109-0099]{Kevin~M.~Huffenberger}
\affiliation{Department of Physics, Florida State University, Tallahassee FL, 32306, USA}
\author{John~P.~Hughes}
\affiliation{Department of Physics and Astronomy, Rutgers University, 136 Frelinghuysen Road, Piscataway, NJ 08854-8019, USA}
\author{Arthur~Kosowsky}
\affiliation{Department of Physics and Astronomy, University of Pittsburgh, Pittsburgh, PA 15260 USA}
\author{Carlos~H.~L{\'o}pez-Caraballo}
\affiliation{Departamento de Matem{\'a}ticas, Universidad de La Serena, Av. Juan Cisternas 1200, La Serena, Chile.}
\affiliation{Instituto de Astrof{\'i}sica and Centro de Astro-Ingenier{\'i}a, Facultad de F{\'i}sica, Pontificia Universidad Cat{\'o}lica de Chile, Av. Vicu{\~n}a Mackenna 4860, 7820436 Macul, Santiago, Chile}
\author{Thibaut~Louis}
\affiliation{Institut d'Astrophysique de Paris, F-75014 Paris, France}
\affiliation{LAL, Univ. Paris-Sud, CNRS/IN2P3, Universit\'e Paris-Saclay, Orsay, France}
\author{Danica~Marsden}
\affiliation{D-Wave Systems, 3033 Beta Avenue
Burnaby, British Columbia
V5G 4M9, Canada}
\affiliation{Department of Physics and Astronomy, University of Pennsylvania, 209 South 33rd Street, Philadelphia, PA 19104, USA}
\author{Kavilan~Moodley}
\affiliation{Astrophysics \& Cosmology Research Unit, School of Mathematics, Statistics \& Computer Science, University of KwaZulu-Natal, Westville Campus, Durban 4041, South Africa}
\author{Michael~D.~Niemack}
\affiliation{Department of Physics, Cornell University, Ithaca, NY 14853, USA}
\author[0000-0002-9828-3525]{Lyman~A.~Page}
\affiliation{Joseph Henry Laboratories of Physics, Jadwin Hall, Princeton University, Princeton, NJ 08544, USA}
\author{Bruce~Partridge}
\affiliation{Department of Physics and Astronomy, Haverford College, Haverford, PA 19041, USA}
\author{Jesus~Rivera}
\affiliation{Department of Physics and Astronomy, Rutgers University, 136 Frelinghuysen Road, Piscataway, NJ 08854-8019, USA}
\author{Jonathan~L.~Sievers}
\affiliation{Astrophysics \& Cosmology Research Unit, School of Mathematics, Statistics \& Computer Science, University of KwaZulu-Natal, Westville Campus, Durban 4041, South Africa}
\author[0000-0002-7020-7301]{Suzanne~Staggs}
\affiliation{Joseph Henry Laboratories of Physics, Jadwin Hall, Princeton University, Princeton, NJ 08544, USA}
\author{Ting~Su}
\affiliation{Dept. of Physics and Astronomy, Johns Hopkins University, 3400 N. Charles St., Baltimore, MD 21218, USA}
\author{Daniel~Swetz}
\affiliation{Department of Physics and Astronomy, University of Pennsylvania, 209 South 33rd Street, Philadelphia, PA 19104, USA}
\affiliation{NIST Quantum Sensors Group, Boulder, CO 80305, USA}
\author[0000-0002-7567-4451]{Edward~J.~Wollack}
\affiliation{NASA/Goddard Space Flight Center, Greenbelt, MD 20771, USA}

\date{Accepted version}

\def\LaTeX{L\kern-.36em\raise.3ex\hbox{a}\kern-.15em
    T\kern-.1667em\lower.7ex\hbox{E}\kern-.125emX}

\begin{abstract}
We present a catalog of 510~radio-loud active galactic nuclei (AGN, primarily blazars) and 287~dusty star-forming galaxies (DSFGs) detected by the Atacama Cosmology Telescope at $>5\sigma$ significance in bands centered on 148~GHz (2~mm), 218~GHz (1.4~mm) and 277~GHz (1.1~mm), from a 480~square degrees strip on the celestial equator with additional (360~square degrees) shallower fields. Combining the deepest available 218~GHz wide-field imaging, 277~GHz data, and multi-band filtering yields the most sensitive wide-field millimeter-wave DSFG selection to date with rms noise referenced to 218 GHz reaching $<2$~mJy. We developed techniques to remove Galactic contamination from the extragalactic catalog, yielding 321 additional Galactic sources. We employ a new flux debiasing method that handles the heterogeneous sample selection due to Galactic cuts. We present spectral properties and source counts of the AGN and DSFGs. The DSFG spectra depart from an optically thin modified blackbody between 218~GHz and 277~GHz, consistent with optically thick emission or an additional cold dust component. For bright AGN, the inter-year RMS fractional deviation in flux density from source variability is $\sim40\%$. We report 8-2870 mJy source counts for AGN and 8-90 mJy source counts for DSFGs, the latter probing both the brighter, lensed population and the fainter, unlensed population. At 277~GHz we report the first source counts measurements at these flux densities, finding an excess above most model count predictions. Finally, we select thirty of the brightest DSFGs for multi-frequency study as candidate high-$z$ lensed systems.
\end{abstract} 


\section{Introduction}
\label{introduction}

Wide-field millimeter-wave surveys open a unique window on the extragalactic universe beyond their traditional association with the primordial cosmic microwave background (CMB). In particular, galaxies are detected in these surveys through their millimeter-wave emission. Strong extragalactic sources of millimeter emission fall into  two categories. The first source category is characterized by  self-absorbing synchrotron radiation extending from radio to millimeter wavelengths. In these sources, jets from active galactic nuclei (AGN)  impart relativistic velocities to electrons that in turn generate synchrotron radiation in the galaxy's magnetic field. Self-absorption of synchrotron radiation is observed when the optically thick emission core of the AGN is within the observer's line of sight. These sources are categorized observationally as blazars, BL Lacertae objects, or flat spectrum radio quasars.\footnote{For the purposes of this paper we will refer to all these synchrotron-source classifications collectively as ``AGN'' or ``blazars''.} Measurements of their synchrotron spectra provide a unique perspective on AGN jets \citep[e.g.,][]{blanford1979,toffolatti1998, deZotti2005,tucci}. The second source category is characterized by thermal radiation from dust extending from millimeter to  far-infrared wavelengths. The dust is heated by UV and optical emission, notably from massive young stars in these  dusty star-forming galaxies (DSFG). Since the first studies of DSFGs as sub-millimeter galaxies at 850~$\mu$m \citep[SMGs; e.g.,][]{Smail97,Hughes98,Barger98}, we have learned that the most prodigious star-formation in the universe generates and is enshrouded by significant dust, making DSFGs important in the history of cosmic star-formation \citep[e.g.,][]{Lilly96,Madau96,Blain02,Chapman05,LeFloch05,Perez-Gonzalez05,Hopkins06,Daddi07,Elbaz07,Casey14,Madau14}.

Current state-of-the-art wide-field ($>100$~sq-deg) millimeter-wave source surveys have been conducted by three observatories: the Atacama Cosmology Telescope \citep[ACT;][]{marriagesources,Marsden2014}, the {\it Planck} Satellite \citep{PlanckSources2011,PlanckSources2014,PlanckSources2016}, and the South Pole Telescope \citep[SPT;][]{Vieira10,mocanu2013}. At longer radio wavelengths, surveys such as the Very Large Array (VLA) Faint Images of the Radio Sky at Twenty centimeters (FIRST) survey \citep{Becker1995} and the Australia Telescope 20~GHz Survey \citep{Murphy2010} provide important complementary data on the millimeter-bright AGN population. At shorter sub-millimeter/far-infrared wavelengths, the {\it Herschel} Space Observatory has undertaken the most comprehensive wide-field source surveys probing the DSFG population \citep[e.g.,][]{Oliver12,Viero14,Valiante2016} with additional contributions from the Submillimetre Common-User Bolometer Array 2 \citep{geach2017, Holland2013}. Complementing these 100+ deg$^2$ surveys, there have been a host of smaller, deeper surveys  by  AzTEC at the JCMT and ASTE \citep[e.g.,][]{austermanncosmos, austermannshades}, by the Max-Planck Millimeter Bolometer Array (MAMBO) and the Goddard-IRAM Superconducting 2-mm Observer (GISMO) on the IRAM 30~m telescope \citep[e.g.,][]{bertoldi2007,Lindner2011,Staguhn2014},  by  Bolocam at the CSO  \citep[e.g.,][]{laurent2005}, and the Large APEX BOlometer CAmera (LABOCA) on the APEX telescope \citep[e.g.,][]{greve2010}.

The wide-field millimeter-wave surveys have modified our understanding of the blazar population. The first catalogs from ACT, {\it Planck}, and SPT provided unprecedented source count data at $150$~GHz (2~mm) and $220$~GHz (1.4~mm), spanning more than three orders of magnitude in flux density down to 10~mJy \citep{Vieira10, marriagesources, PlanckCounts2011}.  To fit the new millimeter-wave data, \cite{tucci} and others have introduced new models that inform classical models of blazar jets. Since then, expanded millimeter-wave catalogs have further constrained these new models \citep{mocanu2013,Marsden2014,PlanckSources2014,PlanckSources2016, datta}.

The second millimeter-bright source population is comprised of DSFGs. A subset of the strictly millimeter-selected DSFGs are local star-forming galaxies, which are bright in optical and infrared catalogs \cite[e.g.,][]{PlanckLocal2011}. However, the majority of the DSFGs detected in the millimeter-wave surveys to date correspond to lensed, high-redshift DSFGs \citep[e.g.,][]{Negrello2007}. The first detections at 1.4~mm were announced by SPT \citep{Vieira10}, and subsequent work by ACT, {\it Planck}, and SPT have expanded the number of published millimeter-selected candidates to many hundreds \citep{mocanu2013,Marsden2014,Canameras15}. As with classical SMGs, the UV and optical light from these galaxies is heavily obscured, leaving nearly all information about the sources in the far-infrared thermal dust spectra and accompanying molecular line spectra. Extensive complementary observations and modeling have established that the millimeter-selected DSFG population is magnified via gravitational lensing by typical factors of $5-10$ with redshifts $z=2-6$, dust temperatures $T=30-60$~K, and significant dust optical depth at the peak of the thermal spectrum ($\lambda\approx100$~$\mu$m in the rest frame) \citep[e.g.,][]{Greve12,Hezaveh13,Vieira13,Weiss13,Canameras15,Harrington16, Strandet16,Spilker16,Su2017}. In addition to the millimeter-wave surveys, far-infrared surveys conducted by {\it Herschel} have  provided extensive samples of DSFGs that are being similarly characterized and studied \citep[e.g.;][]{Negrello2010,Bussmann13,Wardlow13}.

This work is part of a series of publications of millimeter-wave source catalogs from ACT. \cite{marriagesources} and \cite{Marsden2014} provided catalogs of AGN and DSFGs in the ACT southern survey centered at declination $-52\degree$. Recently \citet{datta} published the first polarized source study from the ACTPol survey. Here we describe the detection and initial characterization of sources in the ACT equatorial survey (decl. $0\degree$, range  $\pm2.2\degree)$. New to our approach is the addition of the ACT 277~GHz data together with a multi-frequency matched filter (MMF) to optimize DSFG detection across all three ACT frequency bands. Additionally, the presence of dust and CO emission from the Milky Way forces the introduction of systematic cuts for Galactic contamination. To handle the new source selection methods, we employ a new flux density debiasing technique described in \citet{gralladeboosting} to account for Eddington bias, which is an important consideration for the faintest DSFGs. Enabled by the extra high-frequency channel and MMF, our sensitivity to DSFGs reaches a new level for a wide-field millimeter-wave survey. The rms equivalent 218~GHz standard error is $2-3$~mJy, compared to $3-4$~mJy in \cite{mocanu2013} and \cite{Marsden2014}. At this survey depth, a significant fraction of the recovered DSFGs are predicted to be unlensed high-$z$ systems, similar to those probed by {\it Herschel} \cite[e.g.,][]{Magnelli12,Asboth16,Nayyeri16}.

This paper is organized as follows. Section \ref{sec:data} introduces the ACT equatorial survey. How we processed the data to produce catalogs is presented in Section \ref{sec:methods}, with the catalogs themselves presented in Section \ref{sec:catalog}. The sources are characterized on the basis of their spectral properties, counts, and variability in Section \ref{sec:characterization}, and a sub-sample of the brightest DSFGs that we have chosen for targeted follow-up is presented in Section \ref{sec:dsfgs}. We conclude in Section \ref{sec:conclusion}.
Throughout this work, $\alpha$ denotes the spectral index, relating flux density ($S$) to frequency ($\nu$) according to $S(\nu) \propto \nu^{\alpha}$. In our bands, typical values are $\alpha=-0.7$ for synchrotron emission and $\alpha=3.5$ for dust emission.

\section{Data}
\label{sec:data}

\begin{figure*}[t]
	\centering
	\includegraphics[width=7in]{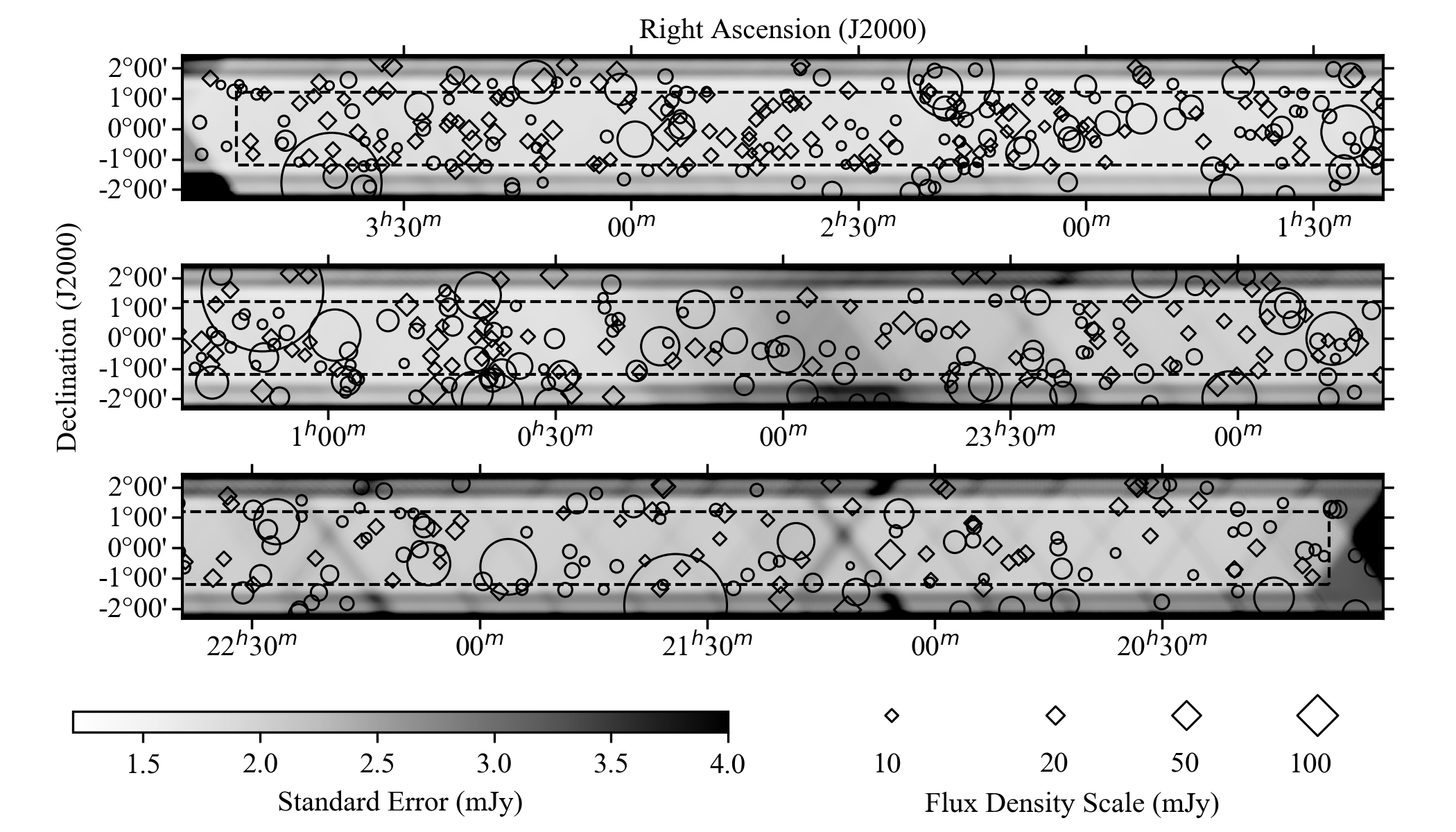}
\caption{ACT equatorial source survey main field. The primary survey area, shown here split in three sections, spans eight hours right ascension along the celestial equator. A dashed rectangle bounds the subregion treated with the multifrequency matched filter optimized to identify DSFGs. The grayscale indicates the noise level of the data (standard error) for the 148~GHz matched-filtered data, which yield a similar  sensitivity to the multifrequency filter (e.g., Figure \ref{fig:filtered_maps}). Diamonds (circles) show the locations of DSFGs (AGN). The size of the symbol indicates the flux density, at 148~GHz for AGN and at 218~GHz for DSFGs, with the scaling given at the lower right corner of the figure. There are additional fields with 148~GHz data centered at R.A. 8$^{\mathrm h}$ and 15$^{\mathrm h}$ (Figure \ref{fig:survey_fig_2}).}
\label{fig:survey_fig}
\end{figure*}

\begin{figure*}[t]
	\centering
	\includegraphics[width=5.5in]{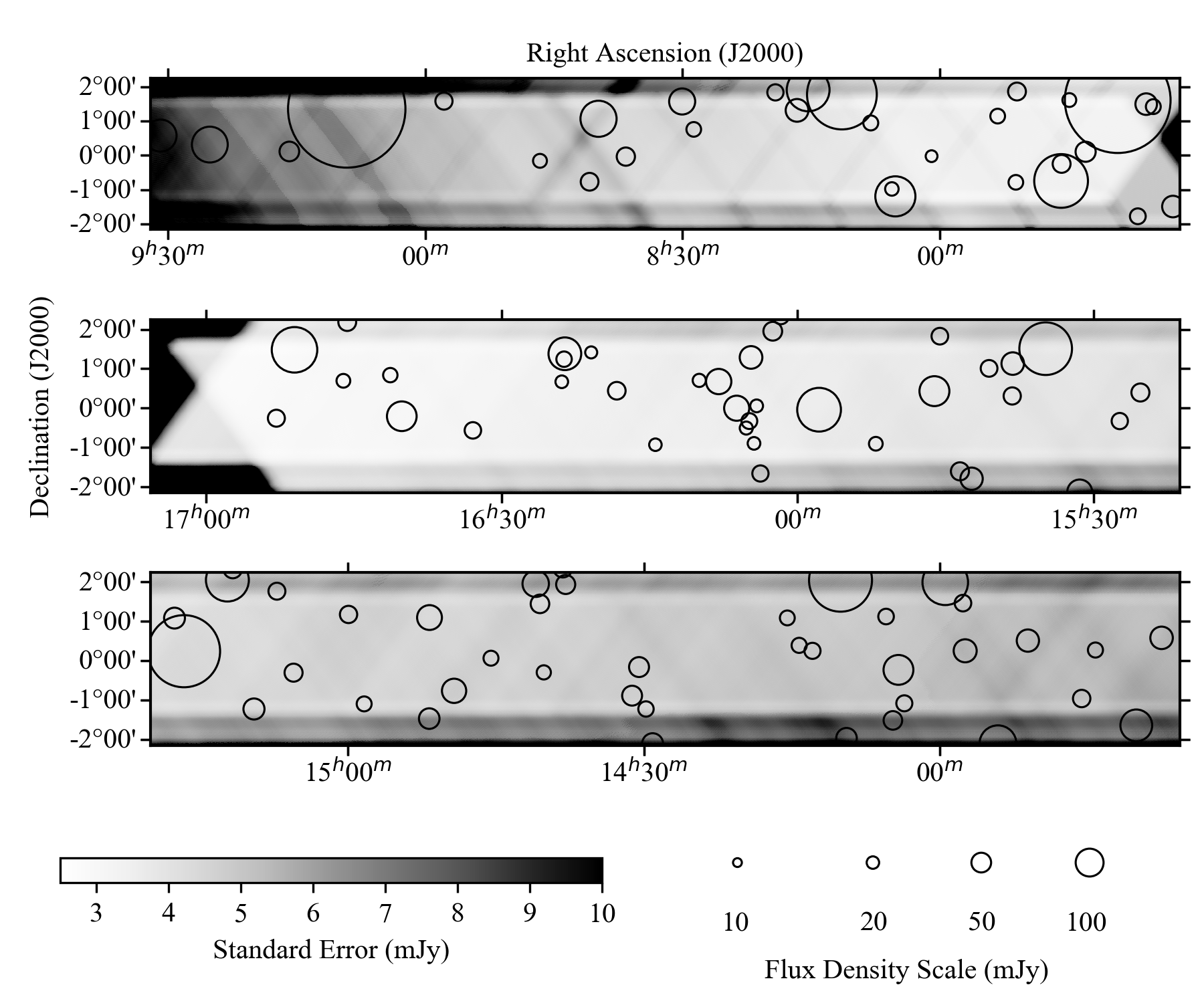}
\caption{ACT equatorial source survey auxiliary fields. Auxiliary survey areas, one centered at R.A.~8$^{\mathrm h}$ (top panel) and the other at 16$^{\mathrm h}$  (bottom two panels), provide extra data at 148~GHz. The sources are marked by circles with flux density indicated by the size of the circle. Although additional frequency bands are not available for spectral classification, at the flux density range probed by these data, the 148~GHz source population is dominated by AGN (see Figure \ref{fig:counts}). The grayscale shows that these fields are shallower than the main field (Figure \ref{fig:survey_fig}). }
\label{fig:survey_fig_2}
\end{figure*}

ACT is a 6~m telescope located in the Atacama Desert at an elevation of 5,190~m \citep{Fowler07}.  ACT's first receiver was the  Millimeter Bolometric Array Camera  \citep[MBAC; ][]{swetz2011}. Using the MBAC, ACT conducted surveys of the southern ($\delta=-52\degree12\arcmin$) and equatorial sky from 2008 to 2010.  For these, ACT observed at three frequencies simultaneously: 148~GHz (2.0~mm), 218~GHz (1.4~mm) and 277~GHz (1.1~mm) with angular resolutions of 1.4$'$, 1.0$'$ and 0.9$'$, respectively \citep{hasselfieldbeam}.  The telescope has since been upgraded with two successive generations of polarization sensitive receivers: ACTPol, described in \citealp{Thornton2016}, and Advanced ACTPol,  outlined in \citealp{Henderson15}.  (MBAC was not polarization sensitive.)  This paper presents point source catalogs from the MBAC-based equatorial survey.  The 148 and 218~GHz data are from the 2009 and 2010 observing seasons, and the 277~GHz data are from the 2010 observing season.   

Figure \ref{fig:survey_fig} shows the main ACT equatorial survey region used in this study.  This region covers approximately 480~$\deg^2$ on the celestial equator with R.A. centered at 0$^{\mathrm h}$ and spanning from 19$^{\mathrm h}$45$^{\mathrm m}$ to 4$^{\mathrm h}$16$^{\mathrm m}$, and with declination ranging from $-2^\circ12\arcmin$ to $2^\circ12\arcmin$. There are, however, notable differences in survey coverage between bands. The survey region for the 277~GHz band, derived from only one year of observations, is smaller than that of the lower frequency bands. For the 148~GHz band, sources were identified in additional equatorial regions centered on R.A. 8$^{\mathrm h}$ and 16$^{\mathrm h}$ (Figure \ref{fig:survey_fig_2}). These additional regions increase the survey area for 148~GHz by 360~$\deg^2$. Each map is produced with a cylindrical equal-area projection with its standard parallel at the equator, making a flat-sky approximation valid to well within errors across the narrow survey region. The square map pixels are $0.5'$ on a side at the equator. The maps used in this study are available on the Legacy Archive for Microwave Background Data Analysis\footnote{https://lambda.gsfc.nasa.gov/}. In particular, the 277~GHz dataset is newly released as of this publication.

For details about ACT observations and mapmaking procedures, see \citet{dunner2013}.  In summary, each detector timestream is analyzed and either retained or rejected based on a number of criteria (i.e., weather, detector performance, etc.).  A preconditioned conjugate gradient solver produces a maximum likelihood map from these timestream data.  We fit for an initial estimate of the point source signals.  Models constructed from these initial estimates are then subtracted from the timestream data, which are then processed into a new, noise-dominated maximum likelihood map.  The source models are then added back into the final map.  This two-step treatment of the map helps prevent source power from biasing noise estimates used to produce the final maps, which in turn prevents biasing the point source signal in the map solution.

The overall flux density calibration of the map for each band is determined from cross-correlation of the CMB power spectrum at $300 < \ell < 1100$ with that measured from the {\it Wilkinson Microwave Anisotropy Probe} (WMAP) using the deepest ACT maps (the 2010 season).  The uncertainty on the absolute temperature calibration of the 148~GHz band to WMAP is 2$\%$ \citep[at $\ell=700$;][]{sievers2013, hajian2011}. \citet{louis2014} cross-correlate the ACT maps with {\it Planck} maps and find excellent agreement.\footnote{The observed calibration factor between {\it WMAP} and {\it Planck} is 0.985, with {\it Planck} lower than {\it WMAP}.} The calibration is then transferred to the 218~GHz data through a cross-correlation with the 148~GHz map. Relative calibration between seasons is performed using cross-correlations between the data of each ACT season.  Details of this process can be found in \citet{das2013}. In addition to the overall calibration uncertainties, errors in the assumed instrument beam and the map-making can propagate to uncertainty in the recovered ACT flux densities. As discussed in \citet{gralla2014}, the overall systematic uncertainties on the flux density measurements for the 148~GHz and 218~GHz bands are 3\% and 5\%, respectively. These uncertainties dominate statistical uncertainties for the brightest sources in our sample. The calibration of the 277~GHz band  derives from observations of Uranus \citep{hasselfieldbeam}. Because this method is less accurate than the CMB-based calibration of the lower frequency bands, the systematic error on the 277~GHz fluxes is 15\%.

Part of the systematic flux density uncertainty is due to the fact that the telescope optical response depends on the source spectrum and whether the source is resolved (like the CMB) or point-like \citep[e.g.,][Table 4]{Page2003,swetz2011}. Publicly available ACT beams   assume a CMB source spectrum,  so their use in recovering point-source flux density is nuanced. In particular, the effective center of the bandpass for a given source depends on the convolution of its intrinsic spectrum with the instrument response over the band. As in \cite{Marsden2014}, we have scaled the beam used in the matched filter and solid angle used in flux recovery to partially account for a shifted effective central frequency (148.65~GHz, 218.6~GHz, and 277.4~GHz for the three ACT bands).  In \citet{datta}, 
a more detailed calculation was done to determine the range of flux density correction factors for different intrinsic source spectra. We have performed a similar analysis for the bands in this paper and estimated an associated systematic error in flux density recovery of $1.5$\%. 

The ACT sensitivity varies throughout the maps according to the depth of coverage.  
Each map was filtered with the beam appropriate for that season and band, as described below. The resulting calibrated, filtered maps were combined into a multi-season map via a weighted average, with the weights set for a given pixel by the integrated time that pixel was observed.  
Typical rms noise levels are 1.8, 2.4, and 5.2~mJy for 148, 218 and 277~GHz, respectively. Figure \ref{fig:survey_fig} shows the noise level across the main survey region for the 148~GHz band.

\section{Methods}
\label{sec:methods}

\subsection{Spatial matched filtering}
\label{sec:mf}

To optimize the signal-to-noise ratio (S/N) of the point-like sources, the ACT data were matched-filtered \citep[e.g.,][]{tegmark98} with the ACT beam \citep{hasselfieldbeam}. The methods used are described fully in \citet{marriagesources} and \citet{Marsden2014}, which present catalogs of sources from the ACT southern surveys.  Here we summarize the main steps of this analysis with an emphasis on unique features of these new catalogs.

The ACT brightness-temperature map is first multiplied by a weighting function $W(\bm{\theta})$ proportional to the square root of the number of observations per pixel to make the white noise per pixel approximately constant across the survey region.\footnote{Over the course of the season,  observations with high and low noise distribute  in similar ratios across the map, making the number of observations per pixel a good proxy for the inverse of the resulting noise variance. This has been confirmed empirically.} Because there are fewer observations towards the edge of the map, this  produces an inverse-noise-weighted map of brightness temperature $T(\bm{\theta })$ with a tapered window function and Fourier transform $\tilde T({\bf k})$.\footnote{${\bf k} = (k_x,k_y)$ is the angular wave vector, with $x$ and $y$ referring to right ascension and declination, respectively.} This map is then filtered in the Fourier domain to produce a (weighted) matched-filtered map $T_{\mathrm{MF}}(\bm{\theta })$:
\begin{equation}\label{eqn:MFC}
T_{\mathrm{MF}}(\bm{\theta }) = \int \exp({2\pi i{\bf k}} \cdot \bm{\theta }) \Phi_{\mathrm{MF}}({\bf k})\tilde T({\bf k})d{\bf k},
\end{equation}
where
\begin{equation}\label{eqn:MF}
\Phi_{\mathrm{MF}}({\bf k}) = \frac{F_{k_{0},k_{x}}({\bf k}) \tilde B^{*}({\bf k}) | \tilde T_{\mathrm{other}}({\bf k}) | ^{-2}}{\int \tilde B^{*}({\bf k'})F_{k_0,k_x}({\bf k'})|\tilde T_{\mathrm{other}}({\bf k'})|^{-2}\tilde B({\bf k'})d{\bf k'}}
\end{equation}
is the matched filter.  Angular features scale as $d\theta \approx \pi/{ k}$ (in radians); e.g.,  ${k}=3000$ corresponds to $3'-4'$ where sources begin to dominate over the CMB power \citep[e.g., see][]{sievers2013}. The function $\tilde B({\bf k)}$ (with units of steradian) is the Fourier transform of the ``effective'' instrument beam ($B(\bm{\theta})$, normalized such that $B(0)=1$), which takes into account the dependence of the beam on the source spectrum and telescope pointing jitter (Section \ref{effbeam}). The beam is well approximated as azimuthally symmetric ($\tilde B({\bf k)} \approx \tilde B(k)$). The Fourier transform of the map data excluding the point sources, $\tilde T_{\mathrm{other}}$, includes atmosphere, detector noise, the CMB, and any other sources of brightness temperature that represent noise for the source signal. Unlike $\tilde B({\bf k)}$, $\tilde T_{\mathrm{other}}$ is not azimuthally symmetric. As in \citet{Marsden2014}, in practice $\mid \tilde T_{\mathrm{other}}\mid^2$ is simply the  power spectrum of the $W(\bm{\theta})$-weighted temperature data, which approximates the noise sources given the low amount of power in the point sources.  A high-pass filter, $F_{k_{0},k_{x}}({\bf k})$, eliminates undersampled modes below $k_0=1000$ and modes with $|k_x|<100$, which are occasionally contaminated by telescope scan-synchronous noise. 

Bright point sources in the maps can cause ringing from the matched filtering, which can introduce spurious low-S/N sources.  We minimize this effect by initially searching for the brightest S/N $>50$ sources, cataloging these, and removing them from the original maps by filling in a $25\arcmin$ radius around each source position with a uniform flux density equal to the typical map noise. For the 148~GHz-selected sample, there are 41 such sources. We then refilter the maps without these bright sources, identifying all sources with S/N $>5$. The source candidates thus identified (both from the initial search and from the re-filtered maps) are passed to the next phase of the analysis in which source location and flux density are reconstructed. 

\begin{figure*}
	\centering
	\includegraphics[width=6.6in]{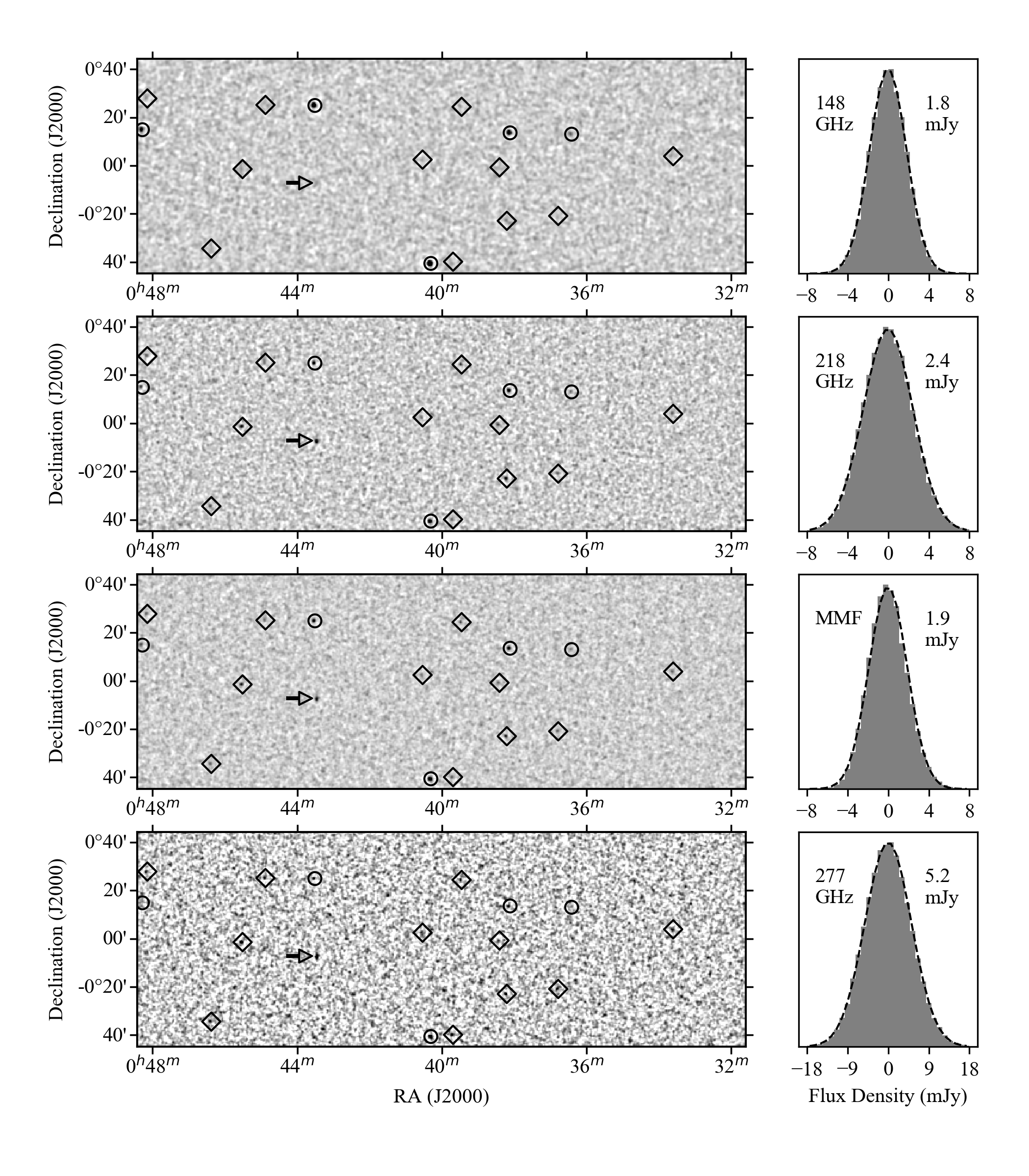}
\caption{Filtered ACT data. From top to bottom on the left, a subregion is shown for the matched-filtered maps corresponding to 148~GHz, 218~GHz, the MMF, and 277~GHz. The MMF is optimized for a thermal dust spectrum and is referenced to 218~GHz. The grayscale limits (white-to-black) of the filtered map are $-5$ and $20$~mJy. Sources detected with S/N$>$5 are marked with circles for synchrotron-dominated emission and diamonds for dust-dominated emission. At right, the pixel flux-density distribution functions are shown for each subregion with corresponding standard deviation values. The dashed line overplotted on each sample distribution is the Gaussian distribution with the  sample standard deviation. A bright dust-dominated source at $00^{\mathrm h}43^{\mathrm m}27.8^{\mathrm s}$ right ascension and $-0^\circ07'30''$ declination (marked by a right-pointing arrow and most apparent in the 218~GHz and MMF maps) was rejected by the cuts for Galactic contamination. This source is the nearby late-type galaxy NGC 237 at $z=0.014$. It was flagged for being more extended than 92\% of all point sources with evidence for rest frame CO(2--1) line emission. Other visible signals in these maps that are not marked are below the detection thresholds of any of the selection methods. See Table~\ref{tab:extragalactic_catalog} for catalog entries for the sources in this figure. }
\label{fig:filtered_maps}
\end{figure*}

Before estimating the flux density from a filtered map, the map is divided by $W(\bm{\theta})$ to undo the weighting applied for source detection and multiplied by the beam solid angle $\Omega_B$ to convert the brightness temperature to units of Jy/beam:
\begin{equation}
    S_{\mathrm{MF}}=W^{-1}(\bm{\theta})
    \frac{\partial I_{BB}}{\partial T} T_{\mathrm{MF}}(\bm{\theta})\Omega_B,
    \label{eqn:mffluxdensitymap}
\end{equation}
where $I_{BB}$ is the Planck intensity function. The partial derivative of $I_{BB}$ evaluated at the CMB temperature converts the map $T_{\mathrm{MF}}$ from brightness temperature to intensity units, which are converted to flux density by the factor of the solid angle. For each detection, we extract a $0.04\degree$-wide sub-map centered on the candidate source. In this submap, we remap the flux density from 0.5$\arcmin$ pixels  using Fourier interpolation (zero padding in {\bf k} space) into 16$\times$ smaller pixels ($\sim2''$ on a side). The effect of averaging the source signal into pixels (``pixel windowing'') is also corrected. The new position and flux density estimation is associated with the maximum in the filtered map, now with finer pixelization. Finding a more accurate position for each source and correcting for pixel windowing  is important for flux density recovery, especially for the higher-resolution 218~GHz and 277~GHz bands and for sources that do not lie near the centers of the larger pixels in the original, filtered map.
Examples of the matched-filtered data are shown in Figure~\ref{fig:filtered_maps}.

\subsubsection{Effective multi-season beams}
\label{effbeam}

The instrument beam transform $\tilde B({\bf k)}$ is measured separately for each observing band and season using observations of Saturn and Uranus \citep{hasselfieldbeam}. The spectral shape of these planets does not match that of most of the compact sources in this analysis, so the effective central frequency of the bands is shifted for each spectral shape (planets, CMB, AGN, DSFGs). As reported by \citet{swetz2011}, the central frequencies are 147.6, 217.6, and 274.8~GHz for AGN and 149.7, 219.6, and 277.4~GHz for DSFGs. To take into account that our source samples include both AGN and DSFGs, we adopt fiducial central frequency values of 148.65, 218.6, and 277.4~GHz. These correspond to frequencies halfway between the central frequencies for steep-spectrum AGN and DSFGs for the 148 and 218~GHz bands. For the 277~GHz band, we simply adopted the DSFG central frequency. The same 148 and 218~GHz central frequencies were used in our previous ACT compact source analysis \citep{Marsden2014}. We rescale the beam widths to account for the shifts in these new effective central band frequencies. For more information on the effects of the beam on the calibration uncertainty, see Section \ref{sec:data}.

The effective instrument beam is also broadened due to variations in pointing.  We include this broadening by multiplying the instantaneous beam transform by $\exp(-k^2 \sigma_{\theta}^2/2)$, where $\sigma_{\theta} = 5 \arcsec$ \citep{hasselfieldbeam}.

For 148 and 218~GHz, data from 2009 and 2010 were combined into single multi-season maps to increase the sensitivity to sources.  These maps were filtered with the 2009 beams, but the choice of beam does not have a large effect. The FWHM of the beam changed by $< 5\%$ from 2009 to 2010 for all bands. To investigate how this change affects flux density recovery, we simulated sources to have the shape of the 2010 beam and added these sources to the 2010 map, but then we filtered this map with the 2009 beam.  The flux densities recovered from this simulated map are lower than those input by 1\%.  Because in the analysis the actual source shapes are some combination of the 2009 and 2010 beams, the flux densities of the sources in the multi-season catalogs are thus only affected at the sub-percent level.  The brightest sources in the catalogs are typically blazars, and their emission varies from one season to the next by more than 1\% (Section \ref{subsec:variability}). The faintest sources' statistical uncertainties are larger than 1\%. Thus, the effect from filtering the combined maps with the 2009 beams is not significant.  Individual season flux densities, for which the 2010 map is filtered with the 2010 beam, are also reported in the catalogs available online.

\subsection{Multi-frequency matched filtering}
\label{sec:mmfdescription}

To improve sensitivity to sources beyond the spatial matched filtering described above, we use the multifrequency information available from ACT.  The methods we use extend beyond what has previously been done with ACT data. In this work, we specifically construct a multifrequency matched filter (MMF) \citep[e.g.,][]{melin2006} to search for faint, dusty galaxies. This choice is motivated by the availability of the  277~GHz data. Blazar detections are statistically dominated by the 148~GHz band, obviating the need for an MMF. However, both the 218~GHz and 277~GHz bands contribute significantly to the filtered DSFG signal, motivating the MMF for DSFG selection.

The MMF optimizes the S/N of a point source using data across multiple bands. As formulated in this work, the MMF produces a single map $T_{\mathrm{MMF}}(\bm{\theta })$ at a reference frequency $\nu_0$ from multiple maps $\{T_{\nu}(\bm{\theta })\}$ at multiple frequencies $\{\nu\}$. By analogy with the single-frequency matched filter (Equation \ref{eqn:MFC}), the MMF map is generated by applying a multi-component filter to the multi-frequency map set $\{T_{\nu}(\bm{\theta })\}$: 
\begin{equation}\label{eqn:MMF}
T_{\mathrm{MMF}}(\bm{\theta }) = \int \exp({2\pi i{\bf k}} \cdot \bm{\theta }) \sum_{\nu} \Psi_\nu({\bf k})T_\nu({\bf k})d{\bf k},
\end{equation}
where, in the fully general case, the filter functions $\Psi_\nu({\bf k})$ account for correlated noise between bands (e.g., correlations due to common modes from the CMB and atmospheric emission in ACT data). In principle, these correlations enter the formulation of the filter. In practice, for source selection at arcminute resolution, the strongly correlated modes at large angular scales can be neglected relative to the more abundant small-angular scale modes that are dominated by uncorrelated detector noise. The inter-band correlations from the CMB and the atmosphere dominate the uncorrelated detector noise for wave vectors ${k}<3000$, corresponding to angular scales $>3'-4'$. Given the arcminute resolution of the ACT 218~GHz and 277~GHz bands, approximately ten ACT beams fit within a correlated patch. In other words, there are ten times more beam-sized modes than modes for which CMB or atmosphere dominate.   The effects of correlations are further minimized by the extra high-pass filter $F_{k_{0},k_{x}}$ in Equation \ref{eqn:MF}, which de-weights the relatively few modes with $k<1000$. There remains residual correlated noise at high-$k$ (e.g., the unresolved cosmic infrared background) in the filtered ACT maps that presents a correlated noise source, but it is subdominant to detector noise. This is reflected by correlation coefficients: 0.13 between 148 and 218~GHz, 0.17 between 218 and 277~GHz, and 0.09 between 148 and 277~GHz. Thus significant gains in S/N are achieved with a simplified filter that assumes independent inter-band noise. 

With independent inter-band noise, the MMF map (Equation \ref{eqn:MMF}) takes a simple, intuitive form. Working in flux-density-per-beam units (Equation \ref{eqn:mffluxdensitymap}), the MMF map is the weighted combination of the single-frequency matched-filtered maps $\{S_{\mathrm{MF};\nu}\}$ corresponding to the linear-least-squares best estimate of the flux density at the reference frequency:
\begin{equation}
S_{\mathrm{MMF},\nu_0}(\bm{\theta }) = \sum_\nu w_\nu(\bm{\theta }) S_{\mathrm{MF};\nu}(\bm{\theta }),
\end{equation}
where the weights $w_\nu$ are
\begin{equation}
w_\nu(\bm{\theta }) = \sigma_{\mathrm{MMF}}^2(\bm{\theta }) \frac{f_\nu}{\sigma_\nu^2(\bm{\theta })}.
\label{eqn:MMF_simple_weight}
\end{equation}
In this equation, $\sigma_\nu^2(\bm{\theta })$ is the (position dependent) flux-density variance in the single-frequency matched-filtered map for frequency~$\nu$. The constant $f_\nu$ encodes the assumed spectral energy distribution (SED) of the source, relating the flux density $S_{\nu}$ at frequency~$\nu$ to the flux density $S_{\nu_0}$ at the reference frequency~$\nu_0$:
\begin{equation}
f_\nu = \frac{ S_\nu }{ S_{\nu_0} } =
\left(\frac{\nu }{\nu_0}\right)^\alpha.
\label{eqn:MMF_scale_factor}
\end{equation} 
Finally, $\sigma_{MMF}^2(\bm{\theta })$ is the variance in the MMF map:
\begin{equation}
\sigma_{MMF}^2(\bm{\theta })=\left(\sum_\nu \frac{f_\nu^2}{\sigma_\nu^2(\bm{\theta })} \right)^{-1}.
\label{eqn:MMF_variance}
\end{equation}
Therefore, given the dominance of independent noise between ACT frequency bands, we can use this simplified formalism, availing ourselves of the tools developed for the single-frequency matched-filter described in Section~\ref{sec:mf}.

For DSFGs, we take the 218~GHz map as the reference dataset and optimize the multi-frequency combination for a typical dusty source spectrum with spectral index $\alpha=3.77$.  Thus the frequency scale factors (Equation \ref{eqn:MMF_scale_factor}) are $f_{148} = 0.232$, $f_{218} = 1.00$ and $f_{277} = 2.47$.
The multifrequency map generated was restricted to the region between 
$20^{\mathrm h}09^{\mathrm m}$ and $03^{\mathrm h}51^{\mathrm m} $ in R.A. 
and between $-1\degr33\arcmin$ and $1\degr24\arcmin$ in declination, where there was adequate coverage in the 277~GHz data.
Filtered data in a subregion of the MMF area are shown in Figure \ref{fig:filtered_maps} for the three single-frequency maps and the MMF map. Equation \ref{eqn:MMF_variance} predicts a noise level in the MMF, referenced to 218~GHz, of 1.6~mJy, however the measured level is 1.9~mJy (compared to the 2.4 mJy noise in the single-frequency map).
The full sensitivity improvement is not achieved due to residual noise correlations between bands, which make Equation \ref{eqn:MMF_simple_weight} suboptimal. In the end, the cleanest comparison between the MMF and 218~GHz single frequency approach is in terms of S/N: for DSFGs in the MMF-derived catalog (Section \ref{sec:catalog}), the median improvement in S/N of the MMF over the 218~GHz data alone is \medianMMFToTwoSNRatio.

\subsection{Detection, selection, localization, and flux-density recovery}
\label{sec:detection_map}

In Section~\ref{sec:catalog}, we introduce a catalog in which sources are detected in multiple maps. As discussed in Section~\ref{sec:mf}, a source is detected in a map if its S/N~$>5$.  By this definition, a source may be detected in a combination of single-frequency filtered maps and the MMF map tuned for a dust-like spectrum. We further identify each source with the dataset in which it is detected with the highest S/N. This ``selection map''  defines the selection function and also provides the most precise location of the source. The dataset in which a source is selected also plays into  flux density debiasing (Section \ref{section:debiasing}).

To estimate flux densities and associated errors, the single-frequency filtered maps $T_{\mathrm{MF},\nu}$ are used. In a $4'$ patch centered on each source, the map $T_{\mathrm{MF},\nu}$ is reprojected with Fourier interpolation to ($\sim2''$) pixels, correcting for the signal-dilution effect of the pixel window function. Then the flux density is obtained from this finer-pixelized map at the selection-map-determined source location (i.e., by ``forced photometry''). Thus even if a source is undetected in a single-frequency map, it will still have an associated (low S/N) flux estimate. 

Note that if the selection map is itself a single-frequency map, then that map will be used to determine the source's selection function, location, and flux density. The MMF map is never used for flux density estimation, only source selection and localization. Conversely, due to its high noise, the 277~GHz map is never used for selection or localization.

\begin{table*}
\caption{A Sample of the Extragalactic Catalog\footnote{The entries in this catalog sample correspond to the sources found in the data shown in Figure \ref{fig:filtered_maps}. Only a subset of columns in the full catalog are shown.}}
\centering
\begin{tabular}{ccccccccc}
\hline \footnotesize
 ACT-S ID\footnote{The ACT-S ID encodes the sexagesimal position of each source (hhmmss$\pm$ddmmss).}  & S/N & Selection\footnote{The selection dataset is that in which the source is detected with the highest S/N (listed at left).} & $\alpha_{148}^{218}$ \footnote{For inter-band spectral indices ($\alpha_{X}^{Y}$) and flux densities ($S_X$), raw (debiased) values are given outside (inside) parentheses.} & $\alpha_{148}^{277}$/$\alpha_{218}^{277}$ \footnote{For 148 GHz-selected sources, we report $\alpha$ between 148~GHz and 277~GHz, whereas for MMF and 218~GHz selection, we report $\alpha$ between 218~GHz and 277~GHz.} & Type\footnote{The type of source is designated as an AGN (DSFG) if the 148$-$218~GHz spectral index is less than (greater than) unity as described in Section \ref{section:specindices}.} & $S_{148}$ & $S_{218}$ & $S_{277}$  \\
  (J2000) &  &  &  & &  & (mJy) & (mJy) & (mJy) \\\hline\hline
003337$+$000353 &  6.4 & MMF & $3.2$  & $3.6$  & DSFG& 3.5$\pm$1.8  & 12.2$\pm$2.8  & 28.7$\pm$5.4 \\ 
 & & &  ($3.0^{+1.6}_{-1.4}$) & ($3.5^{+1.4}_{-1.3}$) & &  (2.7$^{+1.6}_{-1.3}$) & (8.8$^{+3.7}_{-3.2}$) & (22.7$^{+5.3}_{-5.2}$)\\ 
003626$+$001301 &  6.4 & 148 & $0.2$ & $-0.6$ & AGN& 11.9$\pm$1.9  & 12.9$\pm$2.8  & 8.3$\pm$5.7 \\ 
 & & & ($0.1^{+0.7}_{-0.7}$) &  ($-1.1^{+1.2}_{-1.9}$) &&  (10.7$^{+2.2}_{-2.2}$) & (11.4$^{+2.6}_{-2.5}$) & (5.2$^{+5.6}_{-3.6}$)\\ 
003648$-$002052 &  5.9 & MMF & $4.1$  & $3.5$  & DSFG& 2.3$\pm$1.8  & 11.7$\pm$2.8  & 26.6$\pm$5.6 \\ 
 & & &  ($3.5^{+1.7}_{-1.5}$) & ($3.3^{+1.5}_{-1.4}$) & &  (2.0$^{+1.4}_{-1.0}$) & (8.4$^{+3.6}_{-3.0}$) & (20.3$^{+5.3}_{-5.2}$)\\ 
003808$+$001334 &  20.3 & 148 & $-0.3$ & $-0.7$ & AGN& 37.4$\pm$1.8  & 33.3$\pm$2.7  & 23.4$\pm$5.8 \\ 
 & & & ($-0.3^{+0.3}_{-0.3}$) &  ($-0.8^{+0.4}_{-0.5}$) &&  (37.1$^{+2.5}_{-2.5}$) & (32.9$^{+2.7}_{-2.7}$) & (21.8$^{+6.0}_{-6.0}$)\\ 
003814$-$002255 &  10.7 & MMF & $3.0$  & $1.9$  & DSFG& 7.6$\pm$1.8  & 24.6$\pm$2.7  & 39.1$\pm$5.5 \\ 
 & & &  ($3.0^{+0.8}_{-0.7}$) & ($2.0^{+0.9}_{-0.8}$) & &  (7.1$^{+1.8}_{-1.7}$) & (23.5$^{+2.7}_{-3.8}$) & (36.9$^{+5.3}_{-5.3}$)\\ 
003826$-$000044 &  6.0 & MMF & $4.4$  & $1.0$  & DSFG& 2.9$\pm$1.8  & 15.7$\pm$2.7  & 20.2$\pm$5.5 \\ 
 & & &  ($3.7^{+1.5}_{-1.4}$) & ($1.7^{+1.6}_{-1.5}$) & &  (2.6$^{+1.5}_{-1.2}$) & (11.8$^{+3.9}_{-4.7}$) & (15.3$^{+4.8}_{-4.6}$)\\ 
003929$+$002422 &  8.9 & MMF & $3.2$  & $2.1$  & DSFG& 5.7$\pm$1.8  & 20.0$\pm$2.7  & 33.4$\pm$5.5 \\ 
 & & &  ($3.1^{+1.1}_{-1.0}$) & ($2.2^{+1.1}_{-1.0}$) & &  (5.1$^{+1.8}_{-1.6}$) & (17.7$^{+3.9}_{-4.5}$) & (29.9$^{+5.3}_{-5.2}$)\\ 
003943$-$003952 &  6.6 & MMF & $4.4$  & $4.0$  & DSFG& 2.2$\pm$1.8  & 12.1$\pm$2.7  & 31.8$\pm$5.5 \\ 
 & & &  ($3.6^{+1.7}_{-1.5}$) & ($3.9^{+1.4}_{-1.2}$) & &  (2.0$^{+1.4}_{-1.0}$) & (8.6$^{+3.7}_{-3.1}$) & (26.2$^{+5.4}_{-5.3}$)\\ 
004020$-$004035 &  37.1 & 148 & $-0.8$ & $-0.5$ & AGN& 68.5$\pm$1.8  & 49.4$\pm$2.8  & 51.2$\pm$5.8 \\ 
 & & & ($-$)\footnote{Debiasing is not computed for 148~GHz-selected sources with $S_{148}>50$~mJy. Therefore debiased spectral indices are not provided. For ease of catalog use, raw flux densities are reported in the debiased flux density columns. For this class of bright source, the raw and debiased estimates of flux densities are equivalent.} &  ($-$) &&  (68.5$^{+1.8}_{-1.8}$) & (49.4$^{+2.8}_{-2.8}$) & (51.2$^{+5.8}_{-5.8}$)\\ 
004033$+$000228 &  5.6 & MMF & $2.7$  & $2.7$  & DSFG& 4.1$\pm$1.8  & 11.7$\pm$2.7  & 22.0$\pm$5.5 \\ 
 & & &  ($2.6^{+1.5}_{-1.3}$) & ($2.6^{+1.5}_{-1.5}$) & &  (3.1$^{+1.7}_{-1.5}$) & (8.4$^{+3.6}_{-3.0}$) & (15.7$^{+5.0}_{-4.7}$)\\ 
004332$+$002456 &  24.5 & 148 & $-0.8$ & $-0.3$ & AGN& 45.1$\pm$1.8  & 32.8$\pm$2.7  & 37.6$\pm$5.8 \\ 
 & & & ($-0.8^{+0.3}_{-0.3}$) &  ($-0.3^{+0.3}_{-0.3}$) &&  (44.7$^{+2.6}_{-2.5}$) & (32.6$^{+2.7}_{-2.7}$) & (36.6$^{+5.8}_{-5.7}$)\\ 
004454$+$002509 &  5.8 & MMF & $4.7$  & $4.4$  & DSFG& 1.6$\pm$1.8  & 10.1$\pm$2.7  & 28.7$\pm$5.5 \\ 
 & & &  ($3.8^{+1.7}_{-1.5}$) & ($3.7^{+1.4}_{-1.3}$) & &  (1.7$^{+1.2}_{-0.9}$) & (7.8$^{+3.4}_{-2.6}$) & (22.3$^{+5.4}_{-5.2}$)\\ 
004532$-$000127 &  11.1 & MMF & $3.8$  & $2.1$  & DSFG& 5.9$\pm$1.8  & 25.9$\pm$2.8  & 42.4$\pm$5.6 \\ 
 & & &  ($3.8^{+0.9}_{-0.8}$) & ($2.1^{+0.9}_{-0.8}$) & &  (5.6$^{+1.7}_{-1.7}$) & (25.0$^{+3.5}_{-4.3}$) & (40.7$^{+5.4}_{-5.3}$)\\ 
004624$-$003424 &  6.4 & MMF & $4.7$  & $2.1$  & DSFG& 2.4$\pm$1.8  & 15.2$\pm$2.8  & 25.1$\pm$5.5 \\ 
 & & &  ($3.8^{+1.6}_{-1.4}$) & ($2.4^{+1.5}_{-1.4}$) & &  (2.2$^{+1.5}_{-1.1}$) & (11.1$^{+4.0}_{-4.6}$) & (19.7$^{+5.1}_{-5.0}$)\\ 
004810$+$002750 &  6.3 & MMF & $1.8$  & $2.3$  & DSFG& 6.7$\pm$1.8  & 13.3$\pm$2.8  & 23.1$\pm$5.5 \\ 
 & & &  ($1.6^{+1.1}_{-1.1}$) & ($2.5^{+1.5}_{-1.5}$) & &  (5.7$^{+1.8}_{-1.9}$) & (9.5$^{+3.9}_{-3.9}$) & (17.2$^{+5.1}_{-4.9}$)\\ 
004818$+$001452 &  13.7 & 148 & $-1.1$ & $-2.1$ & AGN& 25.4$\pm$1.9  & 16.9$\pm$2.8  & 6.6$\pm$5.8 \\ 
 & & & ($-1.0^{+0.4}_{-0.5}$) &  ($-2.4^{+1.0}_{-1.4}$) &&  (24.6$^{+2.6}_{-2.6}$) & (16.7$^{+2.6}_{-2.5}$) & (5.5$^{+5.0}_{-3.3}$)\\ 

\hline
\end{tabular}
\label{tab:extragalactic_catalog}
\end{table*}

\section{Catalogs}
\label{sec:catalog}

\subsection{Overview}
\label{sec:catalog_overview}
We present the ACT equatorial extragalactic and Galactic source catalogs based on detections in three different maps.\footnote{Catalogs are available for download at the NASA Legacy Archive for Microwave Background Data Analysis (https://lambda.gsfc.nasa.gov/).} As discussed in Section~\ref{sec:detection_map}, sources are said to be detected in a map if they have a S/N~$>5$. The source is said to be ``selected'' in the map in which it has the highest S/N. Thus each source is uniquely identified with one of the following groups.
\begin{itemize}
\item The first subset of sources is selected from the single-frequency matched-filtered 148~GHz data. The 148~GHz selection is typically most sensitive for sources with flat or falling spectra such as AGN. Finally, we note that the 148~GHz selection includes the auxiliary fields shown in Figure~\ref{fig:survey_fig_2}, and that the sources in these fields lack coverage in the other ACT bands.
\item The second group of sources is selected using a three-frequency MMF map. As discussed in Section~\ref{sec:mmfdescription}, the MMF is optimized for sources with a dust SED, and the resulting map is limited by the footprint of the 277~GHz map, which is smaller than the 218~GHz coverage. Informed by the simulations of catalog completeness (Section~\ref{section:debiasing}), we further limit our MMF selection to the most sensitive part of the MMF map, within the declination range  $-1\degr12\arcmin$ to $1\degr12\arcmin$.
\item The third source selection is based on the single-frequency matched-filtered 218~GHz data outside the sky region used for MMF selection. Like the MMF map, the 218~GHz data provide more sensitivity to DSFGs than the 148~GHz data. 
\end{itemize}

Sources in the equatorial catalog are tested for Galactic dust or CO emission according to automated methods (Section~\ref{section:dustremoval}), which flag \ngalactic~detections as being likely of Galactic origin. These sources are provided in the Galactic catalog. Because the Galactic contamination cuts rely on multifrequency data and mainly affect the MMF and 218~GHz selection, all flagged sources are in the main survey field (Figure~\ref{fig:survey_fig}). Excluding these sources, the total number of sources in the extragalactic catalog is \nmaster. Of these, {\masNauxfields}~sources lie in the auxiliary fields (Figure~\ref{fig:survey_fig_2}) with only 148~GHz data. In terms of selection, the 148~GHz, MMF, and 218~GHz methods described above yield \masNone, \masNmmf, and \masNtwo~extragalactic sources, respectively.  

Based on their spectral properties (Section~\ref{section:specindices}), \masNdust~sources are labeled in the catalog as DSFGs, and \masNagn~sources are identified as AGN. Of the \masNdust~DSFGs, \nDSFGmmf~sources were MMF-selected, and \nDSFGtwo~sources were 218~GHz-selected. No sources classified as DSFGs were 148~GHz-selected. Analogously, only 5\% of the AGN-classified sources were selected in the 218~GHz (\nAGNtwo) and MMF (\nAGNmmf) maps. The vast majority (\nAGNone) of the \masNagn~AGN are 148~GHz-selected. 

The remaining sources lack the requisite multifrequency information for spectral classification. They  are flagged in the catalog as AGN* (\masNagnstar~sources; mostly from auxiliary fields) or DSFG* (\masNduststar~source) based on whether they are detected only at 148~GHz or 218~GHz. (MMF-selected sources, by definition, have enough multifrequency data for spectral classification.)

To identify DSFGs that have nearby dusty galaxy counterparts, we visually inspected 50$\arcsec$ cutouts from the Sloan Digital Sky Survey \citep[SDSS;][]{Alam15} of all sources in the sample, including those our automated algorithm identified as Galactic. There is an additional flag in the catalog for these nearby galaxies, and their ACT spectral index distribution is discussed in Section~\ref{section:specindices}. We additionally flag three sources that may correspond to stars with associated radio or far infrared emission. In total, \masNearby~sources in the extragalactic catalog are flagged as likely local galaxies or stars, and \galNearby~sources in the Galactic catalog are flagged. An example of this is shown in Figure~\ref{fig:filtered_maps} for NGC~237 at $z=0.014$.

Source data provided in the extragalactic catalog include source ID, position and S/N based on the selection map, raw and debiased flux densities, raw and debiased spectral indices, AGN/DSFG classification,  statistics associated with Galactic cuts, and the local galaxy flag.\footnote{Debiasing is not performed for 148~GHz-selected sources above 50~mJy where the effect is negligible. In these cases the debiased flux density is not left blank in the catalog but reported as the (equivalent) raw flux density for ease of catalog use.} For 148~GHz-selected sources in the extragalactic catalog, we additionally provide raw flux density estimates from data in 2009 and 2010 separately. All 148~GHz-selected sources are AGN, and the per-year flux densities give a handle on AGN variablility (Section \ref{subsec:variability}). In the Galactic catalog we provide the same data as in the extragalactic catalog excluding debiased flux densities and spectral indices and per-year flux density estimates.   A sample of the extragalactic catalog for sources shown in Figure~\ref{fig:filtered_maps} is given in Table~\ref{tab:extragalactic_catalog}.

\subsection{Removing dusty Galactic emission}\label{section:dustremoval}

In addition to the extragalactic sources of interest in this study, ACT also detects Galactic dust emission, particularly in the 218 and 277~GHz data. Galactic CO emission may also contaminate the 218~GHz band. There is more Galactic contamination at the eastern and western edges of the maps (RA $> $~03$^{\mathrm h}$45$^{\mathrm m}$ and R.A. $< 20^{\mathrm h}15^{\mathrm m}$), where large Galactic structures are clearly visible.  The source finding algorithm, although optimized for point sources, also identifies bright regions of extended emission as sources.  To distinguish true extragalactic sources from the Galactic emission, we characterize the sources by their shapes, SED, and clustering properties. We also use information from {\it Planck} to identify regions of the sky with high Galactic dust emission.

With the exception of nearby galaxies (e.g., NGC~1055), extragalactic sources tend to be unresolved by the ACT beams (FWHM $0.9\arcmin$ to $1.4\arcmin$).  We characterize the shapes of the sources in the following way.  Starting with the filtered detection map, for each source we create a submap with upgraded pixel resolution ($\sim2''$) and Fourier interpolation (zero padding) of the data. In this submap we identify the pixel with the peak brightness and normalize all pixels in the submap by that value. The resulting submap has pixel values that are fractions of the peak value. We then sum the normalized pixel values in a 2.5$\arcmin$ square around that peak, calling the result the ``extended index.''  By this definition, the extended index is a simple metric that increases with the solid angle subtended by the source (albeit in a filtered map). We note that although the extended index is defined within 2.5$\arcmin$, sources that are even larger are also likely to have a large extended index because they are unlikely to have pixel distributions as concentrated as the beam shape within the 2.5$\arcmin$ submap. 

Because of its spectral shape and extended morphology, Galactic dust is not a strong contaminant for low-frequency, high-resolution interferometric radio surveys.  Thus, ACT sources with counterparts in the VLA FIRST survey are more likely to be extragalactic than Galactic.  We calculate the cumulative distribution of the extended indices of sources with FIRST counterparts.  This cumulative distribution defines a scale for judging the extended index.  For each source we define the ``extended percentile'' as the fraction of ACT sources with FIRST counterparts that are more compact (smaller extended index) than that source.  For example, an ACT source with an extended percentile of 0.98 has a larger extended index than 98\% of the FIRST sources in the map. 

\begin{figure*}
	\centering
	\includegraphics{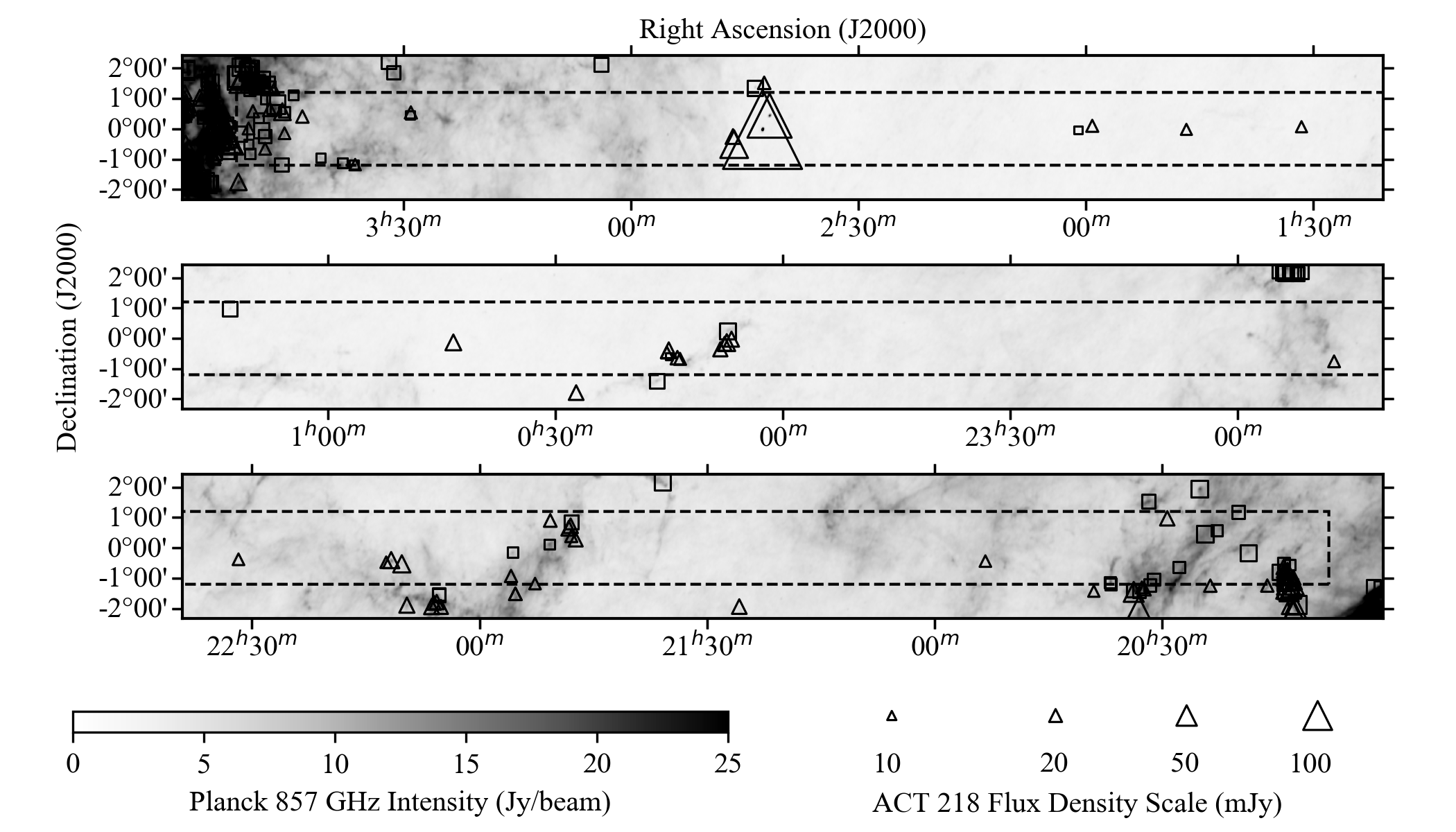}
	\caption{Source detections flagged by Galactic contamination cuts. Sources were selected as potential Galactic contamination based on combinations of criteria (Section \ref{section:dustremoval}) including compactness, clustering, CO(2--1) emission, and location in regions of high {\it Planck} 857~GHz intensity (shown in grayscale). Sources with evidence of CO contamination are shown as triangles whereas other sources, generally (but not exclusively) with dust-like spectra, are shown as squares. A dashed rectangle bounds the subregion treated with the multifrequency matched filter optimized to identify DSFGs. The sizes of the symbols indicate the ACT 218~GHz flux densities. The \ngalactic~sources flagged as Galactic are not included in the extragalactic catalog (Figures \ref{fig:survey_fig} and \ref{fig:survey_fig_2}), but are provided in a separate catalog. }
\label{fig:galactic_map}
\end{figure*}

Because extended Galactic emission is often resolved into clustered groups of sources, many of the Galactic sources have multiple neighbors nearby.  For each source, we compute the number of neighboring sources within a 0.5$\degree$ radius.
To contextualize this metric, we calculated the number of neighbors for random positions throughout the map.   The correlation lengths of lensed DSFGs and AGN are long enough that a random position approximation holds for this purpose. The fraction of randomly located sources with more than 4 neighbors is only \fracNeighborFourSim. On the other hand, the fraction of detected sources with more than four neighbors is \fracNeighborFourAllData, motivating the cuts for contamination described below.

We also investigated the spectral properties of the sources, particularly in the region that has information in all three frequency bands.  Details of the spectral index distributions of the ACT sources are presented in Section \ref{section:specindices}.  Galactic dust contamination is expected to have a dust-like spectrum ($\alpha > 3$).  Additionally, we find  that many of the sources identified as clearly Galactic dust contamination have excess emission in the 218~GHz band.  Thus, their spectral indices from 148 to 218~GHz are positive (and significantly larger than $\alpha = $~3 or 4 characteristic of dust), while their spectral indices from 218 to 277~GHz are negative.  This excess emission is likely caused by bright spectral line emission from CO(2--1) at 230.538~GHz. The ACT band extends from a half-power point of  210~GHz to a half-power point of  230~GHz \citep[see Figure 9 in ][]{swetz2011}, with the response being a steep function of frequency.  
As discussed below, we only use this spectral information in conjunction with other measures (morphology and clustering) to flag sources as Galactic. Thus, compact, isolated extragalactic sources that may have unusual spectra remain in the catalog. 

We visually inspected the sources in the map to develop criteria for removing Galactic dust contamination based on the measurements described above.  These choices are informed by the properties of the clearly Galactic sources in dust complexes such as those at R.A. 03h50m, 00h11m and 20h13m. We apply four different cuts for Galactic contamination based on the MMF or 218~GHz data. (1)
We remove as Galactic contamination all MMF and 218~GHz detected sources with both $\alpha_{148-218} > 1$ and $\alpha_{218-277} < 1$ (a proxy for CO contamination) that also have either more than 4 neighbors \emph{or} have extended percentiles greater than 0.8. There are \galNcutOne~sources that are flagged by this cut. (2) We also remove all MMF and 218~GHz detections that have more than four neighbors \emph{and} extended percentiles greater than 0.8. This cut flags \galNcutTwo~sources. It removes an additional \galNcutTwoNotOne\ beyond the initial cut. (3) For sources that lie within areas of the {\it Planck} 857 GHz map with intensity above 6 MJy/sr (as determined from an adjacent pixel to reduce bias from the source itself), we apply a more restrictive cut that removes sources with MMF or 218~GHz-based extended percentile greater than 0.5 \emph{or} that have more than four neighbors and evidence for CO.  This third set of criteria flags \galNcutThree~sources. It removes an additional \galNcutThreeEx\ sources beyond the first two cuts. (4) We flag sources as Galactic if they are extreme in terms of number of neighbors ($\ge 10$) or extended percentile ($>0.995$). This identifies an additional \galNcutFive~sources as Galactic. In addition to these cuts based on 218~GHz or MMF data, we remove sources detected \emph{only} at 148~GHz that lie within $0.12\degr$ of sources that are detected at either 218~GHz or in the MMF map and are identified as Galactic by the above criteria or that have a 148-218~GHz spectral index $> 2$. This cut removes an additional \galNcutFour\ sources. In all, \ngalactic~sources are flagged as Galactic contamination. These sources are shown superposed on the {\it Planck} 857~GHz data in Figure~\ref{fig:galactic_map}. Among the sources flagged as Galactic, the 148~GHz, MMF, and 218~GHz selection methods (Section~\ref{sec:catalog}) are associated with \galNOne, \galNmmf, and \galNTwo~sources, respectively. The fact that the majority of sources flagged as Galactic originate with the 218~GHz selection follows from the facts that Galactic cirrus is faint at 148~GHz and that many of the cirrus complexes seen in Figure~\ref{fig:galactic_map} fall outside the MMF region. Additionally, the MMF selection is less susceptible to CO emission, because it does not rely solely on the contaminated 218~GHz band. Of the \ngalactic~flagged sources, a majority, \galNCO, show evidence (according to the criterion in cut (1)), of CO contamination.

The Galactic contamination cuts that we have implemented need to be aggressive enough to ensure purity of the DSFG sample, but also surgical enough to leave the vast majority of DSFGs. For instance, after the cuts, the fraction of sources in the extragalactic catalog with more than four neighbors within 0.5$^\circ$ is reduced to \fracNeighborFourMaster~from~\fracNeighborFourAllData~ (compared to \fracNeighborFourSim~for random locations). Similarly, before the cuts, the fraction of sources with an extended percentile exceeding 0.95 is \fracExtendedNinetyFiveAllData, significantly higher than the 0.05 fraction expected if all sources were point-like. After the Galactic contamination cuts, the fraction of sources with extended percentile exceeding 0.95 is reduced to \fracExtendedNinetyFiveMaster. To further check that Galactic cuts are effective, we investigated the effect of the cuts on the number counts of the remaining sources.  As discussed in Section \ref{sec:countDescription}, we calculated the number counts separately for regions of the map that are relatively free of Galactic dust emission and for those that underwent significant cleaning (in which one might expect extra false positives if the Galactic cuts were inadequate). 
We found that the number counts are statistically consistent between these different regions.

Since we have erred on the side of caution with the Galactic contamination cuts, some extragalactic systems will be flagged. For instance, nearby star-forming galaxies (such as NGC~237 in Figure \ref{fig:filtered_maps}) may be cut due to significant CO(2$-$1) emission in the 218~GHz band \citep[e.g., Figure 3 of ][]{gralla2014} and extended brightness profiles. And while sources at higher redshift will neither be extended nor suffer significant CO contamination, they may be removed by chance. Below in Figure~\ref{fig:completeness}, we nevertheless show that we  flagged and removed unresolved extragalactic sources at the few-percent level, based on how many randomly distributed simulated sources are flagged. Summaries of the extragalactic and Galactic catalogs are given in Table \ref{tab:catalog}.

\begin{table}
\centering
\caption{ACT Equatorial Catalog Summary}
\begin{tabular}{ccc}
\hline
   & Extragalactic  & Galactic  \\
   & Catalog\footnote{The full source sample is divided into extragalactic and Galactic catalogs based on cuts for Galactic contamination (Section \ref{section:dustremoval}). } & Catalog \\\hline\hline
  Total Detections & \nmaster & \ngalactic \\\hline
  148~GHz Selection\footnote{Selection methods correspond to the map in which a source has the highest significance (Section \ref{sec:catalog}).} &\masNone & \galNOne \\
  MMF Selection & \masNmmf & \galNmmf \\
  218~GHz Selection & \masNtwo & \galNTwo \\\hline
  In Auxiliary Fields\footnote{Auxiliary fields, with only 148~GHz data, are shown in Figure~\ref{fig:survey_fig_2}.}  & \masNauxfields & N/A \\\hline
  DSFG\footnote{Distinction between DSFG and AGN is based on spectral information (Section \ref{section:specindices}) whereas DSFG* and AGN* denote sources lacking spectral information but sorted into source category based on the detection map (most are from auxiliary fields). Typical source spectral indices are listed in Table \ref{tab:alpha}.} & \masNdust & N/A \\
  AGN  & \masNagn & N/A \\
  DSFG* & \masNduststar & N/A \\
  AGN* & \masNagnstar & N/A \\
  CO\footnote{CO indicates a spectrum in sources flagged as Galactic that is indicative of CO(2--1) emission (Section \ref{section:dustremoval}).}   & N/A & \galNCO\\\hline
  Nearby Galaxy/Star\footnote{This flag denotes sources that lie within 50$''$ of a nearby ($z<.1$) galaxy in SDSS or by one of three radio/far-infrared bright stars.} & \masNearby & \galNearby \\
\hline
\end{tabular}
\label{tab:catalog}
\end{table}

\subsection{Debiasing and Completeness} 
\label{section:debiasing}

\subsubsection{Debiasing method and description of simulations}
\label{sec:debiasing_method}
The flux densities of sources recovered from a population with steeply falling number counts are biased high relative to their intrinsic flux densities. In the sub/millimeter wavelength community, correcting for this bias is referred to as ``deboosting'' \citep[e.g.,][]{Coppin2005}.  Motivated in part by the presence of Galactic dust in these equatorial maps and the methods we developed to remove them from the catalog, we developed a new method to perform this deboosting based on injecting simulated sources into the maps. We refer to our methods as ``debiasing'' rather than deboosting to distinguish our treatment, which can correct flux densities either up or down, from traditional deboosting. We describe in detail these methods in an accompanying paper \citep{gralladeboosting}. In summary, we adopt a similar formalism as outlined in \citet{crawford2010} (hereafter C10), which describes a Bayesian approach to determining the intrinsic flux density of the brightest source within a given pixel. However, we marry the analytic approach of C10 with  simulations to better account for our source selection function.  

To debias the source flux densities of the catalogs presented in this work, three sets of simulations were generated: one set for 148 GHz-selected sources, and two sets for sources selected in the 218~GHz and MMF maps. Each simulation set is composed of a number of ``trials'', and each trial corresponds to 1000 simulated sources injected into the filtered maps. In this way we straightforwardly capture the effects of noise and contamination in the data itself. Care must be taken in constructing the simulations to ensure adequate statistics in the relevant regimes of flux density and spectral behavior to describe the source population. 

To generate the simulations we begin with the Fourier transform of a filtered point source centered on the survey map. We multiply this transform by the appropriate phase function to shift its location in angular space. We repeat this process and accumulate in Fourier space all the sources for the simulation. We then apply an inverse Fourier transform to produce a map (a ``signal template'') with the simulated filtered sources (and nothing else). This template is then added to the filtered data. We developed this procedure as an efficient way to create a signal templates with simulated sources located with sub-pixel accuracy. 

For the 11 trials of 148~GHz simulations, flux densities were selected from a uniform random distribution ranging from 0 to 50~mJy. This range was chosen to extend below the completeness limit of the survey and up to a flux density at which we expect the completeness to be close to 1.0 and the debiasing to be small. 

For the 18 trials of faint 218~GHz and MMF simulations, 218~GHz flux densities were selected from a random normal distribution with mean 10~mJy and standard deviation 5~mJy.\footnote{As noted later in the text, the flux density distribution used for simulated sources is divided out in the final estimation of the likelihood function. The form of the distribution is chosen to provide enough statistical weight to simulate the ACT sample. But this choice does not impact the likelihood if chosen properly.} For MMF simulations, these were then assigned spectral indices randomly selected from a normal distribution with mean 3.4 and standard deviation 1.3, which are the parameters of the Gaussian that best fits the distribution of the DSFGs' $\alpha_{148-218}$ values (listed in Table \ref{tab:alpha}). The flux densities for 148 and 277~GHz were calculated from these, and each band's simulated filtered sources were then added to the filtered data for that band. (As discussed in Section~\ref{sec:catalog_overview}, 92\% of MMF and 218~GHz-selected sources are DSFGs, so this choice of spectral index distribution is appropriate). For sources with measured 218~GHz flux density above 30~mJy, we used a set of 9 trials of simulations populated uniformly in flux density in the range $0-100$~mJy with spectral indices distributed according to the same Gaussian distribution used to debias the faint sample.
For the MMF simulations, the maps for each of the bands were combined in the same way as outlined in Section~\ref{sec:mmfdescription}.  The full source recovery procedures (single band and MMF) were then performed on the associated trials, including the removal of Galactic contamination (Section~\ref{section:dustremoval}).   

As described in Sections~\ref{sec:detection_map} and \ref{sec:catalog_overview}, each source is identified as ``selected'' in the map (148~GHz, 218~GHz, or MMF) in which the source is detected with the highest S/N. Furthermore, for each source we identify a ``primary band'' for debiasing. In the case of 148~GHz and 218~GHz selection, the primary band is simply the corresponding frequency band. For the MMF, however, there is no one-to-one correspondence with a frequency band. We therefore choose 218~GHz as the primary band for debiasing sources selected with the MMF, because this band has similar or better DSFG sensitivity compared to 277~GHz, and it has better calibration. This choice of primary band also makes the MMF-selected debiasing closer to the treatment for 218~GHz-selected debiasing. Other bands are referred to as ``secondary bands''. For instance, the secondary bands associated with MMF selection would be 148~GHz and 277~GHz.

For primary-band debiasing, we use the simulations to estimate the likelihood function of the intrinsic flux density ($S_1$) of the brightest source in a resolution element, $\mathcal{L}(S_1) = P(S_1^m \mid S_1)$, where the superscript $m$ indicates a measured quantity. This likelihood is corrected for the input flux distribution of the simulations. It captures the effects of the detector and confusion noise and selection process on the distribution of recovered flux density. We then multiply the likelihood by an analytic prior $P(S_1)$ that accounts for the expected distribution of source counts (Equation 2 of \cite{gralladeboosting}). This approach has the advantage of allowing changes to the analytic prior  without the need for more simulations. The result is the primary band flux density posterior $P(S_1 \mid S_1^m) \propto \mathcal{L}(S_1)P(S_1)$. We report the median and 16\% and 84\% quantiles of this posterior as the primary band debiased flux.

We take a simplified approach to multi-band debiasing. As shown by C10 (Figure 2), debiasing of robust primary band detections does not significantly benefit from information in other bands, particularly from noisier secondary bands.  Therefore we do not re-estimate the primary band debiasing in the multi-band process, but use  the primary band posterior previously computed to constrain the secondary band flux densities and associated spectral indices. In secondary-band debiasing, we use the low level of noise correlation between bands (Section \ref{sec:mmfdescription}) to decompose the two dimensional likelihood into independent functions of $S_1^m$ and $S_2^m$: $P(S_1^m,S_2^m \mid S_1,S_2) \approx P(S_1^m \mid S_1) P(S_2^m \mid S_2)$. For the MMF selection when the secondary band is 277~GHz, the secondary band likelihood ${P(S_2^m \mid S_2)}$ is approximated as a Gaussian distribution centered on the simulated likelihood and with a width set by the error on the raw flux density. This is needed to capture the selection effects of the MMF. For all other selection method-secondary band combinations, we simply approximate the likelihood as the Gaussian likelihood given the raw flux density and error. To formulate the prior, as in C10 we work in the parameter space of primary-band flux density and spectral index. We then factor the prior: $P(S_1,\alpha)=P(\alpha \mid S_1) P(S_1)$. The conditional probability $P(\alpha \mid S_1)$ is the normalized sum of broad spectral index distributions for AGN and DSFGs (Table \ref{tab:alpha}, Row 8), weighted by the number of expected sources of each type, as established by count models for $S_1$. $P(S_1)$ is the same analytic counts-based prior used in the primary-band debiasing. With these ingredients, we can construct the posterior distribution, also expressed in the ($S_1$,$\alpha$) parameter space:
\begin{align}
    P(S_1,\alpha & \mid  S_1^m,  S_2^m)  \propto P(S_1^m, S_2^m \mid S_1,\alpha) P(S_1,\alpha) \nonumber \\
    & \approx P(S_1^m \mid S_1) P(S_2^m \mid S_1, \alpha)P(\alpha \mid S_1) P(S_1) \nonumber \\
    & \propto P(S_2^m \mid S_1, \alpha)  P(\alpha \mid S_1) P(S_1 \mid S^m_1).   
    \label{eqn:second_debias}
\end{align}
The first line in Equation \ref{eqn:second_debias} is Bayes' theorem. The second line expands the prior and assumes independent noise between flux densities to split the likelihood. The final line combines the $S_1$ likelihood and prior into the previously computed primary-band posterior distribution. This last line shows in practice how we implement the secondary band debiasing. Equation \ref{eqn:second_debias} is transformed from the ($S_1,\alpha$) parameter space to a function of $S_2$ according to $S_1/S_2 = (\nu_1/\nu_2)^{\alpha}$ and marginalized to obtain the posterior ``debiased'' distribution for $S_2$. By marginalizing over $S_1$, this method also produces a posterior distribution for $\alpha$. We report the median and 16\% and 84\% quantiles of the debiased $S_2$ and $\alpha$ distributions. 

Because the debiased secondary band flux densities include a prior on the $\alpha$ distributions, in Appendix~\ref{app:alphaoutliers} we further investigate outliers in these distributions for which the debiased secondary band flux densities could be less optimal. We do not find any evidence for an unusual source population altering the measured sample $\alpha$ distributions on which the priors are based. 

\begin{figure*}[t]
	\centering
	\includegraphics[width=.95\textwidth]{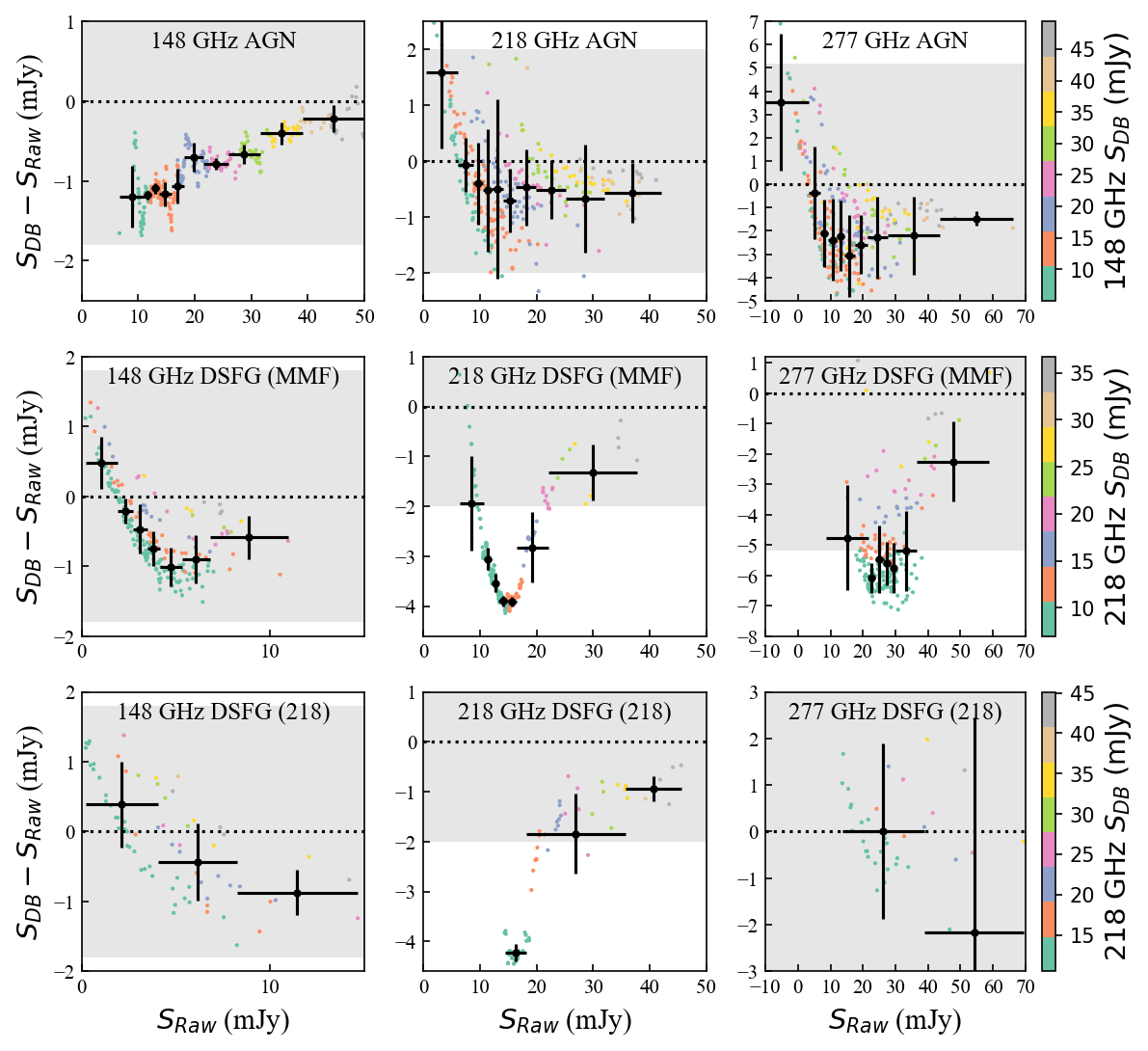}
\caption{Debiased flux densities vs measured flux densities in each ACT band for sources in the extragalactic catalog. The top row of plots shows debiasing for AGN, of which 95\% are selected at 148~GHz. The other plots are for DSFGs, which are selected with the MMF (middle row) or 218~GHz  data (bottom row). In these plots, the debiased flux density minus the raw flux density is shown on the $y$-axis. Points, color-coded by primary-band flux density, show results for individual sources, while the black data with errors show the mean and standard deviation on these points in bins of width indicated by the horizontal bars. Except for the highest flux density bins, each bin is chosen to contain 30 sources. The light gray region indicates the characteristic statistical uncertainty associated with each band, centered at zero (in some cases extending beyond the plotted $y$-axis limits). For top and bottom rows, not all sources have corresponding 277~GHz data. The debiasing shifts the flux densities downward to correct for Eddington bias (i.e. ``deboosting''). This behavior is altered at the lowest flux densities due to inter-band information communicated through the likelihood and priors.
\label{fig:debiasing_diff}}
\end{figure*}


\subsubsection{Debiasing applied to catalogs}

We apply these debiasing  methods to the ACT catalogs. Figure \ref{fig:debiasing_diff} compares the debiased with the measured flux densities in each ACT band for AGN (top row) and DSFGs, both with MMF selection (middle row) and 218~GHz selection (bottom row). We restrict the plotted population to sources with raw flux density less than 50~mJy, because above this flux level, the debiasing has minimal effect. The sources are categorized as AGN or DSFGs according to their raw spectral indices as described in Section \ref{section:specindices}. For DSFGs, the primary debiasing band is 218~GHz. For AGN, only \nAGNtwo~(\nAGNmmf) of the \masNagn~sources classified as AGN were selected in 218~GHz (MMF) maps, so the vast majority of AGN-classified sources have 148~GHz as the primary debiasing band. (If we restrict to the 95\% of AGN selected at 148~GHz there is no significant change to the  plotted results.) The gray band in Figure \ref{fig:debiasing_diff} shows the standard error. Most shifts due to debiasing are at or within the standard error, the exception being DSFGs at 218~GHz and, to a lesser degree, AGN and DSFGs at 277~GHz. 

The primary effect of the debiasing in Figure \ref{fig:debiasing_diff} is a downshift (``deboosting'') in the distribution of flux densities.  At the high flux density end, this effect increases with decreasing flux density as the prior based on source counts becomes more important. At the lowest flux densities, the debiasing of  AGN and of 218~GHz-selected DSFGs behaves differently from MMF-selected DSFG debiasing. For the former, in the primary band (148~GHz for AGN, 218~GHz for DSFGs), the deboosting continues to the lowest fluxes. In the secondary bands at low fluxes, the imposition of an $\alpha$ prior in conjunction with the robust primary-band detection counters the deboosting effect. As the measured flux density in the secondary band reaches zero (or even negative values), this results in a positive correction. Also additional scatter is seen in the secondary-band debiasing plots. This is expected due to the combination of noise and the imposition of independent information from the primary band through the $\alpha$ prior.  

For DSFGs selected with the MMF, even in the primary 218~GHz band, the debiasing diminishes below 15~mJy. All DSFGs in this flux density range are below the 218~GHz 5$\sigma$ detection threshold and so \emph{require} the MMF selection, which has a significant contribution from the 277~GHz data. The simulated MMF likelihood captures the probability that a source with  $S_{218}^m$  below the 218~GHz detection threshold has higher intrinsic $S_{218}$ given the extra information from the 277~GHz band. This is also an effect for the 277~GHz band, which inflects below its nominal threshold of $\sim30$~mJy. In contrast, the story is simpler for the sub-threshold 148~GHz data, which are debiased through the $\alpha$ prior, similar to the secondary bands for the AGN. 

Because the 277~GHz band plays an important (albeit high noise) role in the MMF, we take one more step to estimate the intrinsic 277~GHz flux density. Sources below the 218~GHz detection limit ($<15$~mJy) are only detected if they have a correspondingly strong 277~GHz flux density measurement. This introduces a selection effect whereby, at low 218~GHz flux density, the MMF is biased towards selecting sources with measured 277~GHz flux density scattered high. As noted in Section~\ref{sec:debiasing_method}, we use the simulations  to centroid the 277~GHz likelihood function for MMF-selected DSFGs, analogous to the first step of the primary band debiasing. The resulting 277~GHz flux density posterior thus captures this selection effect and includes a modest, extra debiasing contribution from this simulated likelihood. This illustrates how the use of simulations can capture subtle selection effects introduced by use of non-trivial data filters.

When implementing these methods, one must select an angular separation tolerance with which to match the measured positions of recovered simulated sources with the input positions. If this tolerance radius is too small, sources that should be matched will be missed, and the statistics of the simulated catalogs will suffer and possibly be biased toward brighter sources. If this tolerance radius is too large, spurious matches can be introduced. Because the primary band flux density prior, which is derived primarily from the source counts, is so steep with so much power at low flux densities, spurious matches seen as low outliers in the likelihood function become greatly exaggerated in the resulting posterior distribution. For each set of simulations, the tolerance radius is initially chosen based on inspection of the distribution of distances between the input and measured positions for sources matched with an initially very large radius. After matching using a tolerance radius thus determined, we then also remove sources that appear to be recovered but have input flux densities well below the completeness limit of the survey. Finally, in a plot of the measured vs debiased flux densities, sharp discontinuities occur where misidentified simulated sources bring the debiased flux densities erroneously low. We verify that the primary-band debiased flux density is a smoothly varying function of the measured flux density for each selection, as is evident in the top-left and central lower two panels of Figure~\ref{fig:debiasing_diff}. For the MMF selection, we use a tolerance radius of 0.005$\degree$ (18$\arcsec$) and remove matched sources with input flux density below 4~mJy (below which we are unlikely to recover simulated sources that are true rather than spurious matches; e.g., see the completeness in Figure~\ref{fig:completeness}). For the 218~GHz selection, we use a tolerance radius of 0.007$\degree$ (25.2$\arcsec$) and remove matched sources with input flux density below 4~mJy. For the 148~GHz selection, we use a tolerance radius of 0.004$\degree$ (14.4$\arcsec$) and remove matched sources with input flux density below 6.5~mJy. We calculate the debiasing correction for sources selected at 148~GHz up to 50~mJy. Above these limits, the debiasing becomes very small, and our simulations do not include enough bright sources to robustly determine the debiasing. All MMF and 218 GHz-selected sources are debiased.

\subsubsection{Completeness}
\begin{figure}
\centering
\includegraphics[width=85mm]{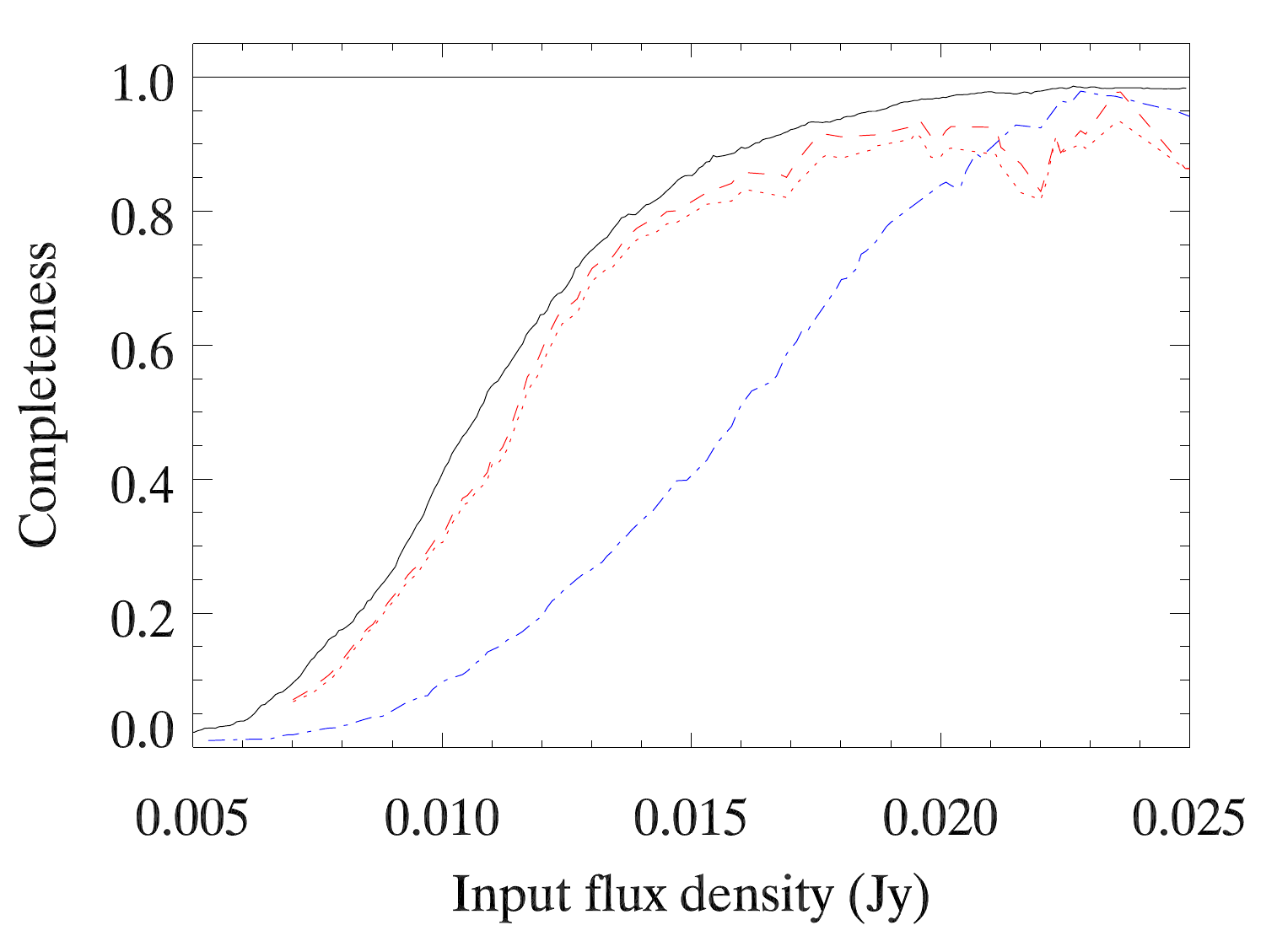}
\caption{The completeness as a function of flux density, as determined by recovering simulated sources. The black solid curve represents the 148~GHz selection, shown as a function of input 148~GHz flux density. The red dashed (dotted) curves represent the MMF selection  before (after) the Galactic dust cuts have been applied. The blue dot-dashed curve represents the 218~GHz selection, after the dust cuts have been applied. For the MMF sample, the completeness is shown as a function of primary-band 218~GHz flux density. The 148~GHz, MMF, and 218~GHz-selected samples are 50\% complete at 10.5~mJy, 11.5~mJy, and 16~mJy, respectively. 
}
\label{fig:completeness}
\end{figure}

These simulations also provide an estimate of the sample completeness as a function of intrinsic flux density.  
We calculate the numbers of simulated sources that are matched to sources recovered from the source finding procedure.  
The completeness is given by the ratio of the number recovered to the number input for the 148~GHz, MMF, and 218~GHz selections. For each of these selection methods we recast the completeness in each band as a function of the flux density in that band.  The results are shown in Figure \ref{fig:completeness}. For 218~GHz and MMF selected samples (which are 92\% DSFGs), we restrict the simulated sources to those with input $\alpha > 1.0$ to mimic our DSFG selection criterion. 
In this way, the MMF and 218~GHz-selection completeness is equivalent to the completeness of DSFG selection, a fact we use when estimating DSFG source counts.  
We also apply the same cuts to remove Galactic dust (described in Section \ref{section:dustremoval}) from the simulated source samples. Figure \ref{fig:completeness} shows the MMF completeness with and without these dust cuts, which do not strongly affect the completeness of extragalactic DSFGs. The MMF completeness never reaches $100\%$ even without Galactic cuts because of our criterion that the input $\alpha$ be greater than 1.0, which excludes $3.2\%$ of the population (see Figure \ref{spectralhistogram} for the $\alpha$ distribution). We do not need to impose a similar criterion on the AGN simulations because the distribution in $\alpha$ is much narrower, and the resulting cut would only exclude $0.007\%$ of the population. In the same way as we equate 218~GHz and MMF completeness to that of DSFGs, the AGN completeness, used for source count estimation, is taken to be the same as 148~GHz completeness. This association is reasonable as 95\% of AGN are selected at 148~GHz.

We made one modification to the simulated source sample when calculating the 148~GHz flux density sample completeness.  In our main source-finding pipeline, sources that are located within 25$\arcmin$ of very bright (S/N $> 50$) sources are excluded from the sample. However, because there are many such bright sources in the simulated samples, given the input distribution of $S_{218}$ and $\alpha_{148-218}$, this practice excludes a non-negligible ($\sim 10\%$) fraction of the catalog. We ran a separate source finding trial on one of the 1000-sample simulated source maps reducing the size of the exclusion region to 1$\arcmin$, and the completeness curve is based on this trial rather than the full sample.  This exclusion would not affect the main catalogs because we first mask any bright sources and rerun the source finding after they have been removed. Shrinking the exclusion region to 1$\arcmin$ does not affect the filter for the simulations because we construct the filter from the maps of the real data, so the filter does not include noise from the simulated sources.

\begin{figure}
\centering
\includegraphics[width=85mm]{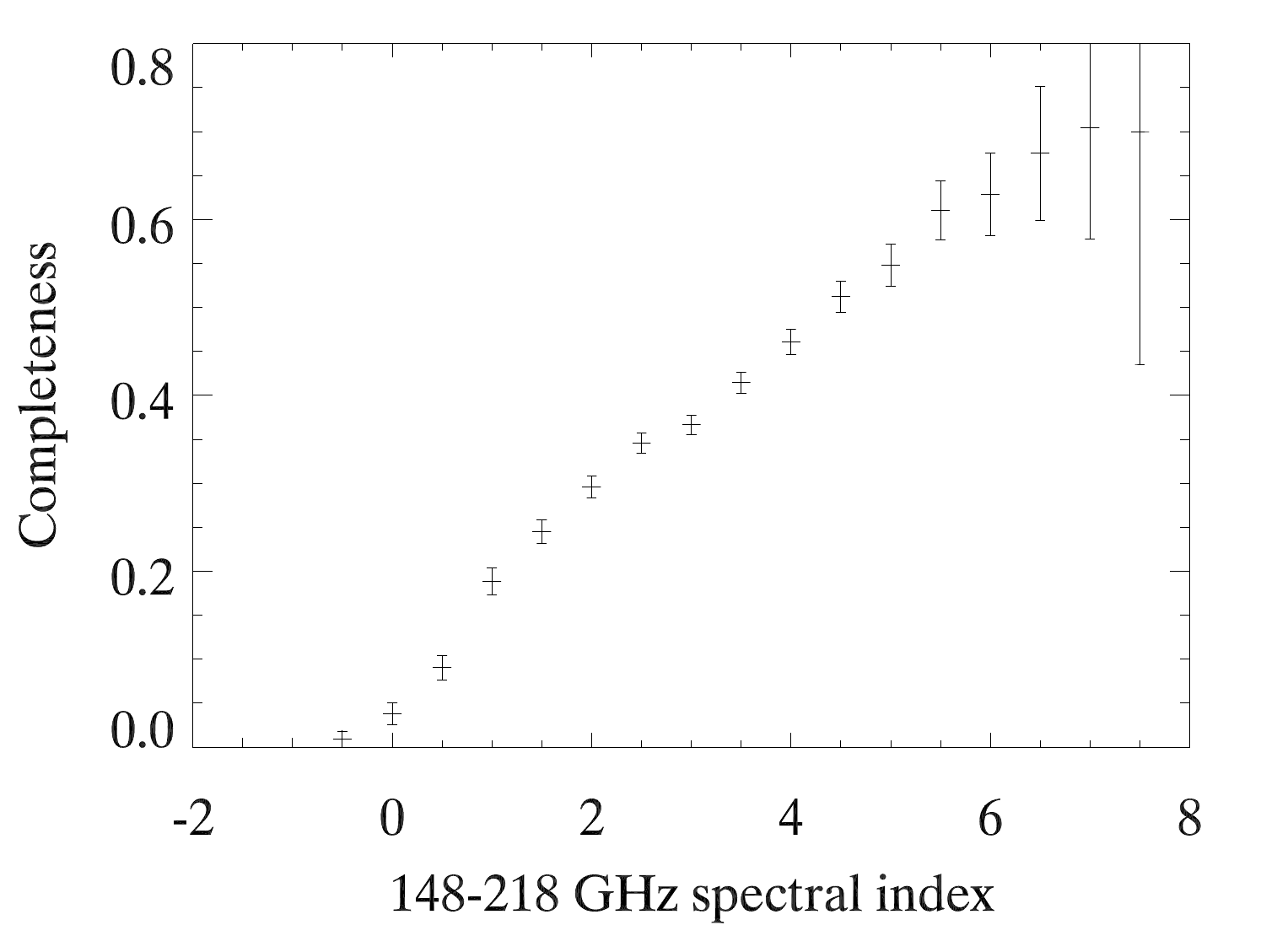}
\caption{The MMF completeness as a function of spectral index, as determined by recovering simulated sources. Note that the DSFG sample is selected using the criterion $\alpha_{148-218}~>~1.0$, so DSFGs with shallower, flat or negative spectral indices would be classified as AGN and removed from the DSFG sample.   \label{fig:completenessalpha}}
\end{figure}

The MMF completeness depends on the spectral index as well as the flux density. We use the same set of MMF simulations to investigate this dependence, shown in Figure \ref{fig:completenessalpha}. We populated the MMF simulations with an input $\alpha$ distribution that is Gaussian with mean 3.4 and standard deviation 1.3 to approximate a DSFG source population. The uncertainty on the completeness reflects this, as the completeness is better determined for the more numerous simulated sources with $\alpha$ near the center of the input distribution. Note the completeness in Figure \ref{fig:completenessalpha} never approaches unity because, for each bin in $\alpha$, it is computed over the full population of sources. With mean $S_{218}=10$~mJy and standard deviation 5~mJy, this population includes many sources well below the detection threshold.

\subsection{Purity} \label{sec:purity}

To calculate the catalog purity, we multiplied the maps by $-1$ and ran our source finding algorithms.  For the 148~GHz map, the Sunyaev Zel'dovich effect associated with galaxy clusters introduces compact negative signals in the map (which become compact positive signals after multiplication by $-1$).  We therefore masked out 6 arcminute regions around the optically confirmed galaxy clusters presented in \citet{hasselfieldclusters}.  Bright sources also introduce negative (i.e., positive after multiplication by $-1$) features nearby due to ringing from the filtering.  We generated a template of the sources by populating a map with the beam profile at the position and amplitude of each source in the catalog, and we subtracted this template from the 148~GHz map before inverting it.  
Three sources (10--17~mJy) are detected in the inverted map, after disregarding the few detections near bright features.
Thus according to this analysis, the estimated departure from purity in the AGN sample is at the $(3/\masNagn) < 1\%$ level. For comparison, based on simulations of the noise in the map, \citet{hasselfieldclusters} expected 1.8 false positives.

For the sources selected via the MMF, we first inverted the multi-band map and ran the source finding algorithm.  We then applied the cuts to eliminate Galactic dust emission, which may also introduce negative sources into the map through ringing from the filter. After applying Galactic cuts and again removing sources near bright features, only one source (13~mJy) is recovered from the inverted map. 
Thus, the estimated departure from purity in the MMF-selected DSFG sample is at the $(1/\nDSFGmmf) < 1\%$ level. 

For sources selected at 218~GHz outside the area of the MMF map, we also inverted the map, ran the source finding algorithm, and applied the Galactic dust cuts.  Because most of this area lacks reliable 277~GHz data, we did not cut based on source spectrum. The dust cuts reduced the number of sources found in the inverted map (after removing a few near bright features in the map) from 85 to 7. Of these, five are clearly associated with extended Galactic dust emission.  
According to this analysis, the estimated departure from purity in the 218-selected DSFG sample is at the $(2/\nDSFGtwo) < 4\%$ level.

We note that the map-inversion method described here does not account for false detections due to Galactic dust contamination. We have taken careful measures to remove such contamination (Section \ref{section:dustremoval}). Using source counts, we further check that the catalog is not significantly biased by  Galactic dust contamination. This check is described in  Section \ref{sec:countDescription} (Figure \ref{fig:dustfreecounts}).

\subsection{Astrometry}

\begin{figure}
	\centering
	\includegraphics[width=84mm]{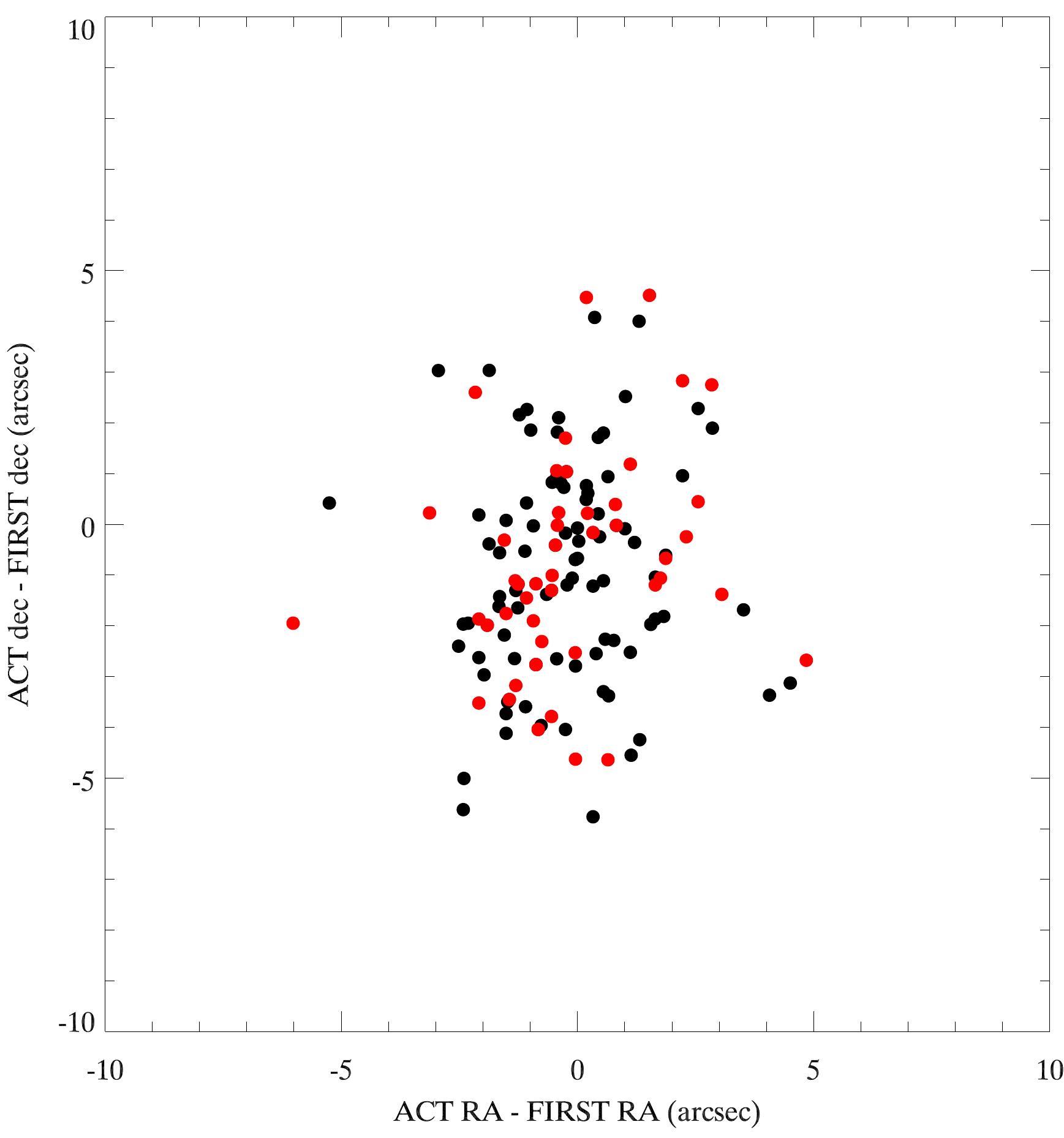}
\caption{The positions of ACT sources with S/N $>16$ relative to FIRST positions. Black points represent locations from the 148~GHz data, and red points represent locations from the MMF data.   \label{fig:astrometry}}
\end{figure}

To determine the accuracy of the catalog astrometry, we matched ACT sources within 1.2$\arcmin$ of radio sources selected from the VLA FIRST survey \citep{Becker1995}.  The positional uncertainty for FIRST sources with 1.4~GHz flux densities above 3~mJy is $<0.5\arcsec$. For a summary of the astrometric offsets, see Table \ref{tab:astrometry}. For the 148~GHz catalog, there are 268 ACT sources with matches in FIRST, 264 of which have integrated 1.4~GHz flux density $>$3~mJy.  The mean difference in R.A. between the ACT sources and the FIRST sources is $-0.02\arcsec$ and the standard deviation of this difference is 5$\arcsec$.  The mean difference in declination between the ACT sources and the FIRST sources is $-0.8\arcsec$ and the standard deviation of this difference is 5$\arcsec$.  Because the FIRST survey has a high angular resolution (5$\arcsec$), some extended radio sources may be resolved into multiple components.  If we restrict the matched sample to ACT sources that only have a single FIRST match, there are 204 matches and the positional differences do not change qualitatively.  If we restrict the sample to ACT sources that have S/N above 16, there are 94 matches. Figure \ref{fig:astrometry} shows the results of this matching. For this sample, the mean difference in R.A. between the ACT sources and the FIRST sources is $-0.2\arcsec$ and the standard deviation of this difference is 2$\arcsec$.  The mean difference in declination between the S/N $>$ 16 ACT sources and the FIRST sources is 0.1$\arcsec$ and the standard deviation of this difference is 2$\arcsec$.  

The angular resolution of the 218~GHz map is higher, although the noise level is also higher.  
If we adopt the 218~GHz positions of the 148~GHz-selected sources, there are fewer matches to FIRST sources (218) and the scatter between the ACT and FIRST positions is higher (6$\arcsec$).  This is likely due to the higher significance of the 148~GHz measurements, which are both more sensitive and in a brighter part of the typical SED for AGN.  For the 40 sources with 218~GHz S$/$N $>$ 16, the scatter between the 218~GHz positions and the FIRST positions is reduced to 2$-$3$\arcsec$.    

\begin{table}[htb]
\centering
\caption{Astrometry}
\begin{tabular}{|cccccc|}
\hline
 & $N_{\mathrm{match}}$\footnote{ACT locations are compared to matched FIRST source locations, which have $0.5''$ precision.} & \multicolumn{2}{c}{R.A.\footnote{Offsets listed in arcsec.}} & \multicolumn{2}{c|}{decl.}\\
 & & mean & $\sigma$ & mean & $\sigma$ \\\hline
148 GHz, S/N $>$ 5 & 264 & $-0.02$ & 5 & $-0.8$ & 5 \\ 
148 GHz, S/N $>$ 16 & 94 & $-0.2$ & 2 & 0.1 & 2 \\ 
MMF, S/N $>$ 5 & 151 & $-0.1$ & 4 & $-2$ & 5 \\
MMF,  $\mathrm{S/N}>16$ & 47 & $-0.1$ & 2 & $-0.8$ & 2 \\
\hline
\end{tabular}
\label{tab:astrometry}
\end{table}

For sources selected through the MMF technique, there are 151 with matches in the FIRST catalog (after the dust cuts are applied to the ACT sample, which removes two sources).  The mean difference in R.A. between the ACT sources and the FIRST sources is $-0.1\arcsec$ and the standard deviation of this difference is 4$\arcsec$.  The mean difference in declination between the ACT sources and the FIRST sources is $-2\arcsec$ and the standard deviation of this difference is 5$\arcsec$.  For the 47 sources with S$/$N $>$ 16 and FIRST matches, the scatter between the ACT positions and the FIRST positions is reduced to 2$\arcsec$ in R.A. and 2$\arcsec$ in declination.  

 We can compare the astrometric accuracy with previous ACT source analyses. \citet{marriagesources} matched ACT sources from the southern survey having $\mathrm{S/N}>20$ with AT20G sources, and found the rms of the offsets to be $3.5\arcsec$ in R.A. and $3.3\arcsec$ in declination. For sources with $\mathrm{S/N}>16$, \citet{Marsden2014} found the rms of the offsets to be $2.1\arcsec$ in R.A. and $1.8\arcsec$ in declination at 148~GHz (and larger, $\sim3.5\arcsec$ at 218~GHz). Thus, the astrometry of the ACT equatorial source catalog is comparable to earlier ACT source catalogs. This is as expected, because the instrument beam did not change significantly between the southern and equatorial surveys, and similar S/N thresholds were considered.

\section{Source characterization}
\label{sec:characterization}

\subsection{Millimeter Spectral Indices} \label{section:specindices}

The spectral behavior of sources across the ACT bands is bimodal due to the presence of two distinct populations: AGN and DSFGs. Table \ref{tab:alpha} summarizes the typical spectral indices for these source populations, as discussed throughout this section, which is organized as follows. First we discuss the spectral properties of all the sources in the catalogs as measured by ACT. We then split the sample into the two dominant populations and report the sample statistics for each. We next describe tests of our recovery of spectral information, and last we discuss our results and compare with other studies.  Sources that are flagged as Galactic (see Section \ref{section:dustremoval}) are not included in the following analyses or figures.

\subsubsection{Spectral properties of the full sample}
\begin{figure}
	\centering
	\includegraphics[width=84mm]{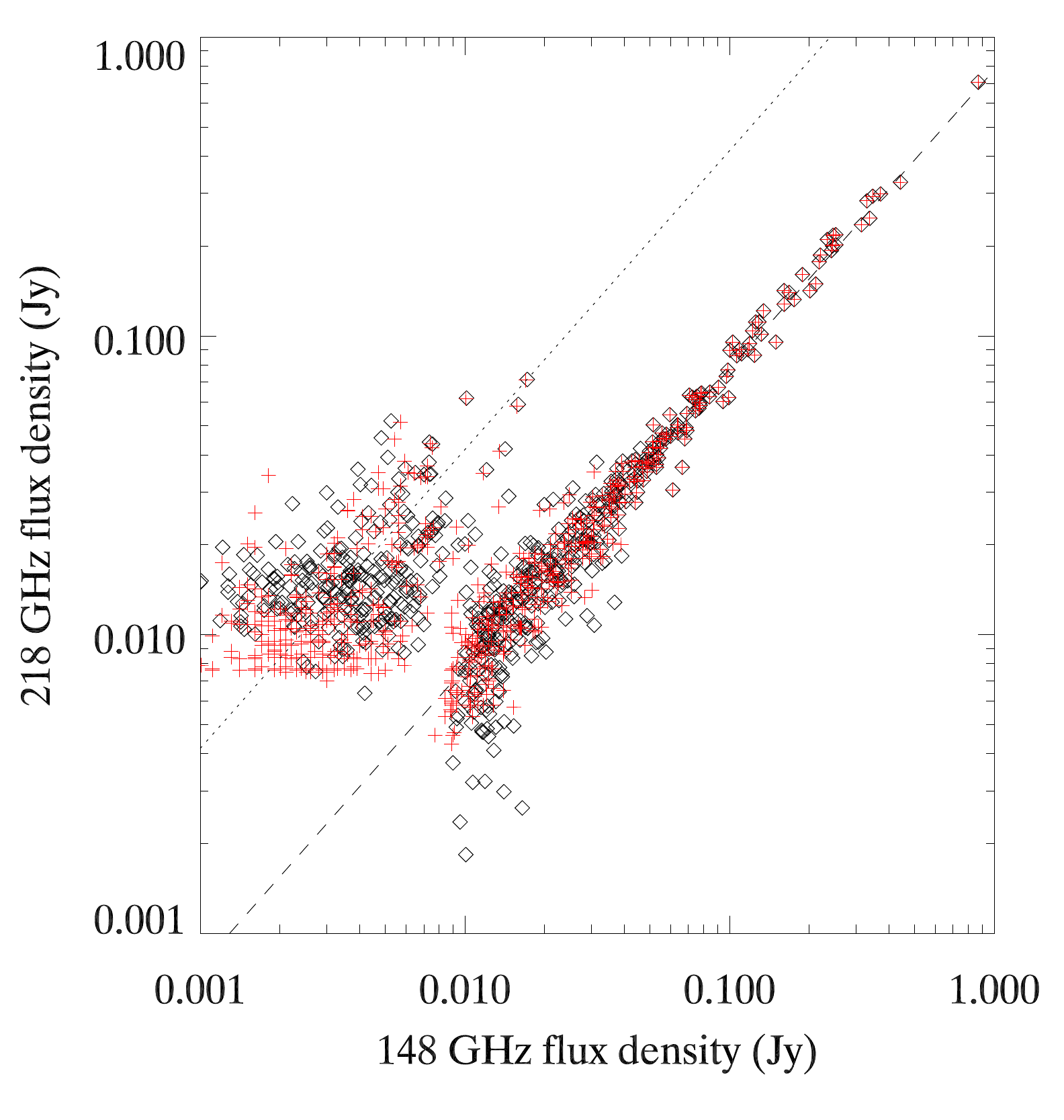}
\caption{148~GHz flux density vs. 218~GHz flux density. Black diamonds (red crosses) are raw (debiased) flux densities. The dashed line indicates the best-fit median AGN spectral index ($-0.8$), and the dotted line indicates the best-fit median DSFG spectral index ($3.4$). The population at the right extending to higher flux densities is composed of AGN, whereas the fainter, 218~GHz-dominated population is composed of DSFGs. \label{mmfluxes}}
\end{figure}

\begin{figure}
	\centering
	\includegraphics[width=84mm]{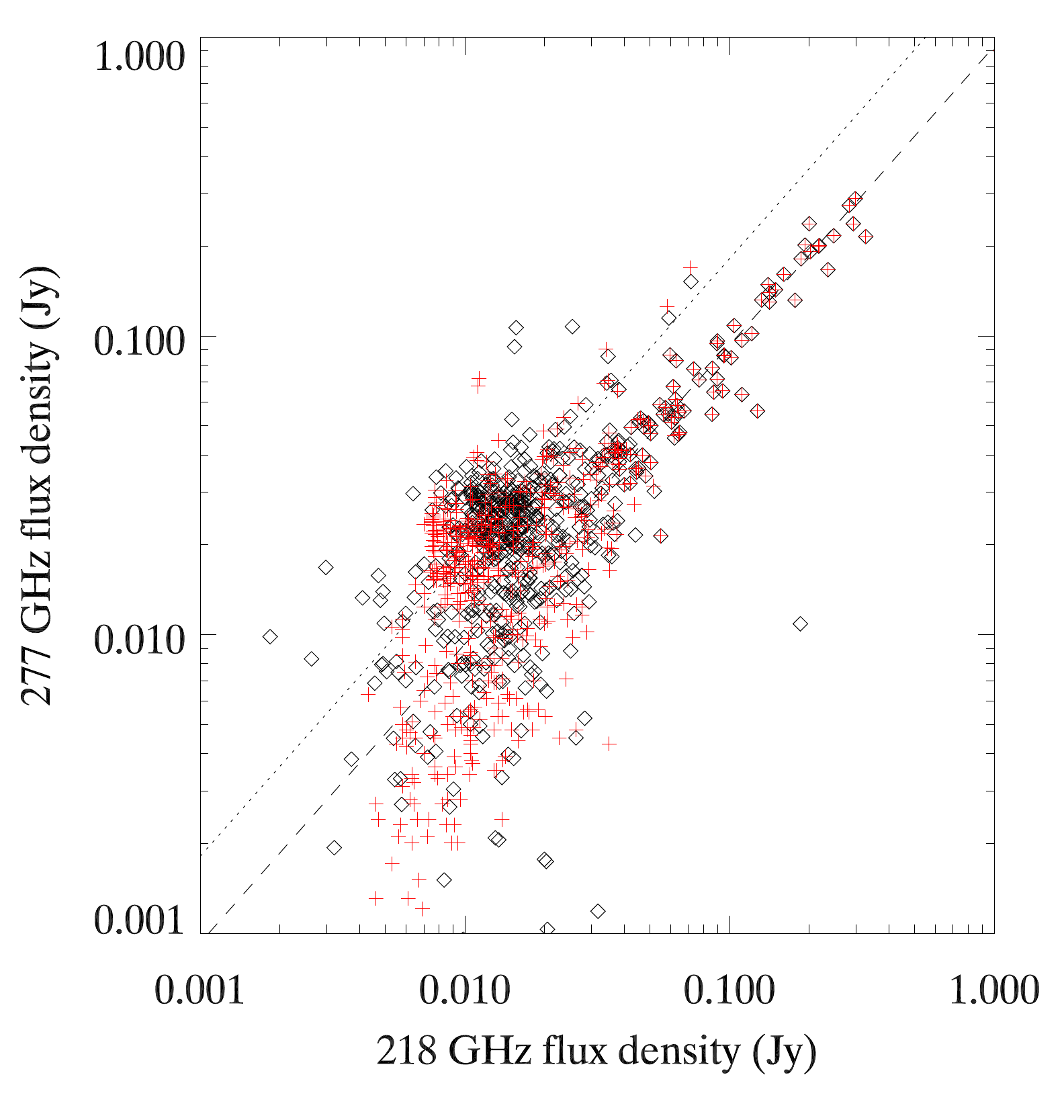}
\caption{218~GHz flux density vs. 277~GHz flux density. The plotting convention is the same as in Figure \ref{mmfluxes}. The dashed line indicates the median AGN spectral index ($-0.33$), and the dotted line indicates the median DSFG spectral index ($2.4$). The AGN and DSFG populations are no longer clearly distinguishable. The MMF allows us to detect DSFGs below the typical 218~GHz 5$\sigma$ noise threshold. \label{mmfluxes2}}
\end{figure}

\begin{figure}
	\centering
	\includegraphics[width=84mm]{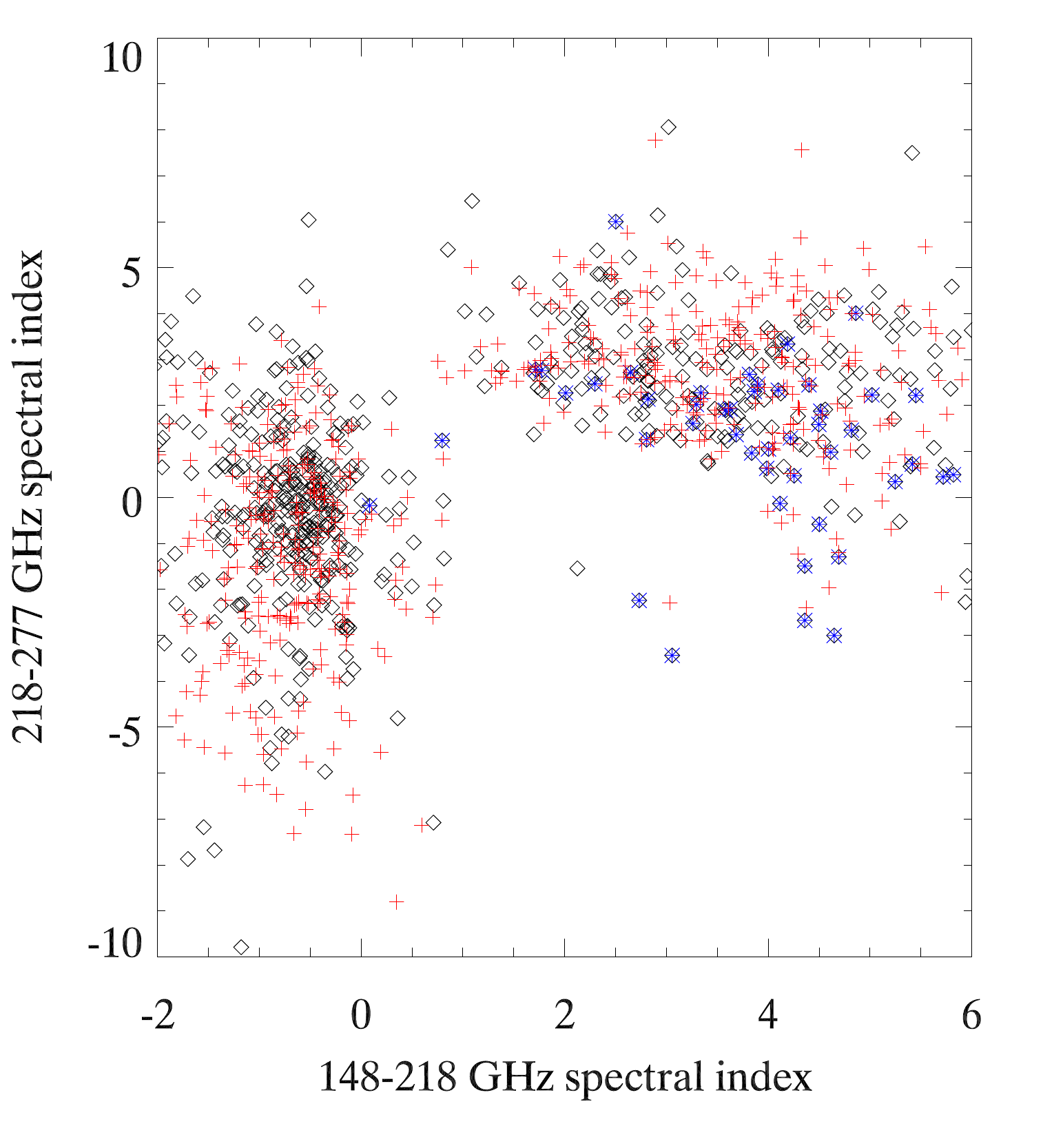}
\caption{A color-color diagram of all sources. Black diamonds (red crosses) are raw (debiased) spectral indices. The blue stars are nearby dusty galaxies. The DSFG (at right) and AGN (at left) populations clearly separate.  As expected, this separation is even more pronounced for the debiased flux densities. \label{colorcolor}
}
\end{figure}

Figure \ref{mmfluxes} plots 148~GHz flux density vs 218~GHz flux density, and Figure \ref{mmfluxes2} plots 218~GHz flux density vs 277~GHz flux density for all sources in the catalog. The DSFG sample populates the higher spectral indices and mostly lower flux densities (and many are undetectable at 148~GHz), while the AGN extend to higher flux densities.  Both the raw measured and debiased (Section \ref{section:debiasing}) flux densities are shown. The debiasing methods tend to concentrate the sources into the two populations, as expected for the lower significance data given the prior probability distribution (see Figure \ref{colorcolor}). 

Figure \ref{colorcolor} plots the 148-218~GHz spectral indices vs the 218-277~GHz spectral indices.  The two populations are clearly discernible, and the 218-277~GHz spectral indices are flatter (closer to 0) than the 148-218~GHz spectral indices for the DSFGs. Spectral indices plotted in Figure~\ref{colorcolor} are estimated directly from flux densities as $\alpha=\log(S_2/S_1)/\log(\nu_2/\nu_1)$. The black points show indices estimated from raw flux densities   while those computed from debiased flux densities are shown in red. Note that these spectral indices derived from debiased flux densities are not exactly the same as what is reported in Table 4; in the latter, spectral indices are calculated directly from a posterior computed from Equation~\ref{eqn:second_debias}. Indeed, for AGN, we do not even compute the 218-277~GHz spectral index through the posterior. The sources that have been visually identified in SDSS optical images as nearby dusty galaxies are shown in blue. Although they lie within the DSFG spectral index distribution, their $\alpha_{218-277}$ typically falls on the flatter side of the distribution, as also seen in the sample medians reported in Table~\ref{tab:alpha}. This could be explained by sources that are extended and thus resolved in the 277~GHz maps or by CO line emission at 230.5~GHz contributing to the 218~GHz flux densities. In Appendix \ref{app:alphaoutliers} we further investigate the sources with spectral indices that lie at the edges of the distributions. 
 
%

\begin{figure}
	\centering
	\includegraphics[width=84mm]{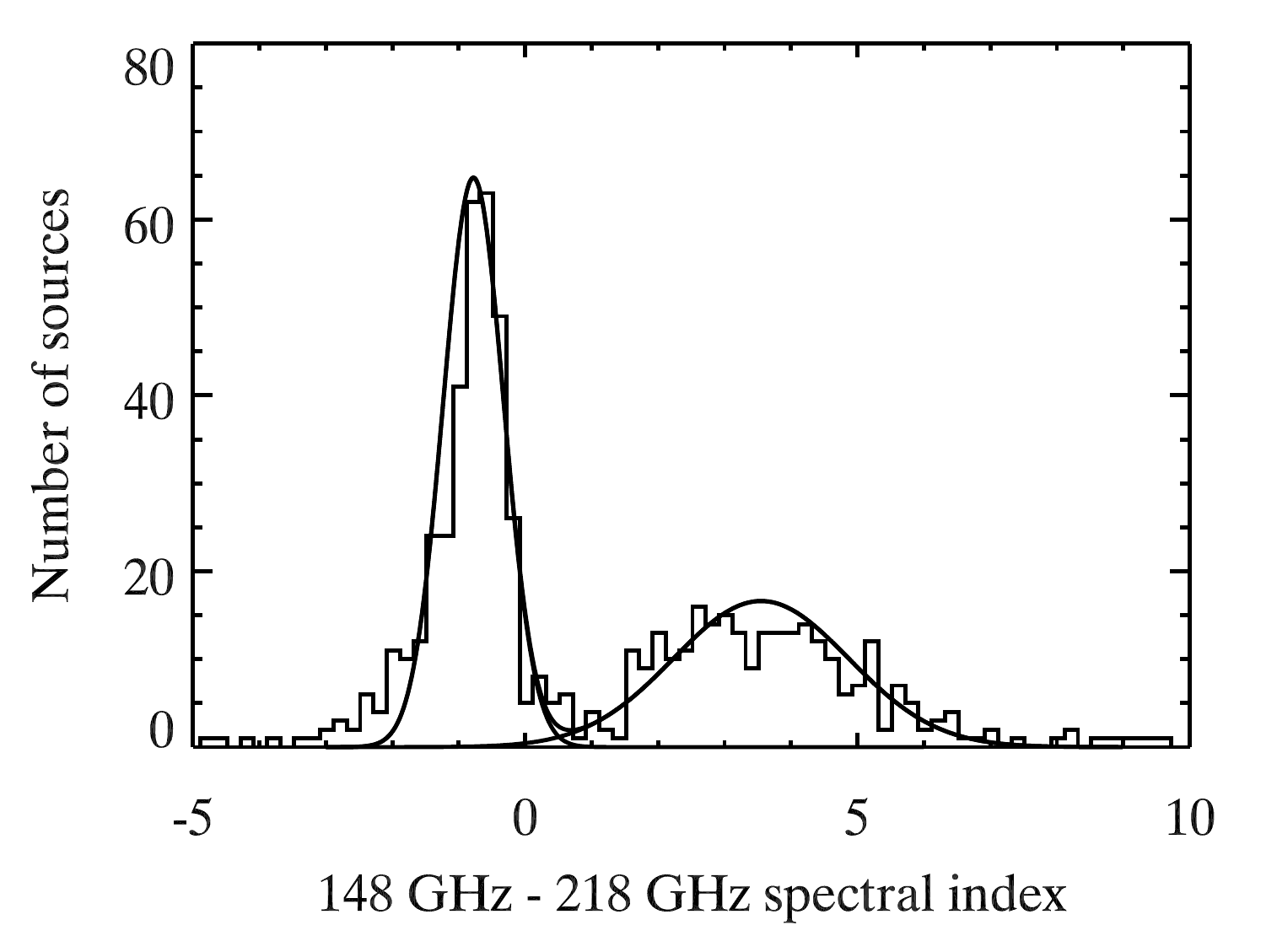}
\caption{Histogram of the spectral index $\alpha$ from 148~GHz to 218~GHz ($S \propto \nu^{\alpha}$). The two source populations, AGN and DSFGs, are discernible. A model composed of two Gaussians, one for AGN and one for DSFGs, is fit to the data and overplotted, with best fit model parameters listed in Table~\ref{tab:alpha}. \label{spectralhistogram}}
\end{figure}

Figure \ref{spectralhistogram} shows a histogram of the measured $148-218$~GHz spectral indices. As seen in the figure, the two populations are clearly distinguishable. The flux densities used to construct this histogram have not been debiased. A simple model consisting of two Gaussians, one describing each population, has been fit to the data and is also shown.  The best fit median spectral index for the AGN is $-0.8$, with standard deviation 0.4.  The best fit median spectral index for the DSFGs is 3.4, with standard deviation 1.3. These values are listed in Table~\ref{tab:alpha}. 

\begin{table*}[htb]
\centering
\caption{Typical millimeter spectral indices for AGN and DSFGs}
\begin{tabular}{|c|cccc|cccc|}
\hline
Median spectral indices & \multicolumn{4}{c|}{AGN} & \multicolumn{4}{c|}{DSFGs}\\
  & median $\alpha_{148}^{218}$ & $\sigma$ & median $\alpha_{148}^{277}$ & $\sigma$ & median $\alpha_{148}^{218}$ & $\sigma$ & median $\alpha_{218}^{277}$ & $\sigma$ \\\hline
Full sample\footnote{The spectral indices reported here are based on raw (not debiased) flux densities, except where otherwise noted.} & $-0.66$ & 1.2 & $-0.54$ & 1.9 & 3.7 & 1.8 &  2.4 & 4.2 \\ 
Restricted to $-1.2\degree< \mathrm{dec}<1.2\degree$ & $-0.65$ & 0.9 & $-0.45$ & 1.1 & 3.6 & 1.6 & 2.7 & 1.5 \\
Restricted to $S>20$~mJy AGN & $-0.62$ & 0.67 & $-0.50$ & 1.0 & & & & \\
Restricted to $S>50$~mJy AGN & $-0.58$ & 0.32 & $-0.49$& 1.3 & & & & \\
Bootstrap samples\footnote{For this row, $\sigma$ denotes the standard deviation of the medians, while for the other rows $\sigma$ denotes the sample standard deviation. The statistics reported here are based on resampling the full catalog 1,000 times.} & $-0.66$ & 0.03 & $-0.54$ & 0.04 & 3.7 & 0.13 & 2.4 & 0.11 \\
Debiased spectral indices\footnote{The debiased spectral index for each source is calculated from the posterior distribution of the spectral index, as discussed in Section \ref{section:debiasing}. For $\alpha_{148}^{277}$ and $\alpha_{218}^{277}$, the sources are restricted to $-1.2\degree<$ dec $<1.2\degree$.} & $-0.52$ & 0.6 & $-0.67$ & 2.6 & 3.8 & 1.1 & 2.8 & 1.1 \\
Parameters describing best-fit Gaussian & $-0.8$ & 0.4 & & & $3.4$ & 1.3 & & \\
Prior used in debiasing secondary bands & $-0.7$ & 1.2 &  $-0.4$ & 3.4 & 3.7 & 2.2 & 2.7 & 2.1 \\
Nearby dusty galaxies & & & & & 4.0 & 1.7 & 1.3 & 4.2 \\
\hline
\end{tabular}
\label{tab:alpha}
\end{table*}

\subsubsection{Spectral properties of AGN and DSFG populations}
We split the sample into AGN and DSFGs according to their measured (not debiased) 148-218~GHz spectral indices. Sources with $\alpha > 1.0$ are identified as DSFGs, and sources with $\alpha < 1.0$ are identified as AGN. Because the populations are less easily separated by the 218-277~GHz spectral indices (Figure \ref{colorcolor}), we identify every source as an AGN or DSFG based solely on its 148-218~GHz spectral index. As listed in Table \ref{tab:catalog}, \masNagn~sources are classified as AGN, and \masNdust~sources are classified as DSFGs.
In Appendix \ref{app:alphaoutliers}, we further investigate the sources with spectral indices that fall near the AGN/DSFG boundary.

The median spectral index for AGN is $-0.66$ for 148-218~GHz and $-0.54$ for 148-277~GHz, as shown in Table~\ref{tab:alpha}. The median spectral index for DSFGs is $3.7$ for 148-218~GHz and $2.4$ for 218-277~GHz. The sample standard deviations are also reported in Table~\ref{tab:alpha}. The medians and standard deviations of the debiased spectral indices, as derived from the debiasing analysis summarized in Section \ref{section:debiasing}, are also presented in Table~\ref{tab:alpha}. In order to estimate the uncertainty on the medians, we randomly select sources from the sample to generate bootstrapped catalogs. The median and standard deviations on the medians for 1000 such catalogs are reported in the line for ``Bootstrap samples'' in Table~\ref{tab:alpha}. 

We compare the distributions of 148-218 spectral indices with the distributions of spectral indices involving 277~GHz for both AGN and DSFGs. For AGN, the median measured $\alpha$ for 148-277~GHz is $-0.54$, compared to $-0.66$ for 148-218~GHz. However, debiasing the flux densities affects the AGN 148-277 $\alpha$ more, such that the debiased 148-277 $\alpha$ is steeper ($-0.67$ if restricted to the MMF area) than both the measured 148-277 $\alpha$ and the debiased 148-218 $\alpha$. For DSFGs, the median 218-277~GHz spectral index is flatter (closer to 0) than the median 148-218~GHz spectral index. The median measured $\alpha$ for 218-277~GHz is 2.4 for DSFGs (2.7 if restricted to the MMF area, 2.8 for the debiased spectral indices), compared to 3.7 for 148-218~GHz (3.8 for the debiased spectral indices), which we interpret as evidence for optically thick emission near the peak of the thermal spectrum or an additional cold dust component (see below, Section~\ref{sec:spectralindicesdiscussion}). 

The increased scatter for spectral indices involving the 277~GHz band  can be explained by the higher noise level at 277~GHz. For example, for a high S/N source with $S_{148} = 15$~mJy, a spectral index of $-1$ results in $S_{277} = 8$~mJy. 
Typical noise at 277~GHz is 5.2~mJy.
Thus $\pm1\sigma$ on the 8~mJy translates to an apparent range in the 148-277~GHz spectral index of $-2.7$ to $-.2$. For both populations, the standard deviation of spectral indices involving 277~GHz is reduced when we constrain the sample to sources in an area of the map with lower noise in the 277~GHz data. 

\subsubsection{Tests of spectral index recovery}

Using mock catalogs, we find that noise in the 277~GHz data does not account for the flatter 218-277~GHz spectral index for the DSFGs, but it does account for the flatter measured spectral index for the AGN.
The mock catalogs are generated as follows. For every source, we assign a mock 277~GHz flux density by scaling the 218~GHz flux density to 277~GHz assuming the 148-218~GHz spectral index measured for that source. We add normally distributed noise with width equal to the 277~GHz measurement error for that source, repeat this mock 277~GHz flux density generation over the entire catalog, and then repeat this analysis to generate 1,000 catalogs. Of these 1,000 catalogs, none return a median 218-277~GHz spectral index as flat as the measured median 218-277~GHz spectral index for the DSFGs. For these mock catalogs, the median 148-277~GHz spectral index is $-0.55$ for AGN ($-0.44$ when restricted in declination to $-1.2\degree<\mathrm{dec}<1.2\degree$), which agrees well with the median measured 148-277~GHz spectral index. Thus the mock AGN 148-277~GHz spectral index is measured to be shallower than the input in a way that reproduces what is observed. For DSFGs, the median 218-277 spectral index of the mock samples is $3.4$, which is steeper than the median measured 218-277~GHz spectral index of the catalog (2.4, or 2.7 when restricted in declination).

\subsubsection{Discussion of spectral indices} \label{sec:spectralindicesdiscussion}
We interpret the flattening of the dust spectrum from 218 to 277~GHz as indicating that a non-negligible optical depth is likely introducing curvature to the spectrum. Alternatively, emission models that include a cold dust component \citep[e.g., two temperature models such as ][]{dunne2001} may also be able to reproduce the spectral flattening that we observe at the higher ACT frequencies. However, without more observations at other frequencies, it is difficult to distinguish between optically-thin dust emission at different temperatures and optically-thick emission models. We note that the our use of $\alpha$ to describe the dust spectrum from 218 to 277~GHz is only a convenient parameterization rather than a description of a true underlying power law spectrum. \citet{Su2017} combine the ACT data with flux densities from {\it Herschel} and model the SEDs of a subset of nine of the DSFGs in our sample (as further discussed in Section \ref{sec:dsfgs}). The best fit models indicate that the spectra become optically thick at higher frequencies. This introduces a factor of $1-e^{-\tau}$ into the spectrum \citep[see Equations $3-5$ in][]{Su2017}. The median 148-218~GHz $\alpha$ and the median 218-277~GHz $\alpha$ of the subsample presented in \citet{Su2017} are 3.5 and 2.6, respectively. These agree with the median values of $\alpha$ for the full DSFG sample. Given that these sources are likely strongly lensed, it is not surprising that their SEDs are representative of their fainter counterparts in the full DSFG sample.\footnote{Alternatively, one could expect that the lensed sources might be drawn from a higher redshift population. In the sub-millimeter wavelengths at these bright flux densities, and given the high redshift nature of most unlensed sub-millimeter galaxies, the redshift distribution is actually expected to be similar between the lensed and unlensed populations \citep[e.g.,][]{Hezaveh11}.} 
For nearby ($z < 0.09$) dusty galaxies, even flatter 218-277 spectral indices (and steeper 148-218~GHz spectral indices) are likely due to contamination by CO $J(2-1)$ line emission. High redshift galaxies are not similarly contaminated by this CO line because it is redshifted out of the ACT 218~GHz band. In \citet{gralla2014}, we stacked ACT and {\it Herschel} data for radio-selected nearby star-forming galaxies and modeled their median SEDs. Our best-fit model contained a 2.8~mJy contribution to the 218 GHz flux densities from CO line emission. This was found to be in agreement with an example star-forming galaxy.
To investigate the potential contribution of CO to the SEDs of nearby galaxies in this sample, we subtract 2.8~mJy from their measured 218 GHz flux densities. This brings the median 148-218~GHz $\alpha$ for the nearby galaxies (4.0 before subtracting, 3.7 after) into agreement with the full sample (3.7). Similarly, subtracting the 2.8~mJy brings the nearby galaxies' 218-277~GHz $\alpha$ (1.9 before subtracting, 2.6 after) into agreement with the full sample (2.7). 

For the AGN, to compare our spectral index results with previous studies, we restrict our sample to match these studies' flux density limits. If we restrict the sample to sources whose 148~GHz and 218~GHz flux densities both exceed 20~mJy (50~mJy), then the best fit median spectral index for the AGN is $-0.62$ ($-0.58$).
For comparison, the median $\alpha$ for sources with $S_{148} > 50$~mJy in \citet{Marsden2014} is $-0.6$. However, the median $\alpha$ for fainter sources in \citet{Marsden2014} ($-0.51$) is shallower than for $S<50~$mJy sources in this sample ($-0.7$), at low ($2\sigma$) significance. \citet{mocanu2013} also find that the fainter AGN in their sample have relatively steeper spectral indices.

\subsection{Source counts}\label{section:counts}

\subsubsection{Description of source counts}
\label{sec:countDescription}

\begin{figure*}
\centering

\subfigure{\includegraphics[width=8.9cm]{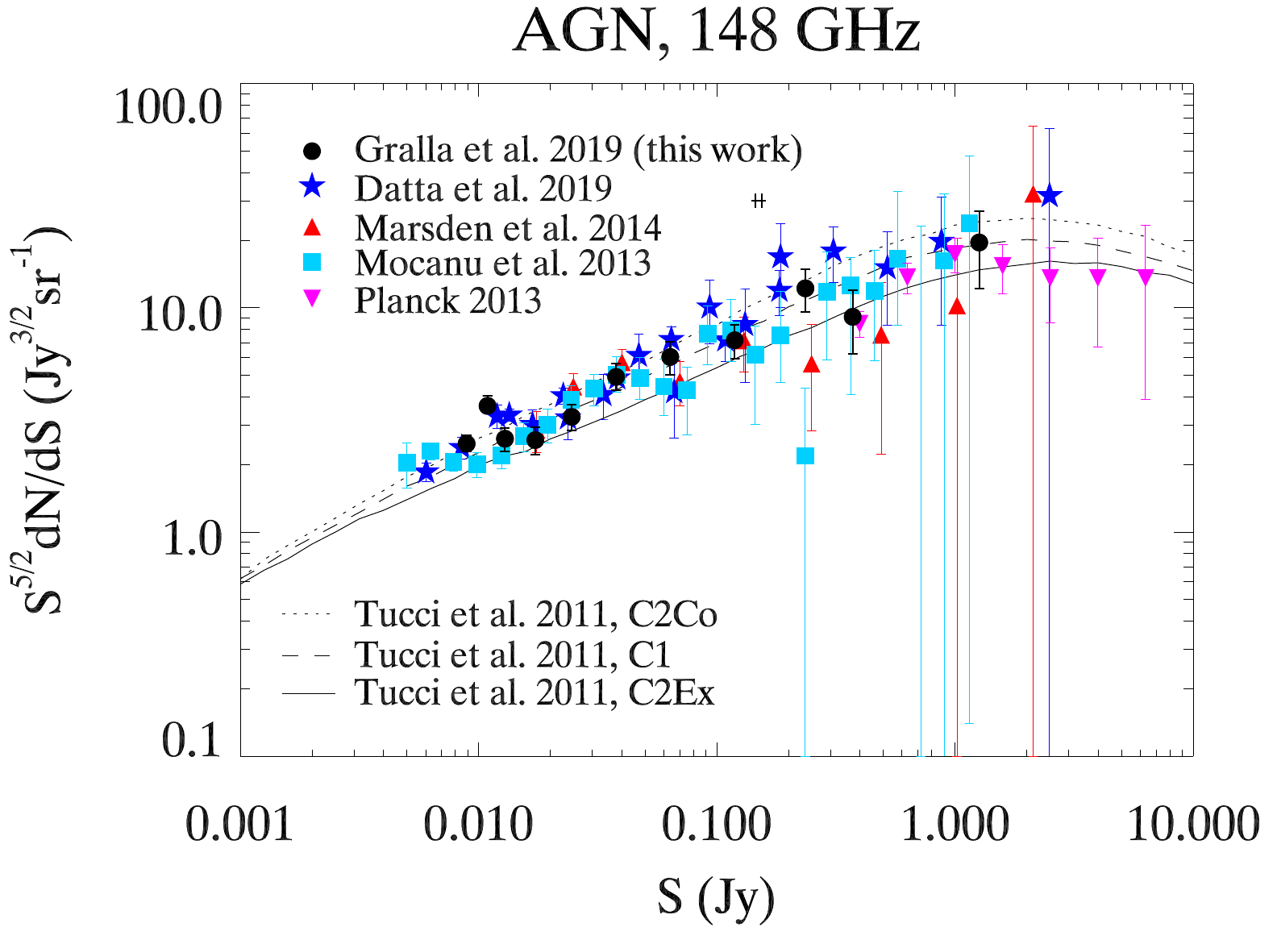}}
\hfill
\subfigure{\includegraphics[width=8.9cm]{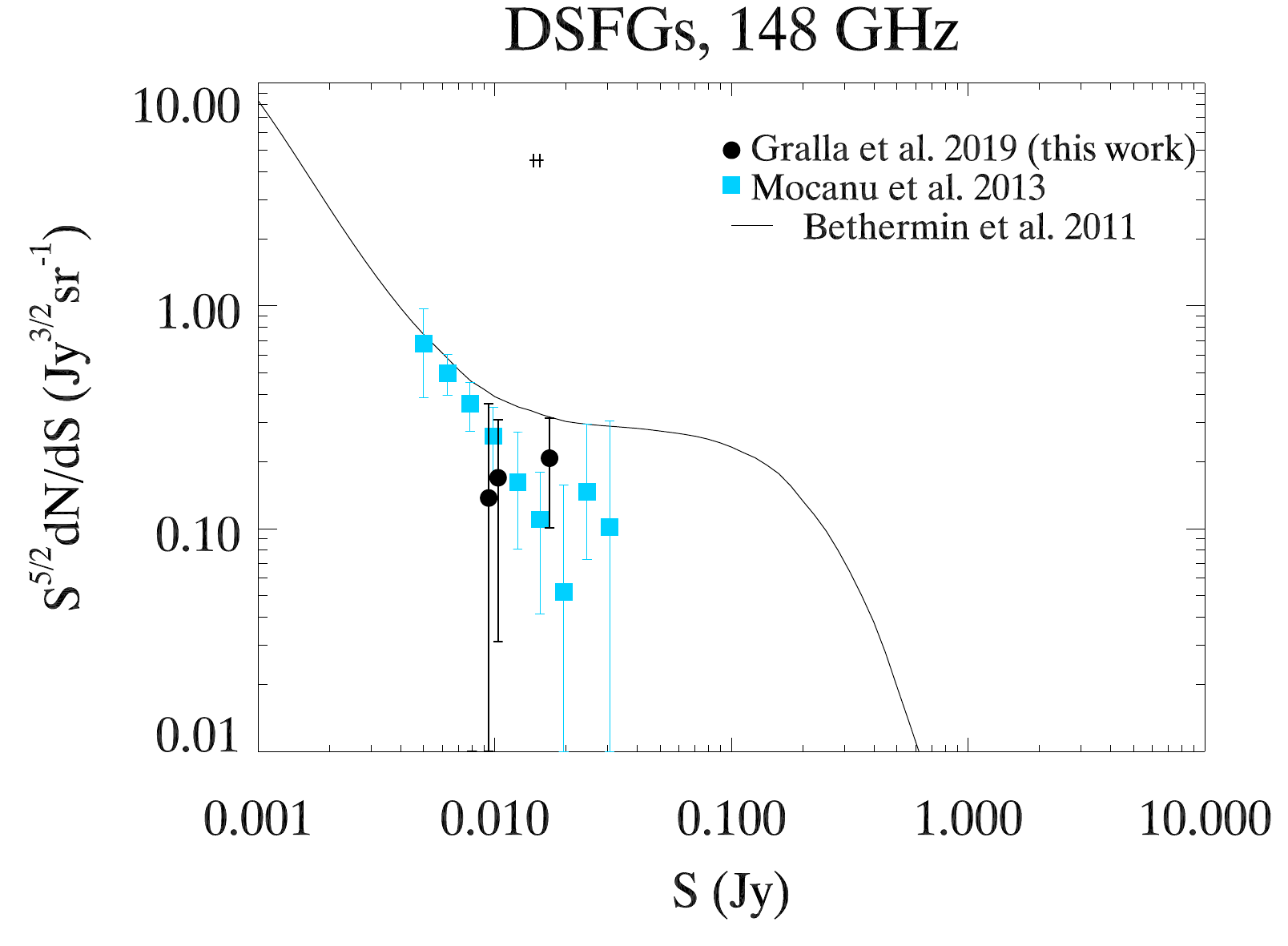}}
\hfill
\subfigure{\includegraphics[width=8.9cm]{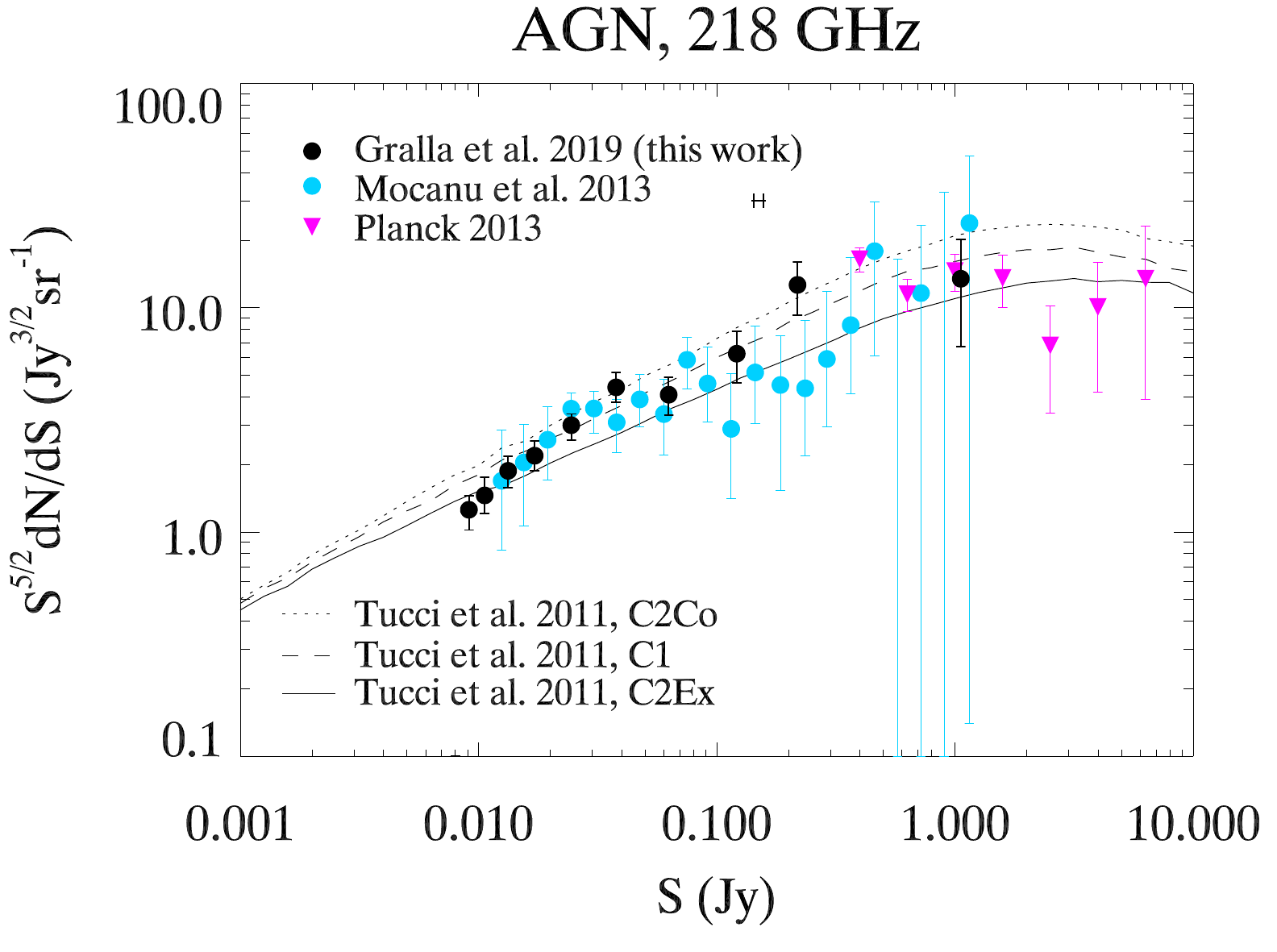}}
\hfill
\subfigure{\includegraphics[width=8.9cm]{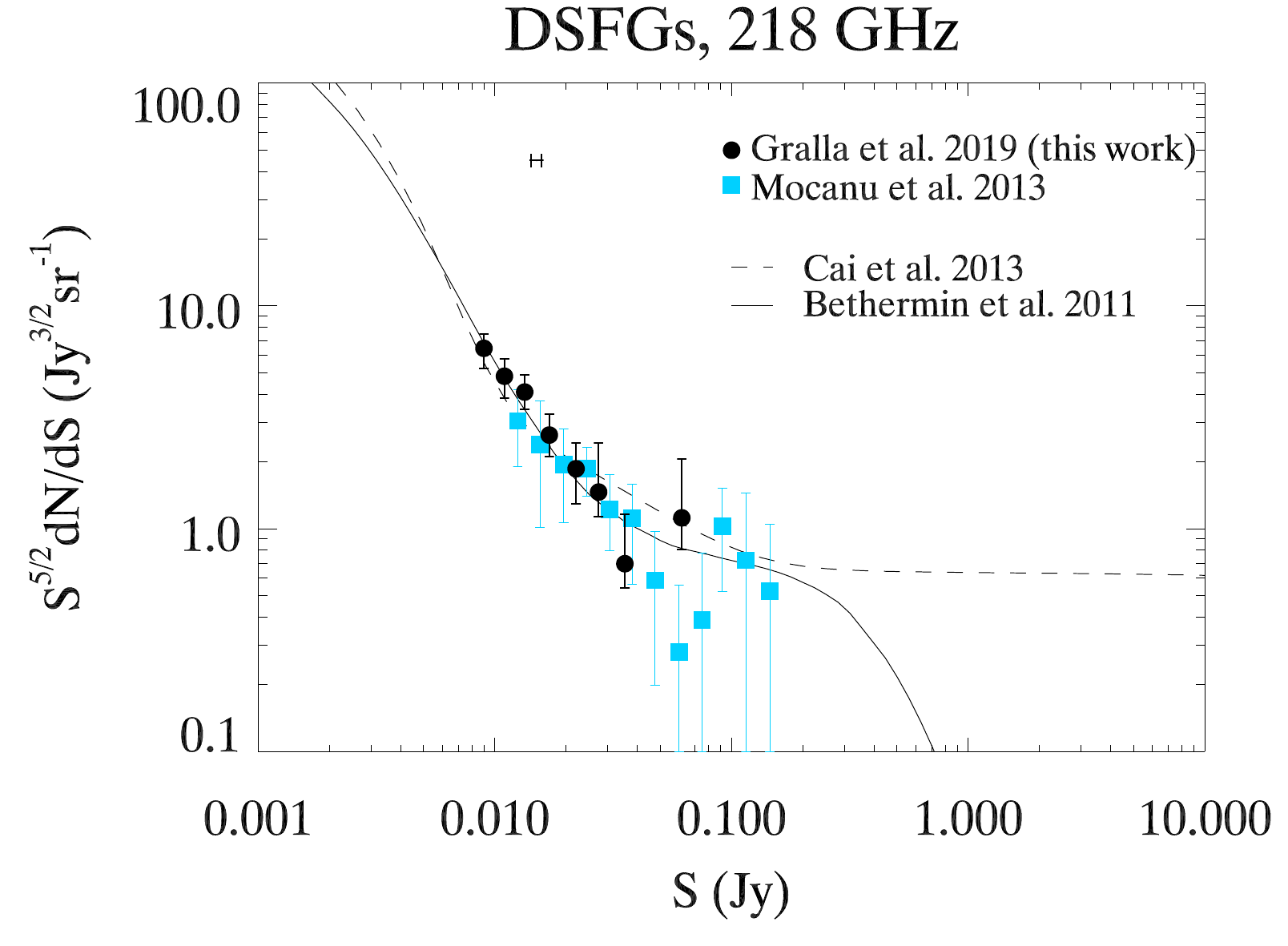}}
\hfill
\subfigure{\includegraphics[width=8.9cm]{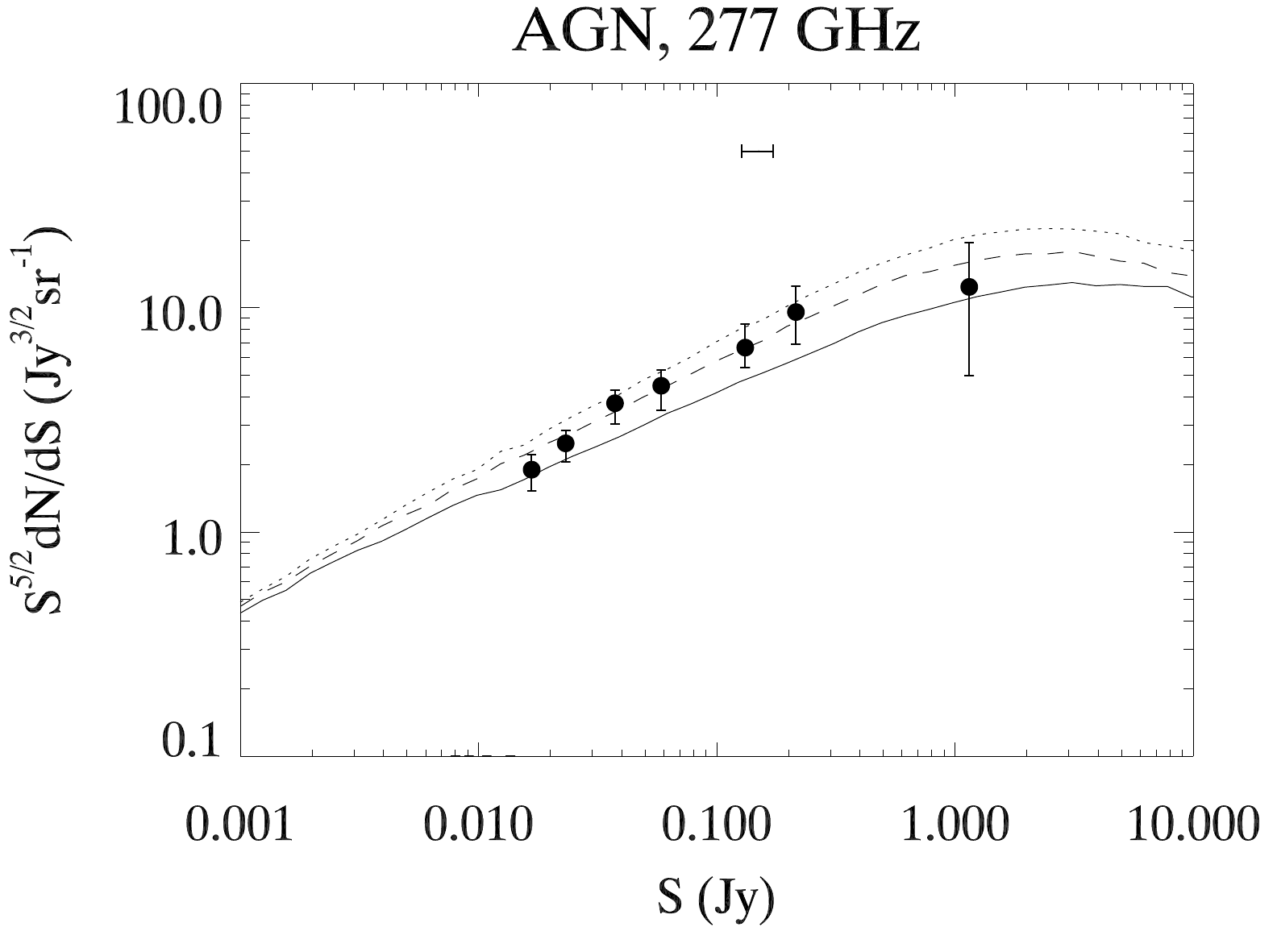}}
\hfill
\subfigure{\includegraphics[width=8.9cm]{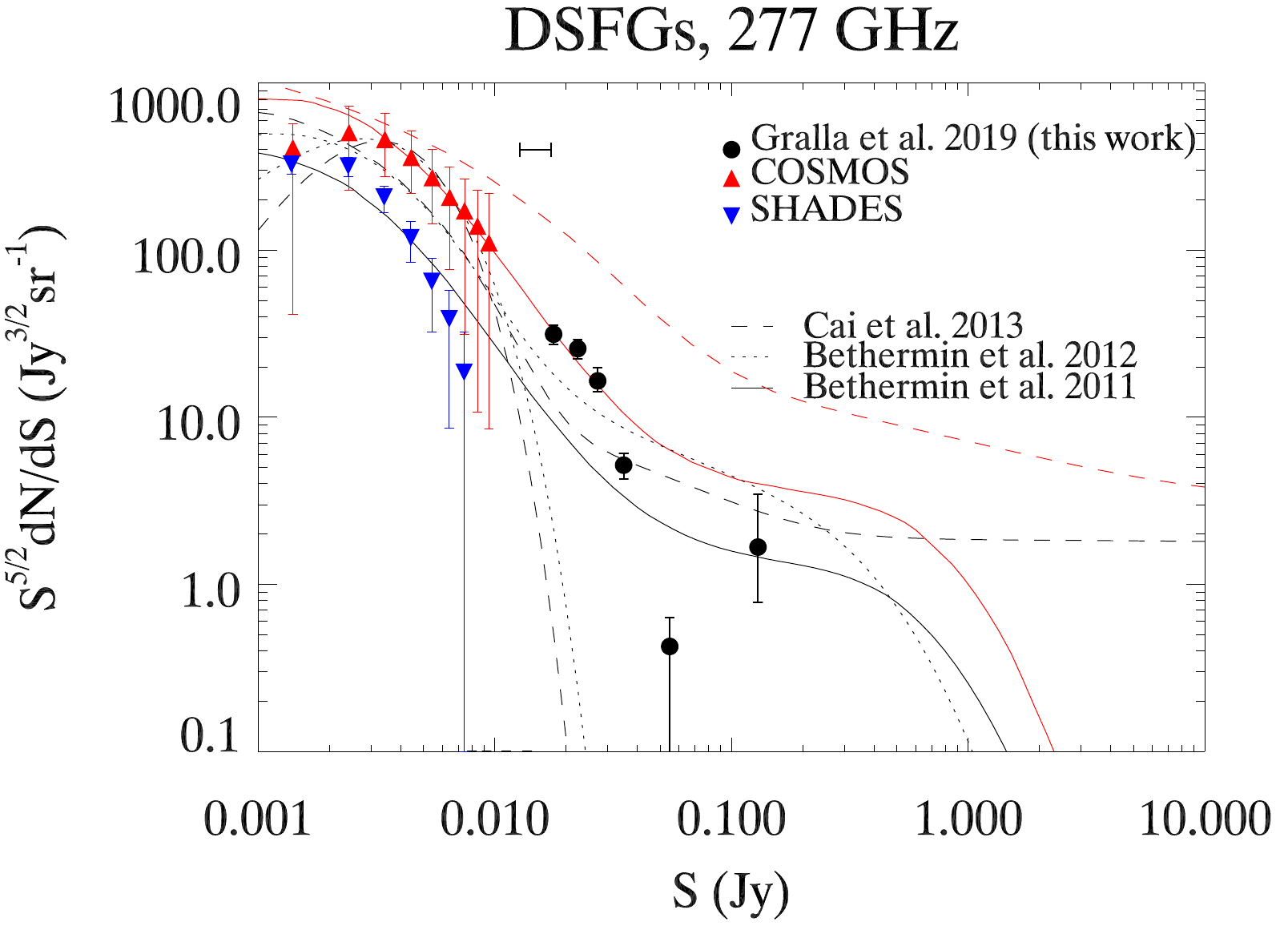}}

\caption{Source number counts for the ACT AGN and DSFG source samples. Black circles represent the source counts of the catalog presented in this paper, calculated by sampling the posterior distributions of the debiased flux densities (as described in Section \ref{section:counts}) for all but the bright ($S>90$~mJy) AGN, which are calculated directly from the catalogs. Other data sets and models are presented as labeled in the legends. A horizontal black error bar at the top of each plot indicates the band-dependent calibration uncertainty. In all plots, the counts have been corrected for completeness, values for which are listed in Table \ref{tab:counts} for each flux density bin. On the left, AGN source counts are shown in each ACT band. The primary ACT selection band for AGN is 148~GHz. In the 277~GHz plot, the model lines are calculated by scaling the 218~GHz models by a factor that corresponds to the median 218-277 spectral index ($-0.33$) for all AGN. 
On the right, DSFG source counts are shown in each ACT band. The DSFGs are selected from the MMF data. 
In the 277~GHz plot, the model lines shown in red are calculated by scaling the 218~GHz models by a factor that corresponds to the median 218-277 spectral index (2.7) for MMF sources. 
\label{fig:counts}}
\end{figure*}

The differential source number counts for each ACT band and for each subpopulation (AGN and DSFGs) are shown in Figure \ref{fig:counts} and listed in Table \ref{tab:counts}.  The sources have been classified as AGN if their spectral index from 148 to 218~GHz (based on the measured flux densities, not the debiased flux densities) is less than 1, and as DSFGs if this spectral index is greater than 1, as discussed in Section \ref{section:specindices}. For 148~GHz sources below 90~mJy and for all MMF-selected sources, the source counts are constructed using full posterior distributions (as described in detail below) of the debiased flux densities (when available, for sources with $S_{148}>50$~mJy the measured flux densities are used).  Above 90~mJy for the 148~GHz AGN counts, we use all the available survey area (949 square degrees).
Below 90~mJy, we restrict the 148~GHz AGN counts to the 505 square degrees in the main survey field (Figure \ref{fig:survey_fig}). We further restrict the DSFG counts to the MMF selection and map, which is 277.2 square degrees.

The source counts are corrected for incompleteness in the following manner. For a given source in the catalog, we select sources from the simulations described in Section \ref{section:debiasing} that have similar input flux densities. The tolerance used is the median 1$\sigma$ error on the source flux densities for the band and source category (AGN/DSFG) under consideration. 
For the set of simulated sources, we compute the fraction that are recovered ($f_{r, i}$). When computing the source counts, we weigh each $i$th source by $1/f_{r, i}$. 

The debiasing method provides posterior probability distributions for the intrinsic flux densities of all the sources in the catalogs. These full distributions should be taken into account when calculating the number counts from the debiased flux densities. 
In the case of the DSFGs, which have very steep source counts, using only the median debiased flux densities instead of the full probability distributions would introduce both a bias and scatter in the counts.  

To bring the full posterior probability distributions of the source flux densities into the source counts computation, 
we calculate the source counts using mock catalogs generated in the following way \citep[similar to what is done in ][]{Coppin2006, austermanncosmos,mocanu2013}. We make a list of mock flux densities by varying the debiased flux densities around their (asymmetric) errors. For AGN with $S_{148}>50$~mJy, for which debiased flux densities are not calculated, we vary the measured flux densities around their raw measurement errors. We generate a list that is 1,000 times the catalog sample size, calculating the completeness for every source by interpolating the completeness vs. flux density to the value for each of these flux densities. We then generate 1,000 mock samples, drawing from this flux density list. The size of each mock sample is drawn from a normal distribution with mean equal to the size of the catalog and error equal to the square root of the size of the catalog. Sources are permitted to fall below the 8~mJy threshold of the lowest flux density bin, so in practice the mock samples are smaller than the catalog.  We calculate the source counts for these 1,000 trials. For each flux density bin (excluding AGN with $S_{148}>90~$mJy), the source counts shown in Figure \ref{fig:counts} and reported in Table \ref{tab:counts} are the median source counts across these trials. The errors are calculated from the 16 and 84 percentiles of the distribution of the source counts from the trials, which are generally in agreement with the median of the Poisson, completeness-corrected errors. The value used as the central flux density for each bin is the median (across all mock catalogs) of the median (across all sources within the bin for each mock catalog) flux densities. This procedure, based on the debiased flux densities, naturally corrects for sources that should lie below the the lowest flux density bin being boosted into the catalog by noise.

Finally, there is an overall calibration uncertainty (as discussed in Section \ref{sec:data}) that also applies to the source counts. The uncertainty this introduces on the source counts is relatively small and is represented by the horizontal error bars in Figure \ref{fig:counts}.

One way our approach differs from the previous analyses that compute source counts in this way \citep{Coppin2006, austermanncosmos,mocanu2013}  is that we use forced photometry in secondary bands. This has the following implication: although our source-recovery simulations, based on primary-band source selection, may indicate that the secondary bands are reasonably complete in count bins below what would be that band's detection threshold, our ability to \emph{constrain} the secondary-band flux density of sources that fall in these bins is limited due to higher noise in the secondary band. This effect is most pronounced in the 277~GHz band. The 5~mJy noise in this band is poorly matched to  the $2-3$~mJy width of source-count bins at low flux densities. 
As pointed out in \citet{mocanu2013} (Section~5), for a 277~GHz catalog selected based directly on S/N in the 277~GHz data, the fraction of sources that falls out of the lowest count bins corrects the counts in those bins for the fraction of the sources that scattered up into the bin from lower flux densities. However, our 277~GHz catalogs are not selected with 277~GHz data, but with the lower-noise 148~GHz data for AGN and with the MMF for DSFGs. Because the completeness and impurity are not established by 277~GHz selection, the broad 277~GHz flux-density posterior distributions that spill out of the lowest count bins overcompensate for sources that may have scattered up into our bins from lower flux densities. We have run simulations that show this affects our bootstrapped 277~GHz counts that lie $1-2\sigma$ from the formal completeness limit (6~mJy for AGN and 10~mJy for DSFGs) at 277~GHz. In this regime, the counts are biased low. The simulations show that this impacts the three lowest 277~GHz count bins for AGN and  DSFGs, all of which we exclude from the results. This effect may also impact the lowest count bin for AGN at 218~GHz, which we include, but we caution against over-interpretation of any apparent downward trend in the low end of these counts. We note that in future work the simulations could be used to quantify the bias and correct the counts, but in this work we conservatively remove the affected flux density bins.

\begin{figure}
	\centering
	\includegraphics[width=84mm]{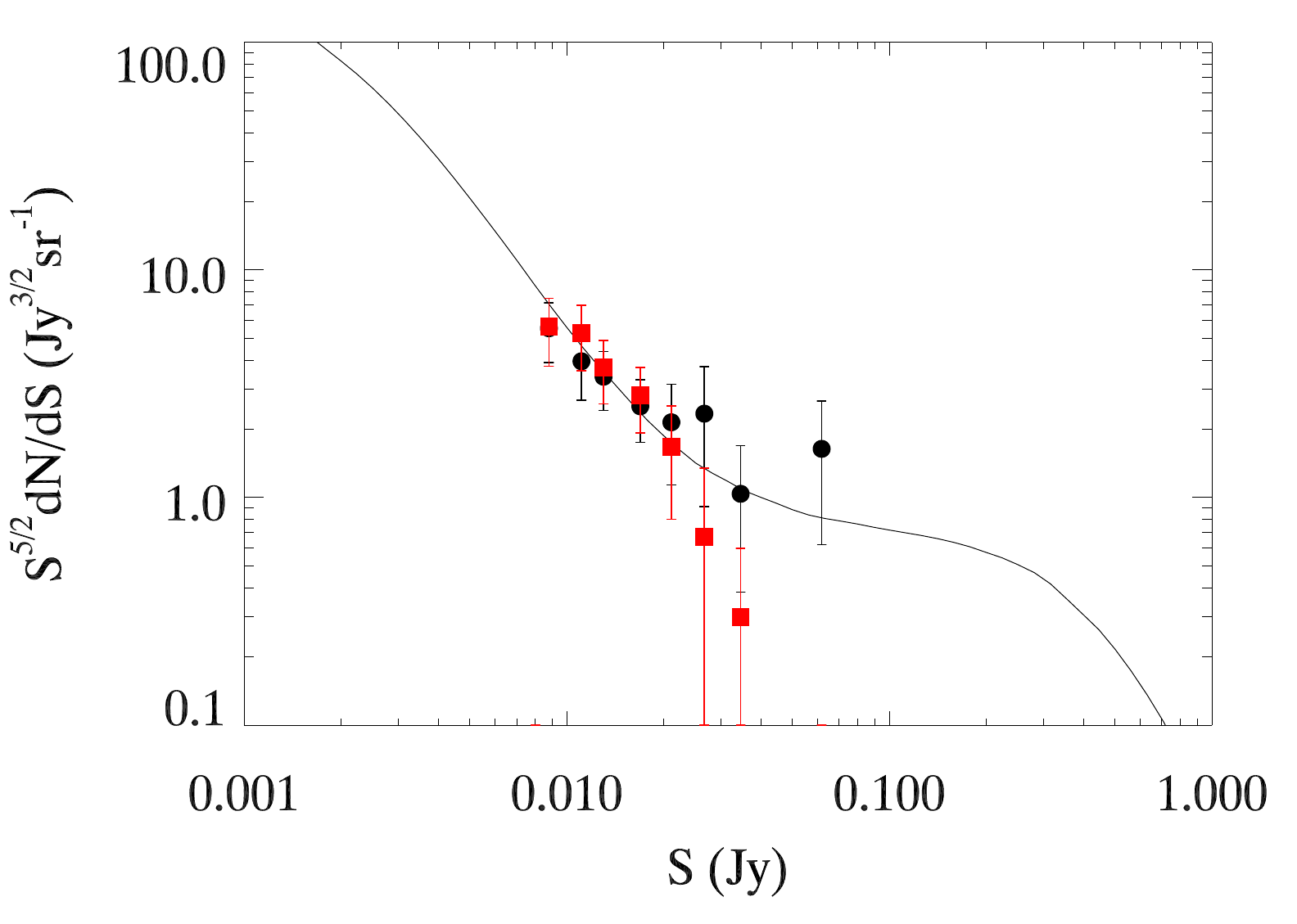}
\caption{A comparison of ``Galactic dust cleaned'' and ``Galactic dust free''  218~GHz source counts of dusty galaxies. The red squares represent the counts of ACT sources located in regions of the map from which we have removed Galactic emission using the criteria described in Section \ref{section:dustremoval}. The black circles represent the counts of ACT sources in regions of the map that appear to be free of Galactic emission, and thus no sources have been removed from these regions. Statistical agreement between the two samples indicates that Galactic contamination does not significantly impact the source counts.   \label{fig:dustfreecounts}}
\end{figure}

We have addressed the sample purity in three ways. First, in Section~\ref{sec:purity}, we considered false detections in maps multiplied by $-1$. We concluded that accounting for the few  spurious detections in the inverted maps does not impact source counts in a statistically meaniful way.  
Second, some sources that have intrinsically fainter flux densities could enter into the sample by being boosted above the detection threshold by noise. This is addressed by our use of the full posterior distributions of the debiased flux densities, as discussed above. Third, 
contamination from Galactic dust could add spurious sources to the catalog that may not be captured by finding sources in an inverted map. We discuss in Section~\ref{section:dustremoval} the criteria we use to remove sources from the catalog that are likely to be Galactic dust emission. To test for the possibility that residual Galactic dust can mimic sources in our catalogs, we also perform the following analysis. We select regions of the map where we have removed sources that are likely to be Galactic (``Galactic dust-removed regions''). We calculate the source counts from these regions and compare them with the source counts from regions of the map where we find no Galactic sources (``Galactic dust-free regions''). For this test, all sources lie within a declination range of $\pm 1.2\degree$ about the celestial equator.  
The Galactic dust-removed regions lie in the ranges $03^{\mathrm h}~20^{\mathrm m} < \mathrm{R.A.} < 03^{\mathrm h}~51^{\mathrm m}$, $0^{\mathrm h}~4^{\mathrm m}  < \mathrm{R.A.} < 0^{\mathrm h}~40^{\mathrm m} $, $21^{\mathrm h}~40^{\mathrm m}  < \mathrm{R.A.} < 22^{\mathrm h}~40^{\mathrm m}$, and $20^{\mathrm h}~9^{\mathrm m}  < \mathrm{R.A.} < 21^{\mathrm h}$. The Galactic dust-free regions are within ranges of $02^{\mathrm h}~48^{\mathrm m}  < \mathrm{R.A.} < 03^{\mathrm h}~20^{\mathrm m}$, $0^{\mathrm h}~40^{\mathrm m}  < \mathrm{R.A.} < 01^{\mathrm h}~44^{\mathrm m} $, $23^{\mathrm h} < \mathrm{R.A.} < 0^{\mathrm h}~4^{\mathrm m}$, and $21^{\mathrm h} < \mathrm{R.A.} < 21^{\mathrm h}~40^{\mathrm m}$. Figure \ref{fig:dustfreecounts} shows the source counts from these dust-free and dust-removed regions.  Below 25~mJy, The dust-removed regions tend to have higher source counts than the dust-free regions, but still lie within the 1$\sigma$ uncertainties. This implies that Galactic dust does not significantly contaminate the source counts of the dusty galaxies. Above 25~mJy, there is marginal tension (2$\sigma$) in the distribution of 13 sources between the two regions, favoring more sources in the dust free regions. Therefore this is still consistent with no contamination from Galactic emission. Interpretation of this as evidence for incompleteness due to Galactic cuts is disfavored by simulations (Figure \ref{fig:completeness}, red curves).

\begin{table*}[htb]
\centering
\caption{Source counts}

\begin{tabular}{|c|cccccc|}
\hline
  
  Flux density\footnote{Source counts presented here are calculated by sampling the posterior distributions of the debiased flux densities for all but the bright ($S>90$~mJy) AGN (as described in Section \ref{section:counts}).} & \multicolumn{6}{c|}{AGN} \\
  
    (mJy) & N/Compl.$_{148}\footnote{For each flux density bin, N corresponds to a completeness-corrected median number of sources for that flux density bin. The Compl. value corresponds to the median completeness for the sources in that bin.  To calculate the number of sources that would typically be measured in a bin, multiply N by Compl. }$  & N/Compl.$_{218}$ & N/Compl.$_{277}$  & ($\frac{dN}{dS}$ S$^{5/2})_{148}$ & ($\frac{dN}{dS}$ S$^{5/2})_{218}$ & ($\frac{dN}{dS}$ S$^{5/2})_{277}$  \\\hline
    8  -    10 & 26/0.26 & 31/0.62 & 20/0.64 &$ 2.5_{-0.2}^{+0.2 }$ &$ 1.3_{-0.2}^{+0.2 }$ & \\
  10  -    12 & 46/0.54 & 28/0.84 & 19/0.72 &$ 3.7_{-0.3}^{+0.3 }$ &$ 1.5_{-0.2}^{+0.3 }$ & \\
  12  -    15 & 46/0.71 & 38/0.92 & 27/0.83 &$ 2.6_{-0.3}^{+0.3 }$ &$ 1.9_{-0.3}^{+0.3 }$ & \\
  15  -    20 & 46/0.90 & 43/0.97 & 36/0.98 &$ 2.6_{-0.4}^{+0.4 }$ &$ 2.2_{-0.4}^{+0.3 }$ &$ 1.9_{-0.3}^{+0.3 }$\\
  20  -    30 & 53/0.98 & 49/1.00 & 42/0.99 &$ 3.2_{-0.4}^{+0.4 }$ &$ 3.0_{-0.4}^{+0.4 }$ &$ 2.5_{-0.4}^{+0.4 }$\\
  30  -    50 & 56/1.00 & 49/1.00 & 40/1.00 &$ 4.9_{-0.7}^{+0.6 }$ &$ 4.4_{-0.6}^{+0.5 }$ &$ 3.7_{-0.6}^{+0.5 }$\\
  50  -    90 & 37/1.00 & 26/1.00 & 29/1.00 &$ 6.0_{-1.0}^{+1.0 }$ &$ 4.0_{-0.8}^{+0.8 }$ &$ 4.5_{-0.9}^{+0.8 }$\\
  90  -   170\footnote{Above 90~mJy, the 148 GHz AGN counts are reported using the full 950 square degree survey area and measured (not debiased) flux densities. For 218 and 277~GHz, and for 148~GHz below 90~mJy, the counts are reported from a survey area of 505 square degrees.} & 34/1.00 & 15/1.00 & 13/1.00 & 7.2$\pm$1.2 & 6.2$\pm$1.6 & 6.6$\pm$1.8 \\
 170  -   330 & 21/1.00 & 14/1.00 & 11/1.00 & 12.2$\pm$2.7 & 12.7$\pm$3.4 & 9.6$\pm$2.9 \\
 330  -   650 & 10/1.00 &    &     & 9.1$\pm$2.9 &   &   \\
 650  -  2870 &  7/1.00 &  4/1.00 &  3/1.00 & 19.6$\pm$7.4 & 13.5$\pm$6.7 & 12.4$\pm$7.2 \\
\hline\hline
  Flux density & \multicolumn{6}{c|}{DSFGs}\\
  \hline

     8  -    10\footnote{DSFG counts are reported from the MMF survey area, which covers 277 square degrees.} &  2/$<0.01$& 32/0.22&  5/$<0.01$ &$ 0.1_{-0.2}^{+0.2 }$ &$ 6.5_{-1.2}^{+1.0 }$ & \\
  10  -    12 &  2/$<0.01$& 28/0.46&  8/0.02 &$ 0.2_{-0.1}^{+0.1 }$ &$ 4.8_{-1.0}^{+1.0 }$ & \\
  12  -    15 &   & 31/0.61& 20/0.06 &  &$ 4.1_{-0.7}^{+0.8 }$ & \\
  15  -    20 &  2/0.86& 25/0.84& 47/0.17 &$ 0.2_{-0.1}^{+0.1 }$ &$ 2.6_{-0.5}^{+0.6 }$ &$ 31.4_{-4.1}^{+4.1 }$\\
  20  -    25 &   & 10/0.91& 48/0.38 &   &$ 1.9_{-0.6}^{+0.6 }$ &$ 25.8_{-3.3}^{+3.3 }$\\
  25  -    30 &   &  4/0.91& 31/0.52 & &$ 1.5_{-0.3}^{+1.0 }$ &$ 16.5_{-2.2}^{+3.3 }$\\
  30  -    50 &   &  4/0.90& 31/0.91 & &$ 0.7_{-0.2}^{+0.5 }$ &$ 5.2_{-0.9}^{+0.9 }$\\
  50  -    90 &   &  3/0.90&  2/1.00 & &$ 1.1_{-0.3}^{+0.9 }$ &$ 0.4_{-0.4}^{+0.2 }$\\
  90  -   170 &    &    &  1/1.00 &  &  &$ 1.7_{-0.9}^{+1.8 }$\\
\\
\hline
\end{tabular}
\label{tab:counts}
\end{table*}

%

\subsubsection{Discussion of source counts}
The models that describe the synchrotron source counts are based on statistical extrapolations from lower frequency radio source counts (i.e., 5~GHz). \citet{tucci} provide models using different physically motivated prescriptions for the spectral behavior of the sources. They include three populations of sources, as classified by their 1--5~GHz spectral indices: steep spectrum ($\alpha < 1$), inverted  spectrum ($\alpha > 1$), and flat spectrum ($\alpha\sim0$). The latter include sub-populations of both BL Lacs and flat spectrum radio quasars (FSRQs). 
The models differ in the frequency at which there is a break in the synchrotron spectrum for the flat spectrum sources. The break frequency indicates the transition from optically thick to optically thin synchrotron emission and thus is related to the size of the region at which this transition occurs. For the \citet{tucci} C1 model, BL Lacs and FSRQs both have the same size transition region, between 0.01 and 10~pc. For their C2Co model, the transition region is more compact in FSRQs than in BL Lacs. For their C2Ex model, the region is more extended in FSRQs than in BL Lacs. Although they do not provide 277~GHz models, we have scaled their 218~GHz models up to 277~GHz using the median spectral index for the sample.

We find that the preferred \citet{tucci} model, the C2Ex model, under-predicts the 148~GHz synchrotron source counts, most noticeably in the intermediate flux density range of $\sim10$ to $\sim100$~mJy (see Figure \ref{fig:counts}). Our results in this range agree with the number counts in \citet{mocanu2013}, who noted this discrepancy with the C2Ex model. When we calculate how many sources the model would predict given our survey area and completeness and compare with the counts in each flux density bin, the $\chi^2$ of the fit of the C2Ex model to our 148~GHz data is 51.3, with 11 degrees of freedom. Other models presented by \citet{tucci} perform better. The fit of the C1 model and the C2Co model produce $\chi^2$ of 21.5 and 22.3, respectively. None of the models are formally good fits, but the C1 and C2Co are much better than the C2Ex.
The preference for the C2Ex model in \citet{tucci} is driven by the better fit to the $> 0.5$~Jy number counts from \citet{PlanckSources2011} at 148 and 220~GHz. We note that these were based on the early release catalogs, which agree (but with larger uncertainties) with the counts of the intermediate catalogs \citep[][shown in Figure \ref{fig:counts}]{planck2013}. These results may indicate differences in the spectral behavior of the AGN populations that contribute to the intermediate vs the brightest number counts.

The models for the dusty sources are from \citet{bethermin2011}, \citet{bethermin2012}, and \citet{cai2013}. \citet{bethermin2011} use a parametric backward evolution model to describe the cosmological evolution of the luminosity function of infrared galaxies. \citet{cai2013} combine a descriptive backward parametric model for late-type galaxies and AGN below a  redshift of 1.0 with a physically motivated forward model for the evolution of spheroids and AGN above a redshift of 1.5. At the bright end, a population of lensed dusty galaxies is expected to dominate the source counts. Toward fainter flux densities, the unlensed component becomes more important. At 277~GHz, we also show (in red) models from 218~GHz that have been scaled up to 277~GHz by the median 218-277~GHz spectral index for the sample.

\begin{figure*}[th!]
	\centering
	\includegraphics[width=.32\textwidth]{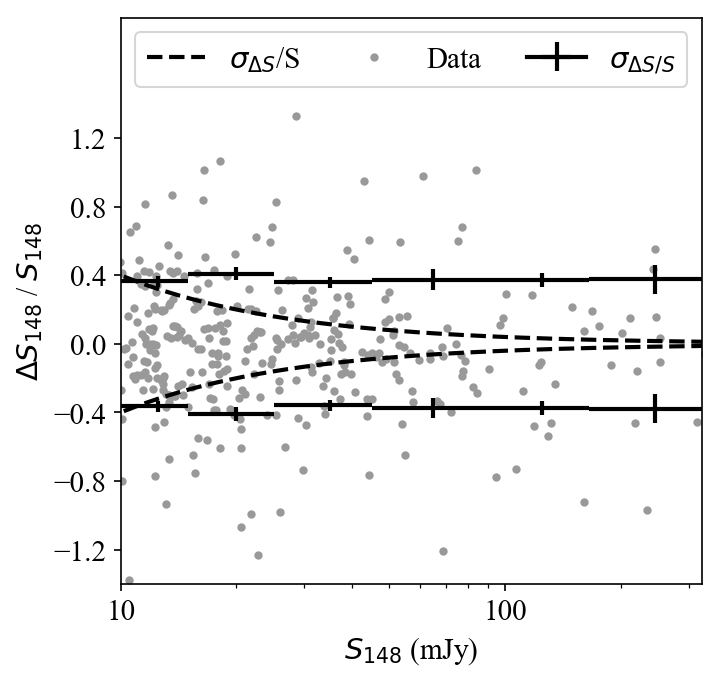} 
	\includegraphics[width=.32\textwidth]{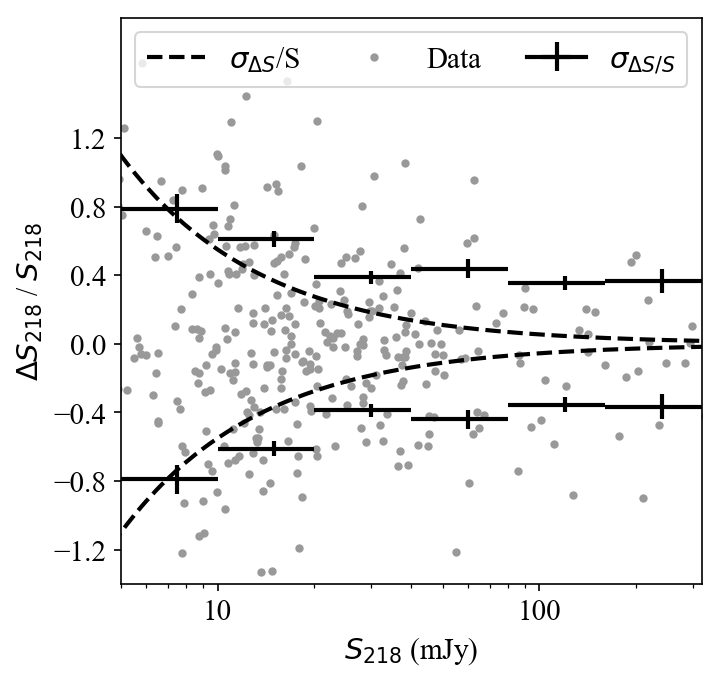}
	\includegraphics[width=.32\textwidth]{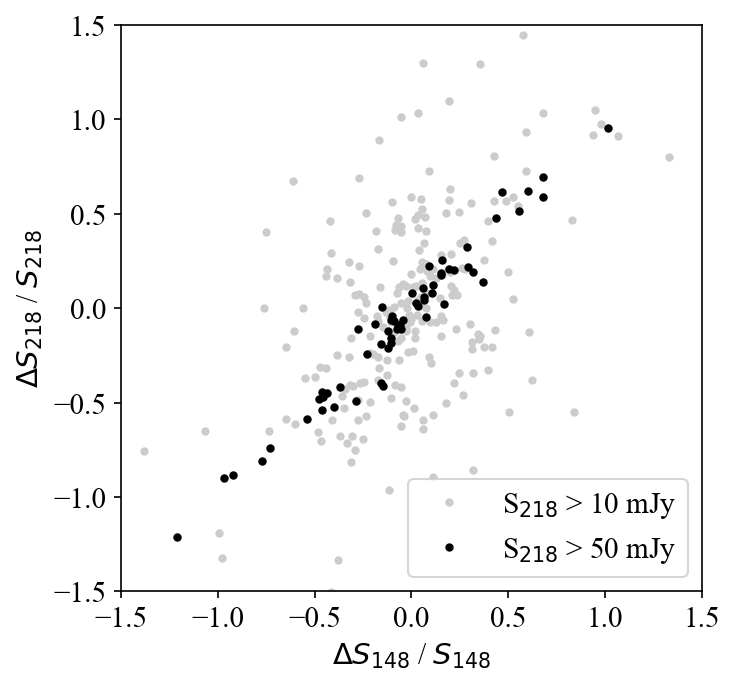}
\caption{Variability of AGN emission. The left two plots show, for 148~GHz and 218~GHz, the fractional deviation in  flux density between 2009 and 2010. The dashed curve shows the prediction for the RMS deviation due to map noise. The points with error bars show flux-density-binned estimates of the RMS deviation of the sample. At low flux densities the inter-year variation is consistent with noise, but at higher flux densities the intrinsic source variability dominates at 40\%. The right plot shows that the variations in the two bands closely track one another in the flux density regime ($>50$~mJy) where intrinsic variability dominates. }
\label{fig:variability}
\end{figure*}

We find good agreement between the DSFG source counts and the models for sources with 218~GHz flux densities above $\sim20$~mJy (see Figure \ref{fig:counts}). These brighter sources are expected to be drawn from a strongly lensed source population, and the lensed nature of these sources is being tested and confirmed with follow-up observations (see Appendix \ref{app:dsfg}). The ACT source counts also agree well with the $>20$~mJy counts from SPT \citep{mocanu2013}\footnote{In addition to SPT counts of DSFGs, counts at $\sim218$~GHz are provided in \citet{planck2013}.  They lie well above both the model predictions and the ACT and SPT counts and are at higher fluxes.  Note, however, that the Planck counts were based on the Early Release Compact Source Catalog, which may have been contaminated by Galactic cirrus.  Planck's later PCCS2 catalog \citep{PlanckSources2016} has improved upon the identification and removal of Galactic dust emission, but the collaboration has not published revised source counts.}. At lower flux densities ($\sim8-13$~mJy at 218~GHz), the ACT sample is likely starting to probe the bright end of the unlensed source population. The 218~GHz source counts data continue to agree well with the models at the faint end, where the more sensitive ACT data push below the limits of current surveys. 

The 277~GHz DSFG source counts presented here are the first published in this flux density range. The source counts at fainter flux densities are constrained by observations from AzTEC in two regions: COSMOS \citep{austermanncosmos} and SHADES \citep{austermannshades}. These earlier studies were in some tension with one another, and the dispersion among the models reflects this. Our higher flux density data prefer the higher models, but none lie within perfect agreement in this new regime.

\subsection{Variability}
\label{subsec:variability}

The emission from blazars shows significant variability correlated across the electromagnetic spectrum  that can give insights into the  processes (e.g., turbulence) in the AGN jet \citep[e.g.,][]{Burbidge1992,Hughes1992,Aller1999,Tingay2003,Ciprini2007,Abdo2010,Fan2018, ORiordan2017}. To probe AGN variability of this sample, we include in the extragalactic catalog  per-year flux density estimates for AGN-classified sources selected at 148~GHz (i.e., $\sim$95\% of AGN). Specifically 148~GHz and 218~GHz raw flux densities (not debiased) are provided separately for the observing seasons 2009 and 2010. Figure \ref{fig:variability} summarizes these data. The left two plots show the RMS (68\% c.l.) fractional deviation in flux densities between years (indicated by data points) along with the  expected RMS deviation due to map noise (indicated by the dashed line). The error on the RMS deviation data is computed with a bootstrap Monte Carlo within each bin. At the lowest flux densities (and thus lowest S/N), the observed RMS deviation is consistent with the expectation from noise. However, above 30~mJy, the inter-year scatter becomes dominated by source variability with an RMS deviation at the 40\% level. This is true for both the 148~GHz and the 218~GHz frequency bands.  The 40\% RMS deviation shows no dependence on flux density. 

We note miscalibration can produce an overall shift between years; however, the systematic uncertainties at 148~GHz and 218~GHz are at the percent level, negligible compared to the 40\% variation observed. Furthermore, a significant systematic would produce an asymmetry around zero flux density in the left two plots of Figure \ref{fig:variability}, an asymmetry that is not observed.

The right plot in Figure \ref{fig:variability} shows that the flux density variations are strongly correlated between bands at flux densities ($>50$~mJy) where the source variability completely dominates. For the fractional flux density variation in these brightest sources, the inter-band correlation coefficient is 0.98.

%

\begin{table*}[t]
\centering
\caption{Partial Catalog Entries for ACT DSFGs Selected for Detailed Study}
\begin{tabular}{ccccccccc}
\hline
 ACT-S ID\footnote{The ACT-S ID encodes the sexagesimal position of each source (hhmmss$\pm$ddmmss).} & S/N & $\alpha_{148}^{218}$\footnote{For inter-band spectral indices ($\alpha_{X}^{Y}$) and flux densities ($S_X$), raw (debiased) values are given outside (inside) parentheses.} & $\alpha_{218}^{277}$ & $S_{148}$ & $S_{218}$ & $S_{277}$  \\
(J2000)&  &  & & (mJy) & (mJy) & (mJy) \\\hline\hline
001133$-$001835 & 6.6 & 3.4 (3.3$^{+1.2}_{-0.9}$) & 0.7 (1.1$^{+1.2}_{-1.3}$)& 5.9$\pm$2.2 (5.3$^{+2.1}_{-1.9}$) & 22.1$\pm$3.4 (20.0$^{+3.2}_{-3.7}$) & 26.3$\pm$8.2 (24.4$^{+6.9}_{-6.4}$)\\ 
002220$-$015523 & 6.5 & 4.3 (4.2$^{+1.1}_{-0.9}$) & 1.8 (2.2$^{+0.9}_{-0.8}$)& 5.2$\pm$2.2 (5.0$^{+2.0}_{-1.8}$) & 27.2$\pm$4.2 (26.3$^{+3.3}_{-4.2}$) & 41.5$\pm$5.4 (41.9$^{+5.2}_{-5.1}$)\\ 
003814$-$002255 & 10.7 & 3.0 (3.0$^{+0.8}_{-0.7}$) & 1.9 (2.0$^{+0.9}_{-0.8}$)& 7.6$\pm$1.8 (7.1$^{+1.8}_{-1.7}$) & 24.6$\pm$2.7 (23.5$^{+2.7}_{-3.8}$) & 39.1$\pm$5.5 (36.9$^{+5.3}_{-5.3}$)\\ 
003929$+$002422 & 8.9 & 3.2 (3.1$^{+1.1}_{-1.0}$) & 2.1 (2.2$^{+1.1}_{-1.0}$)& 5.7$\pm$1.8 (5.1$^{+1.8}_{-1.6}$) & 20.0$\pm$2.7 (17.7$^{+3.9}_{-4.5}$) & 33.4$\pm$5.5 (29.9$^{+5.3}_{-5.2}$)\\ 
004410$+$011818 & 13.3 & 2.8 (2.8$^{+0.5}_{-0.5}$) & 2.9 (3.1$^{+0.6}_{-0.5}$)& 12.1$\pm$1.7 (11.7$^{+1.8}_{-1.7}$) & 35.7$\pm$2.7 (34.7$^{+3.6}_{-4.7}$) & 71.0$\pm$5.1 (70.9$^{+5.0}_{-5.0}$)\\ 
004532$-$000127 & 11.1 & 3.8 (3.8$^{+0.9}_{-0.8}$) & 2.1 (2.1$^{+0.9}_{-0.8}$)& 5.9$\pm$1.8 (5.6$^{+1.7}_{-1.7}$) & 25.9$\pm$2.8 (25.0$^{+3.5}_{-4.3}$) & 42.4$\pm$5.6 (40.7$^{+5.4}_{-5.3}$)\\ 
010729$+$000114 & 7.3 & 6.2 (4.7$^{+1.5}_{-1.3}$) & 1.6 (1.8$^{+1.3}_{-1.2}$)& 1.6$\pm$1.8 (2.2$^{+1.4}_{-1.0}$) & 18.5$\pm$2.8 (15.8$^{+3.6}_{-4.2}$) & 27.1$\pm$5.5 (23.0$^{+5.0}_{-5.0}$)\\ 
011640$-$000457 & 10.8 & 2.8 (2.8$^{+0.8}_{-0.8}$) & 2.5 (2.6$^{+0.9}_{-0.9}$)& 7.7$\pm$1.8 (7.2$^{+1.7}_{-1.7}$) & 22.8$\pm$2.7 (21.2$^{+3.8}_{-4.0}$) & 42.0$\pm$5.4 (39.9$^{+5.3}_{-5.3}$)\\ 
013857$+$021420 & 8.9 & 2.8 (2.9$^{+0.6}_{-0.5}$) & -- & 14.2$\pm$2.9 (13.5$^{+2.9}_{-2.8}$) & 41.9$\pm$4.7 (41.1$^{+3.2}_{-3.6}$) & -- \\ 
020941$+$001557 & 35.5 & 3.7 (3.7$^{+0.3}_{-0.3}$) & 3.1 (3.6$^{+0.2}_{-0.2}$)& 17.2$\pm$1.8 (17.0$^{+1.8}_{-1.7}$) & 71.5$\pm$2.7 (71.1$^{+2.6}_{-3.5}$) & 152.0$\pm$5.4 (169.1$^{+5.3}_{-5.3}$)\\ 
022830$-$005226 & 14.1 & 3.1 (3.0$^{+0.7}_{-0.6}$) & 3.0 (3.4$^{+0.5}_{-0.5}$)& 8.5$\pm$1.8 (8.1$^{+1.7}_{-1.8}$) & 28.7$\pm$2.9 (26.7$^{+2.0}_{-3.0}$) & 58.8$\pm$5.3 (59.5$^{+5.2}_{-5.2}$)\\ 
023120$+$011636 & 6.5 & 3.3 (3.0$^{+1.2}_{-1.2}$) & 1.6 (2.9$^{+1.7}_{-1.2}$)& 5.1$\pm$1.7 (4.4$^{+1.7}_{-1.6}$) & 18.1$\pm$2.8 (14.2$^{+4.2}_{-6.8}$) & 26.8$\pm$4.6 (26.6$^{+4.5}_{-4.5}$)\\ 
025331$-$000318 & 27.7 & 3.4 (3.4$^{+0.3}_{-0.3}$) & 2.8 (3.2$^{+0.3}_{-0.3}$)& 15.9$\pm$1.8 (15.7$^{+1.8}_{-1.8}$) & 59.0$\pm$2.9 (58.1$^{+3.0}_{-2.8}$) & 114.8$\pm$5.4 (125.5$^{+5.3}_{-5.4}$)\\ 
025512$-$011456 & 7.2 & 3.0 (2.9$^{+1.0}_{-1.0}$) & 2.2 (2.9$^{+1.3}_{-1.0}$)& 6.1$\pm$1.7 (5.5$^{+1.7}_{-1.6}$) & 19.4$\pm$2.7 (16.8$^{+4.2}_{-6.4}$) & 32.7$\pm$5.0 (32.6$^{+4.8}_{-4.8}$)\\ 
030410$+$013225 & 6.6 & 5.0 (4.0$^{+1.5}_{-1.5}$) & 4.1 (4.6$^{+1.2}_{-1.1}$)& 2.5$\pm$1.7 (2.3$^{+1.5}_{-1.1}$) & 17.6$\pm$2.6 (13.4$^{+4.5}_{-6.8}$) & 46.7$\pm$7.6 (44.6$^{+7.4}_{-7.4}$)\\ 
031019$-$000215 & 8.6 & 8.3 (5.5$^{+1.4}_{-1.2}$) & 1.9 (1.9$^{+1.0}_{-0.9}$)& 0.8$\pm$1.8 (2.1$^{+1.2}_{-0.9}$) & 21.1$\pm$2.9 (19.4$^{+3.3}_{-3.7}$) & 33.1$\pm$5.1 (29.9$^{+4.9}_{-4.8}$)\\ 
031127$+$013639 & 12.8 & 4.1 (4.1$^{+0.8}_{-0.7}$) & 2.9 (3.1$^{+0.7}_{-0.7}$)& 7.0$\pm$1.9 (6.8$^{+1.8}_{-1.8}$) & 34.4$\pm$2.7 (33.5$^{+3.6}_{-3.2}$) & 69.5$\pm$10.4 (69.3$^{+10.0}_{-9.8}$)\\ 
032104$+$012934 & 6.2 & 5.7 (4.2$^{+1.6}_{-1.5}$) & 0.7 (2.9$^{+1.7}_{-1.5}$)& 1.8$\pm$1.7 (2.0$^{+1.2}_{-1.0}$) & 16.6$\pm$2.7 (12.2$^{+4.6}_{-5.0}$) & 19.6$\pm$6.1 (19.8$^{+5.6}_{-5.3}$)\\ 
032121$-$000221 & 11.9 & 2.8 (2.8$^{+0.8}_{-0.7}$) & 3.1 (3.2$^{+0.8}_{-0.8}$)& 7.9$\pm$1.8 (7.4$^{+1.8}_{-1.7}$) & 23.7$\pm$2.9 (22.3$^{+4.4}_{-4.1}$) & 49.6$\pm$5.2 (48.7$^{+5.0}_{-5.1}$)\\ 
032351$+$012801 & 8.9 & 4.1 (4.0$^{+1.0}_{-0.9}$) & 0.7 (1.7$^{+1.2}_{-1.1}$)& 4.8$\pm$1.6 (4.6$^{+1.5}_{-1.5}$) & 23.8$\pm$2.7 (22.1$^{+4.0}_{-4.4}$) & 27.9$\pm$5.8 (29.3$^{+5.3}_{-5.3}$)\\ 
034003$+$001627 & 8.0 & 4.1 (3.7$^{+1.3}_{-1.2}$) & 2.3 (2.4$^{+1.2}_{-1.1}$)& 3.7$\pm$1.8 (3.4$^{+1.6}_{-1.4}$) & 18.1$\pm$2.8 (15.3$^{+3.8}_{-4.5}$) & 31.1$\pm$5.1 (27.1$^{+4.9}_{-4.8}$)\\ 
034228$-$005644 & 8.9 & 2.5 (2.3$^{+1.0}_{-1.0}$) & 3.0 (3.2$^{+1.2}_{-1.0}$)& 6.8$\pm$1.8 (6.0$^{+1.8}_{-1.7}$) & 17.7$\pm$2.8 (14.5$^{+4.1}_{-4.9}$) & 36.6$\pm$5.2 (33.0$^{+5.1}_{-5.0}$)\\ 
202955$+$012054 & 7.4 & 2.9 (2.8$^{+1.0}_{-0.8}$) & 3.3 (3.7$^{+0.9}_{-0.8}$)& 7.2$\pm$2.0 (6.6$^{+2.0}_{-2.0}$) & 22.0$\pm$3.0 (19.8$^{+3.2}_{-4.1}$) & 48.6$\pm$5.8 (48.0$^{+5.6}_{-5.6}$)\\ 
212740$+$010921 & 8.7 & 5.2 (4.6$^{+1.4}_{-1.1}$) & 1.7 (1.8$^{+1.0}_{-1.0}$)& 3.0$\pm$2.1 (3.3$^{+1.7}_{-1.4}$) & 22.8$\pm$2.9 (21.1$^{+3.9}_{-3.9}$) & 34.2$\pm$6.0 (31.3$^{+5.7}_{-5.6}$)\\ 
213511$-$010255 & 14.9 & -- (7.5$^{+1.2}_{-1.1}$) & 3.8 (4.0$^{+0.5}_{-0.5}$)& -3.2$\pm$2.1 (1.8$^{+1.0}_{-0.7}$) & 34.8$\pm$3.0 (34.1$^{+3.7}_{-3.6}$) & 85.4$\pm$6.8 (90.3$^{+6.7}_{-6.7}$)\\ 
213713$+$011156 & 8.2 & 2.8 (2.8$^{+1.0}_{-0.8}$) & 1.4 (1.5$^{+1.1}_{-1.1}$)& 7.1$\pm$2.1 (6.4$^{+2.0}_{-2.0}$) & 21.1$\pm$3.0 (19.5$^{+3.3}_{-3.8}$) & 29.9$\pm$5.7 (26.6$^{+5.3}_{-5.1}$)\\ 
230239$-$013333 & 6.8 & 3.7 (3.6$^{+1.2}_{-1.0}$) & 3.2 (3.5$^{+0.9}_{-0.7}$)& 6.1$\pm$2.5 (5.6$^{+2.3}_{-2.1}$) & 25.1$\pm$3.7 (23.6$^{+3.2}_{-4.2}$) & 53.8$\pm$6.6 (53.3$^{+6.3}_{-6.4}$)\\ 
231252$-$001524 & 7.3 & -- (6.2$^{+1.4}_{-1.3}$) & 1.7 (1.8$^{+1.1}_{-1.0}$)& -2.2$\pm$2.1 (1.6$^{+1.0}_{-0.7}$) & 21.3$\pm$3.1 (19.6$^{+3.3}_{-3.7}$) & 31.9$\pm$6.1 (28.7$^{+5.7}_{-5.5}$)\\ 
231356$+$010910 & 7.7 & 4.5 (4.0$^{+1.4}_{-1.2}$) & 1.9 (2.0$^{+1.2}_{-1.1}$)& 3.3$\pm$2.0 (3.2$^{+1.7}_{-1.4}$) & 19.4$\pm$2.9 (16.8$^{+3.8}_{-4.4}$) & 30.6$\pm$5.7 (26.8$^{+5.3}_{-5.3}$)\\ 
231643$-$011325 & 6.4 & 8.6 (4.6$^{+1.7}_{-1.7}$) & 2.6 (3.5$^{+1.4}_{-1.1}$)& 0.6$\pm$1.9 (1.6$^{+1.3}_{-0.9}$) & 18.2$\pm$2.9 (14.2$^{+4.2}_{-7.0}$) & 34.0$\pm$6.1 (33.2$^{+5.9}_{-5.8}$)\\ 

\hline
\end{tabular}
\label{tab:dsfg_catalog}
\end{table*}

\section{DSFGs: a closer look}
\label{sec:dsfgs}

A subset of the brightest lensed DSFG candidates was selected from an early version of the ACT equatorial source catalog for multi-wavelength follow-up. Specifically, ACT maps at 218~GHz were matched filtered (Section \ref{sec:mf}), and the brightest 36 sources with dust-like spectra (Section \ref{section:specindices}) and without clear contamination from Galactic dust  were selected. The size of the original sample was chosen to be large enough to enable statistical inferences about source properties, but not so large as to render impractical the extensive (and thus formidable) multi-wavelength follow-up observations required to understand each source. In this initial selection, we also vetoed any sources cross-identified with nearby star-forming galaxies resolved in optical imaging from SDSS. Later, the introduction of systematic cuts for Galactic contamination (Section \ref{section:dustremoval}) and the availability of the 277~GHz data were used to discard six candidates that were unlikely true extragalactic sources. The resulting sample of the brightest 30 lensed DSFG candidates are all significant (S/N~$>6$) unresolved detections with 218~GHz flux greater than 16~mJy and a dust-like spectrum spanning all three ACT frequency bands. Partial catalog entries for these DSFGs are given in Table \ref{tab:dsfg_catalog}. The three-band selection indicates with high confidence that these are all real DSFGs. Indeed two are well studied, lensed DSFGs. The first is ACT-S~J2135--0102,  the so-called ``Cosmic Eyelash'', with lensing magnification $\mu=32$ at $z=2.359$ \citep{Swinbank2010}. Interestingly, we also see evidence for the 150~GHz Sunyeav-Zel'dovich decrement of the massive lens associated with the Eyelash (Table \ref{tab:dsfg_catalog}). The second is ACT-S~J020941+001557 with $\mu\approx10$ at $z=2.553$ \citep{Geach2015,Su2017,Geach2018,Rivera2018}. A third, ACT-S~J202955+012054, we have also confirmed as a lensed system at $z=2.64$ with a CO spectral line energy distribution characteristic of a ULIRG/AGN with a potential galaxy-scale outflow \citep{Roberts-Borsani2017}.  In the remainder of this section and in Appendix \ref{app:dsfg}, we discuss additional data characterizing this brightest sample of DSFGs.

In \cite{Su2017}, we took a preliminary step towards characterizing the redshift distribution and physical properties of ACT-selected DSFG candidates. We modeled ACT and {\it Herschel}-SPIRE photometry to derive characteristics for nine of our 30 sources that fell in the footprint of either the Herschel Stripe 82 Survey \citep[HerS;][]{Viero14} or the HerMES Large Mode Survey \citep[HeLMS;][]{Oliver12}. We emphasize that, while these sources were  detected by {\it Herschel} \citep{Asboth16,Nayyeri16}, only ACT data were used in the subsample selection. When fitting the far-infrared SED of the thermal emission for warm extragalactic dust, one has a range of models from which to choose, each with different physical implications. In \cite{Su2017} we explored four different models with different assumptions about dust temperature distribution and  opacity. Independent of model choice, we found that the subsample has a  redshift range $z=2.5-5.5$ covering the era leading up to ``cosmic noon'' when the star-formation density of the universe peaked \citep[e.g.,][]{Burgarella13}. High apparent infrared luminosities $\mu L_{\rm IR} = 0.3-1.4 \times 10^{14}$~${\rm L}_{\odot}$ (as above, $\mu$ here is lensing magnification) imply apparent star formation rates significantly greater than 1000~${\rm M}_{\odot}{\rm yr}^{-1}$ (neglecting possible AGN contribution). These luminosities are consistent with other samples that have apparent brightness enhanced by strong lensing with $\mu\sim10$ \citep[e.g.,][]{Bussmann13,  Weiss13, Canameras15, Harrington16, Strandet16}. Finally, our modelling showed that the common assumption of optically thin dust is not a good match at the peak of the SED (rest-frame wavelength $\sim$100~$\mu$m) for ACT-selected sources. 

Beyond SED modeling, we have been compiling a suite of multifrequency data both from our own follow-up observations and archival datasets. In Appendix \ref{app:dsfg} we present optical and infrared imaging for the 30 brightest DSFG candidates along with locations of radio, millimeter, and submillimeter counterparts. Based on these results, compelling evidence already exists that  a number of sources, identified at higher resolution with a combination of radio, millimeter, and submillimeter data are lensed by galaxies or galaxy clusters detected in the optical and/or infrared data. A number of these systems with well-measured redshifts are undergoing detailed spectral line imaging and will be the subject of future publications.

\section{Conclusions}
\label{sec:conclusion}

We have presented an extragalactic source catalog of AGN and DSFGs from the ACT 2009-2010 survey of the celestial equator. The multifrequency (148~GHz, 218~GHz, and 277~GHz) dataset is unique in its combination of survey area covering hundreds of square degrees and sensitivity to DSFGs. For AGN, the 277~GHz data provide new constraints.  Therefore, the catalog  extends previous galaxy population studies by ACT \citep{marriagesources,Marsden2014,datta}, SPT \citep{Vieira10,mocanu2013} and Planck \citep{Canameras15,PlanckSources2016}. It complements deeper, degree-scale millimeter surveys \citep[e.g.,][]{austermanncosmos,austermannshades,Staguhn2014} and submillimeter surveys \cite[e.g,][]{Negrello2010,Bussmann13,Wardlow13,Nayyeri16,Asboth16,Nettke2017}. In addition to the galaxy catalogs, all maps, including the equatorial 277~GHz dataset for the first time, are publicly available.

The heterogeneous selection in the presence of Galactic emission presents challenges to maintain sample purity and handle Eddington bias, especially for DSFGs with the lowest flux densities. We have overcome these challenges by developing custom tests for Galactic contamination and a method for debiasing flux densities that accounts for arbitrary selection effects. 

Based on the resulting extragalactic catalog, we presented spectral properties and source counts for the AGN and DSFGs. For AGN, we have shown that the previously measured spectral slope between 148~GHz and 218~GHz extends to 277~GHz. For DSFGs, we have shown that the average spectrum departs from a single-temperature, optically thin greybody above 218~GHz. This may be due to the emission becoming optically thick or indicate an additional cold dust component. In terms of source counts, we present the first blazar source counts at 277~GHz and find  consistency with count models extrapolated from models built for data at lower frequencies. For DSFGs, we extend the existing 218~GHz  counts to lower flux density where unlensed sources dominate; we find good agreement with source models here. At 277~GHz the DSFG counts extend to higher flux densities than previously published and are higher than most models predicted for 277~GHz.

We have estimated the inter-year fractional deviation in flux density of the blazar population and found it to be 40\% for sources with flux densities above 50~mJy in both the 148~GHz and 218~GHz bands. Furthermore, we find this variability to be tightly correlated between the bands, with a  correlation coefficient of 0.98.

Thirty of the brightest DSFGs  are described in detail. These have been the subject of more detailed study \citep{Su2017,Roberts-Borsani2017,Rivera2018}. Appendix~\ref{app:dsfg} describes radio to optical data on these sources. Interestingly, an apparent SZ decrement at 148~GHz is coincident with two of the sources (likely associated with massive lenses).

Looking ahead, the ACTPol receiver has been upgraded as Advanced ACT with additional bands that will span 30--250~GHz. The resulting survey, covering thousands of square degrees to  sensitivies surpassing the current work, will contribute to the next generation of wide-field millimeter-wave extragalactic catalogs.

\section*{Acknowledgements}
This work was supported by the U.S. National Science Foundation through awards AST-0408698 and AST-0965625 for the ACT project, as well as awards PHY-0855887 and PHY-1214379. Funding was also provided by Princeton University, the University of Pennsylvania, and a Canada Foundation for Innovation (CFI) award to UBC. ACT operates in the Parque Astron\'omico Atacama in northern Chile under the auspices of the Comisi\'on Nacional de Investigaci\'on Cient\'ifica y Tecnol\'ogica de Chile (CONICYT). Computations were performed on the GPC supercomputer at the SciNet HPC Consortium. SciNet is funded by the CFI under the auspices of Compute Canada, the Government of Ontario, the Ontario Research Fund -- Research Excellence; and the University of Toronto. In our use of observations from the SMA, the authors are grateful to the people of Hawai'ian ancestry on whose sacred mountain we are
privileged to be guests. Some of the observations reported in this work were obtained with the Southern African Large Telescope (SALT). Funding for SALT is provided in part by Rutgers University, a founding member of the SALT consortium. M.G. and T.M. acknowledge support from Johns Hopkins University. A.J.B. and J.R. acknowledge support from NSF grant AST-0955810. KM acknowledges support from the National Research Foundation of South Africa (grant number 93565). D.C. acknowledges the financial assistance of the South African SKA Project (SKA SA)
towards this research (www.ska.ac.za). R.D., P.A., F.R., and G.M. received funding from the Chilean grants FONDECYT 11100147, and BASAL (CATA). R.D. acknowledges CONICYT for grants FONDECYT 1141113, Anillo ACT-1417, QUIMAL 160009 and  BASAL PFB-
06 CATA. C.L. thanks CONICYT for grant Anillo ACT-1417. This research made use of Astropy, a community-developed core Python package for Astronomy \citep{Astropy2013} and the Astronomy IDL Library \citep{astroidl}.

\bibliography{mn-jour,actequsources}
\bibliographystyle{apj}

\appendix

\section{Auxiliary Data for Brightest DSFGs}
\label{app:dsfg}
This appendix provides a closer look at the 30 sources selected as candidate lensed DSFGs for a campaign of multifrequency follow-up (Section \ref{sec:dsfgs}). For each source we show optical and infrared imaging that, in many cases, reveals a putative lens galaxy and, in fewer cases, shows evidence for the light from a lensed DSFG. The optical images are from the Panoramic Survey Telescope and Rapid Response System (Pan-STARRS) 1 data release, in the r ($\lambda = 622$~nm; $m=23.2$ 5$\sigma$ depth; 1.19$''$ seeing), i ($\lambda = 763$~nm; $m=23.1$ 5$\sigma$ depth; 1.11$''$ seeing), and z ($\lambda = 905$~nm; $m=22.3$ 5$\sigma$ depth; 1.07$''$ seeing) bands \citep{Chambers2016}. We have chosen to show Pan-STARRS imaging instead of the deeper stacked imaging from the Sloan Digital Sky Survey (SDSS) Stripe~82 because the better seeing in Pan-STARRS enables clearer distinction of lensing features. The near infrared Ks-band ($\lambda \approx 2.1$~$\mu$m) images are either from the VISTA Hemispheres Survey (VHS) \citep[VHS;][]{Mcmahon2013} with a 5$\sigma$ detection limit of $m=18.1$ (Vega) or from our own follow-up observations with the NICFPS camera \citep{vincent03} on the ARC 3.5~m telescope at the Apache Point Observatory (APO) with a Ks-band 5$\sigma$ detection limit of $m=19.5$ (Vega). At longer wavelengths, imaging is provided either by the {\it Wide-field Infrared Survey Explorer} all-sky survey in bands W1 ($\lambda \approx 3.5$~$\mu$m) and W2 ($\lambda \approx 4.5$~$\mu$m) \citep{Wright10} or, where available, from the {\it Spitzer} Heritage Archive\footnote{http://sha.ipac.caltech.edu/applications/Spitzer/SHA/}, for which the analogous bands (same wavelengths) are called S1 and S2. Most {\it Spitzer} data derive from the  {\it Spitzer} IRAC Equatorial Survey \citep{Timlin16} and the {\it Spitzer}-HETDEX Exploratory Large-Area Survey (S1, S2) \citep{Papovich16}.

The images are annotated to indicate the locations of detections from the radio to the sub-millimeter. The sizes of the annotations indicate the astrometric errors associated with the corresponding detections ($5''$ for ACT, $3''$ for {\it Herschel}, $1''$ for radio data and $<0.5''$ for SMA \citep{Su2017}). Where the sample overlaps the Herschel Stripe 82 Survey \citep[HerS;][]{Viero14} and the HerMES Large Mode Survey \citep[HeLMS;][]{Oliver12}, there are {\it Herschel} detections (solid black diamonds) for nine sources at $\lambda=250$~$\mu$m, 350~$\mu$m, and 500~$\mu$m \citep{Su2017}. Millimeter-wave detections are provided by ACT (dashed circle; $\lambda = 2$~mm, 1.4~mm, and 1.1~mm) and by our follow-up observations with the Submillimeter Array (SMA; red rectangle; $\lambda = 1.4$~mm). The subarcsecond astrometry of the SMA imaging makes it the most reliable indicator of the DSFG position. Details of the SMA imaging will be given in a future publication. Finally, nearby radio detections (blue pentagons) are selected from (in order, as available) (1) our own deep follow-up with the JanskyVLA (JVLA) at ($\lambda=6$~cm; rms 25~$\mu$Jy/beam; VLA/13A-478, PI M. Gralla); (2)  the VLA SDSS Stripe 82 Survey at 1.4 GHz (herein referred to as VLASS821P4; $\lambda=21$~cm; rms 52~$\mu$Jy/beam) \citep{Hodge2011}; or (3) the VLA Faint Images of the Radio Sky at Twenty Centimeters (FIRST; rms 150~$\mu$Jy/beam) survey \citep{Becker1995,White1997}.

Accompanying the annotated images, we also provide  a written synopsis of what is known about each source and its lens candidate. As the data permit, these synopses clarify radio associations, give details of  SED modeling, provide redshift estimates for the source and/or lens, and point out unique source properties. To establish redshifts for a subset of putative lens galaxies, we use SDSS \citep{Alam15} and our own spectroscopy with the South African Large Telescope \citep[SALT;][]{Buckley06} \citep[PI J. Hughes; for details of the observational setup, data reduction and analysis of the SALT/RSS spectra, see][]{Roberts-Borsani2017}. These images and synopses provide an initial look at the sources of this sample, and further studies (e.g., CO mapping) are underway.

\newcommand{\dsfgentry}[3]{\begin{minipage}{.95\textwidth} \begin{center}
\includegraphics[width=6in]{ACT#1.png}\end{center}
{\bf ACT-S J#2}#3  \vspace{5mm}\end{minipage}}

\begin{minipage}{.95\textwidth}
\begin{center}
    \includegraphics[width=5in]{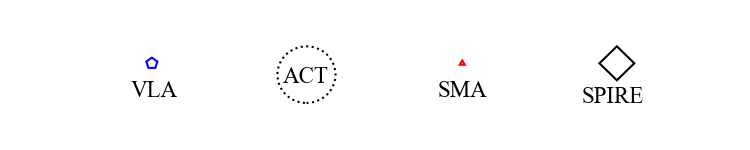}
\end{center}
{\bf DSFG Symbol Legend.} Symbols indicate the location of detections in the centimeter (VLA), millimeter (ACT, SMA), and submillimeter (SPIRE) bands. Sizes of symbols indicate positional uncertainty ($1\sigma$) associated with the detection. Sources may lack detections by VLA, SMA, and/or SPIRE.
\vspace{5mm}
\end{minipage}

\dsfgentry{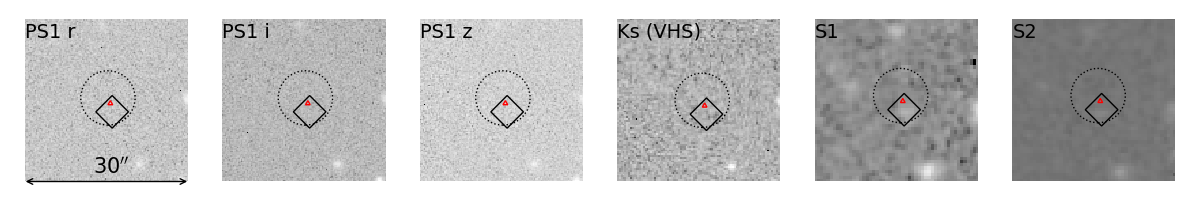}{001133$-$001835}{ has no associated radio emission (only FIRST data available). Futhermore, the optical and infrared imaging show no  evidence for a lens near the SMA location. A fit to the ACT and {\it Herschel} photometry gives a DSFG redshift $z\approx3.3$, temperature $T\approx46$~K, and apparent luminosity $\mu L_{\rm IR} \approx 10^{13.7}$~L$_\odot$ \citep[][where the source is listed as ACT-S~J0011$-$0018]{Su2017}.}

\dsfgentry{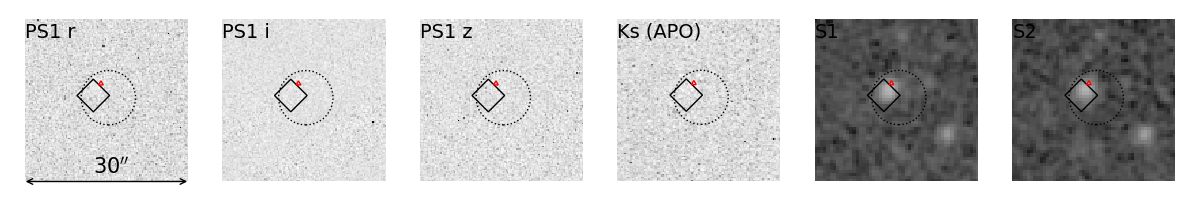}{002220$-$015523}{ has no clear radio emission (only FIRST data available). While the optical and near infrared images show no associated  signal, there is a 3.5~$\mu$m and 4.5~$\mu$m bright source, a lens candidate, near the SMA and {\it Herschel}-based locations.  A fit to the ACT and {\it Herschel} photometry gives a DSFG redshift $z\approx4.4$ , temperature $T\approx42$~K, and apparent luminosity $\mu L_{\rm IR} \approx 10^{13.8}$~L$_\odot$ \citep[][where the source is listed as ACT-S~J0022$-$0155]{Su2017}.}

\dsfgentry{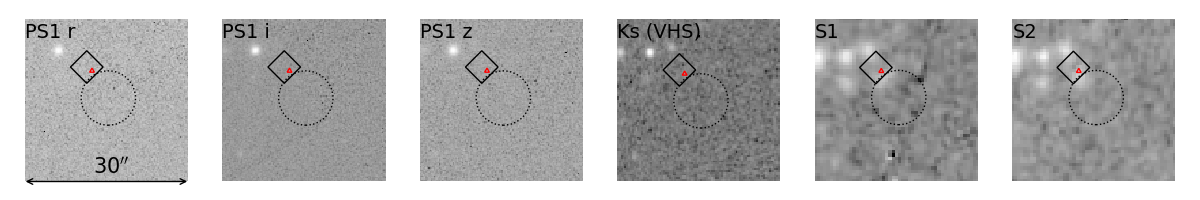}{003814$-$002255}{ has no radio detection (only FIRST data available). The SMA position for this source is consistent with that from {\it Herschel}. Sources to the north-east are detected in the optical and infrared, one or more of which may be a lens. A fit to the ACT and {\it Herschel} photometry gives a DSFG redshift $z\approx4.3$, temperature $T\approx45$~K, and  apparent luminosity $\mu L_{\rm IR} \approx 10^{13.9}$~L$_\odot$ \citep[][where the source is listed as ACT-S~J0038$-$0022]{Su2017}. }

\dsfgentry{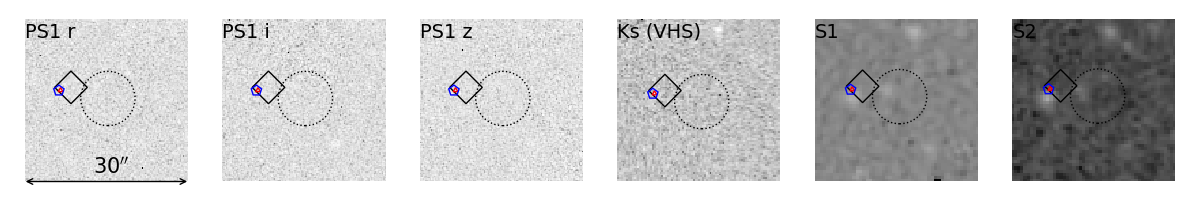}{003929+002422}{ features faint 6-cm radio emission ($94\pm21$~$\mu$Jy) coinciding with the SMA position. Nearby infrared emission comes from a candidate lens galaxy. A fit to the ACT and {\it Herschel} photometry gives a DSFG redshift $z\approx4.3$, temperature $T\approx45$~K, and apparent luminosity $\mu L_{\rm IR} \approx 10^{13.8}$~L$_\odot$ \citep[][where the source is listed as ACT-S~J0039$+$0024]{Su2017}.}

\dsfgentry{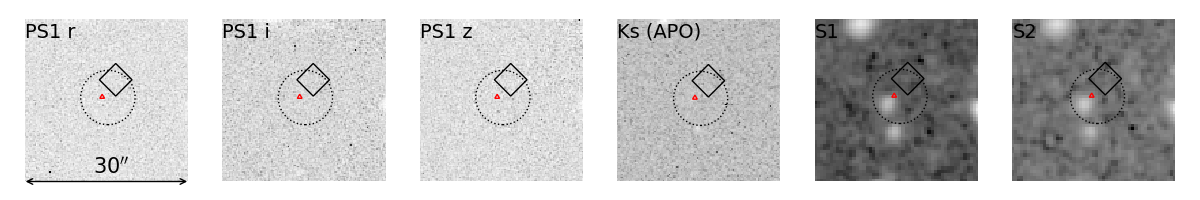}{004410+011818}{ has no detectable radio flux in VLASS821P4 or FIRST. While no optical or near-infrared counterparts are found, the {\it Spitzer} data show a 3.5~$\mu$m and 4.5~$\mu$m bright source, a lens candidate, associated with the SMA detection. A fit to the ACT and {\it Herschel} photometry gives a DSFG redshift $z\approx4.9$ and temperature $T\approx47$~K \citep{Su2017}. Additionally, this source is one of the brightest of the sample with a 218~GHz flux of 35~mJy. The high flux and high redshift of this source imply a high apparent luminosity $\mu L_{\rm IR} \approx 10^{14.2}$~L$_\odot$ \citep[][where the source is listed as ACT-S~J0044$+$0118]{Su2017}.}

\dsfgentry{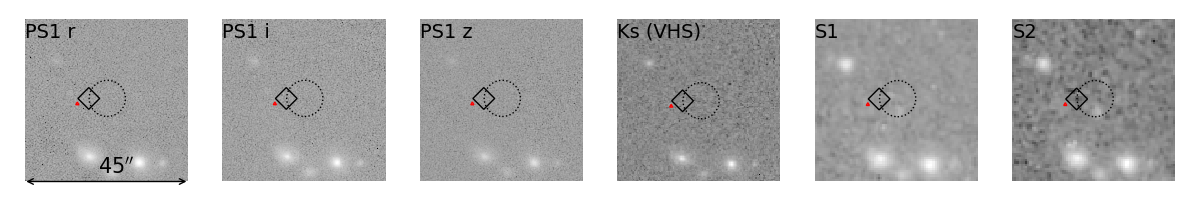}{004532$-$000127}{ has no radio emission detected (VLA 6-cm, VLASS821P4, FIRST). While there is no clear optical or infrared emission near the well-determined location from the SMA, there is a cluster of galaxies to the south, which may be a lens. The redshift of this cluster was measured with SALT to be $z=0.234$.  A fit to the ACT and {\it Herschel} photometry gives a DSFG redshift $z\approx3.8$, temperature $T\approx37$~K, and apparent luminosity $\mu L_{\rm IR} \approx 10^{13.7}$~L$_\odot$ \citep[][where the source is listed as ACT-S~J0045$-$0001]{Su2017}.}

\dsfgentry{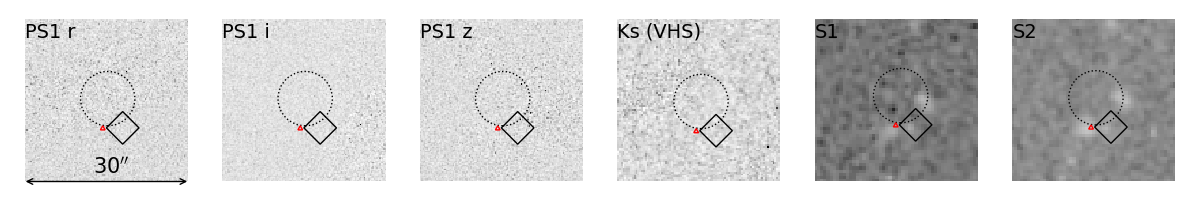}{010729+000114}{ has no detected  radio emission (VLA 6-cm, VLASS821P4, FIRST). This source has a redshift of $z=3.332$ measured with CO lines by the Redshift Search Receiver \citep[RSR;][]{Erickson07}. Modeling of the SED from ACT \citep{Su2017} and {\it Herschel} provides estimates of temperature $T\approx35$~K, and apparent luminosity $\mu L_{\rm IR} \approx 10^{13.4}$~L$_\odot$ \citep[][where the source is listed as ACT-S~J0107$+$0001]{Su2017}. A lens candidate is detected at 3.5~$\mu$m and 4.5~$\mu$m  by {\it Spitzer}.}

\dsfgentry{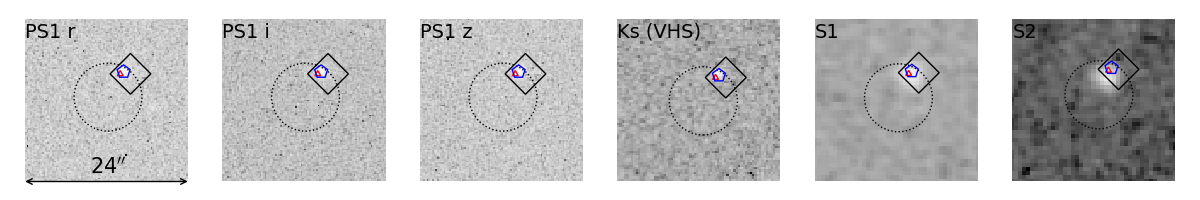}{011640$-$000457}{ has associated 21-cm detections in FIRST (1.4~mJy) and VLASS821P4 (0.8~mJy), but no detection in our 6-cm follow-up.  There is an optical and infrared-bright lens candidate near the SMA position. The photometric redshift of the lens candidate from SDSS is $z=0.45$. A fit to the ACT and {\it Herschel} photometry gives estimates of redshift $z\approx3.9$, temperature $T\approx41$~K, and apparent luminosity $\mu L_{\rm IR} \approx 10^{14.0}$~L$_\odot$ \citep[][where the source is listed as ACT-S~J0116$-$0004]{Su2017}.}

\dsfgentry{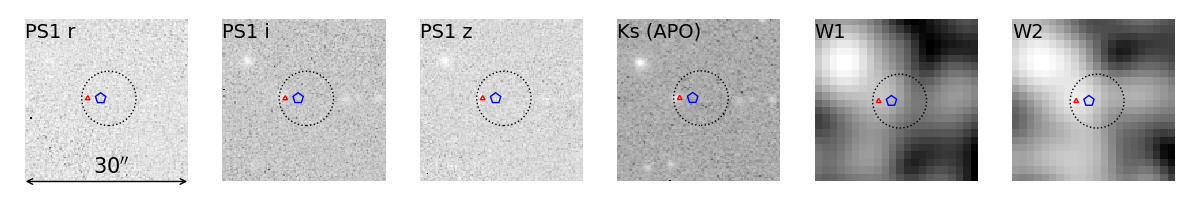}{013857+021420}{ shows radio emission in FIRST (2.3~mJy) separated by a few arcseconds from the SMA position. Whether this emission is associated with the candidate DSFG is unclear, especially given the FIRST 5$''$ resolution. (No other radio data are available.) There is a lens candidate associated with this source in the optical and infrared images with a photometric redshift, $z=0.8$, from SDSS. Unlike other sources in the $00$h$-02$h R.A. range, ACT-S~J013857+021420 is outside of the {\it Herschel} survey footprints.}

\dsfgentry{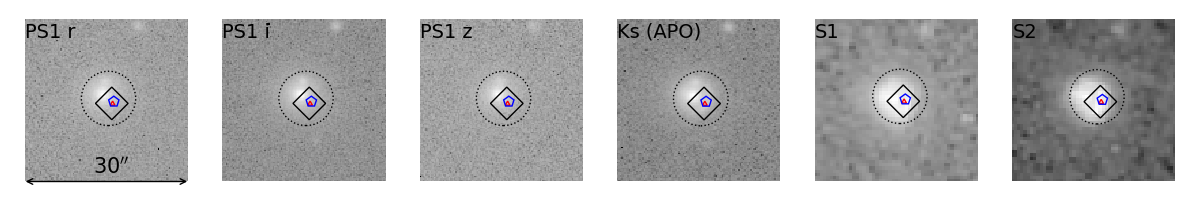}{020941+001557}{, which was first published in \cite{Geach2015}, shows clear lensing structure (a resolved Einstein ring) from the radio to the optical. Our GBT CO-line observations give a redshift of the source of $z=2.553$ \citep[Appendix A of][]{Su2017}.  The lens redshift is $z = 0.202$ \citep{Geach2015}. Recent work with ALMA and NOEMA has produced detailed maps of molecular line emission that provide insight to the dynamics and state of the interstellar medium \citep{Geach2018,Rivera2018}. A fit to the ACT and {\it Herschel} photometry gives estimates of temperature $T\approx42$~K and apparent luminosity $\mu L_{\rm IR} \approx 10^{14.3}$~L$_\odot$ \citep[][where the source is listed as ACT-S~J0210$+$0016]{Su2017}.}

\dsfgentry{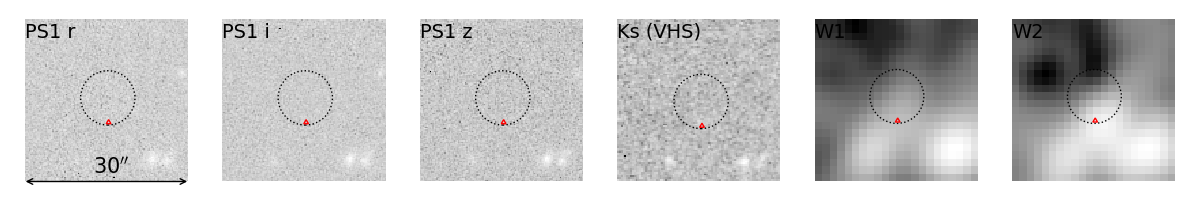}{022830$-$005226}{ was not detected in FIRST. However, 21-cm emission is visible in the FIRST intensity map near the SMA position. There is also faint  infrared emission from a galaxy, a potential lens, near the SMA position.}

\dsfgentry{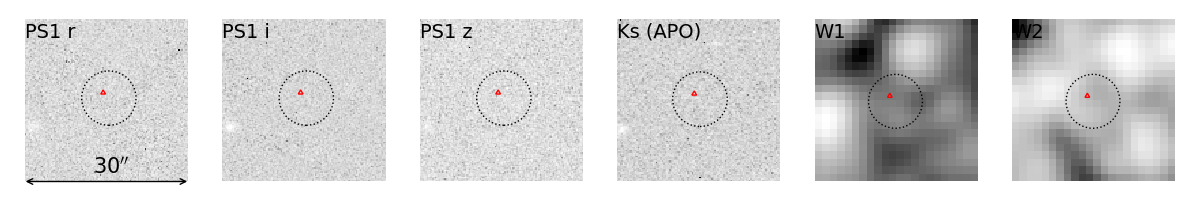}{023120+011636}{ was not detected in FIRST. The SMA position agrees with ACT, but there are no clear associations in the optical or infrared images.}

\dsfgentry{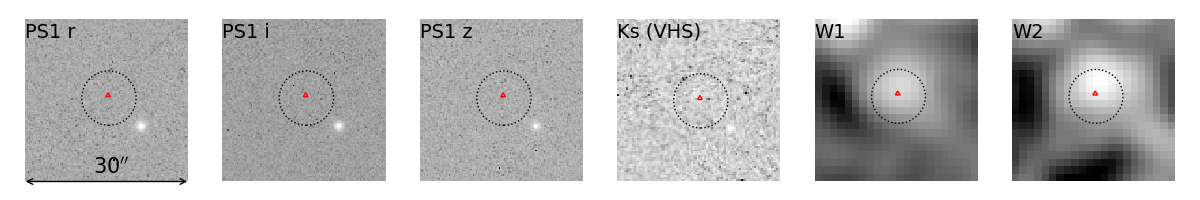}{025331$-$000318}{ was not detected in FIRST, but there is visible 21-cm emission in the FIRST map at the SMA location. There is also associated infrared emission from a lens candidate. This is the second brightest ACT high-$z$ lensed DSFG candidate, with a flux of 58~mJy at 218~GHz.}

\dsfgentry{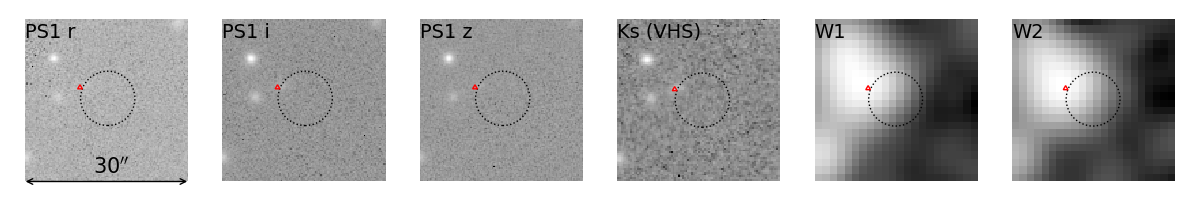}{025512$-$011456}{ was not detected in FIRST, but there is visible 21-cm emission in the FIRST map at the well determined SMA location. There is also associated optical and infrared emission from a candidate lens.}

\dsfgentry{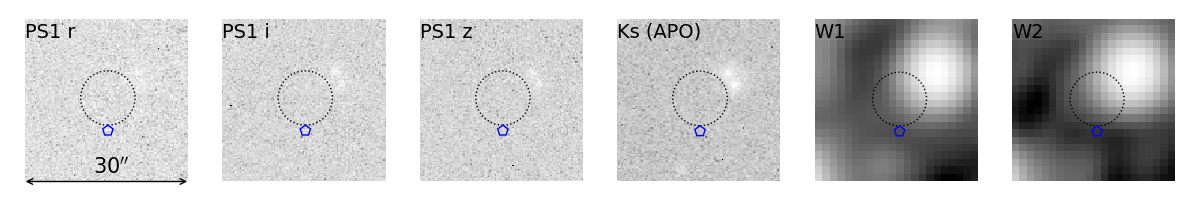}{030410+013225}{ has  radio emission ($336\pm62$~$\mu$Jy) detected to the south in our 6-cm VLA follow-up. There is also an infrared bright double source to the north-west. Without more precise astrometry for the DSFG, however, associations remain ambiguous.}

\dsfgentry{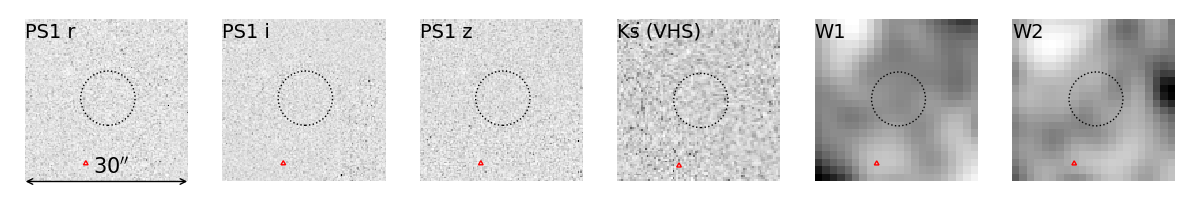}{031019$-$000215}{ was detected by the SMA south of the ACT position. No radio associations were detected either in FIRST or in our 6-cm VLA imaging. There is no associated optical or infrared emission detected near the SMA position.}

\dsfgentry{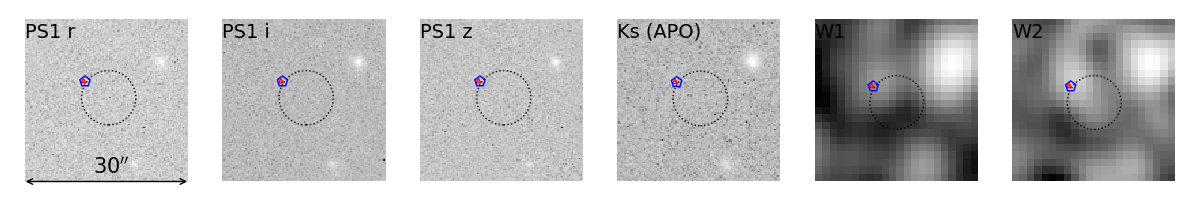}{031127+013639}{ was detected by the SMA and in our 6-cm VLA imaging ($174\pm22$~$\mu$Jy). There is also faint infrared emission near the SMA position.}

\dsfgentry{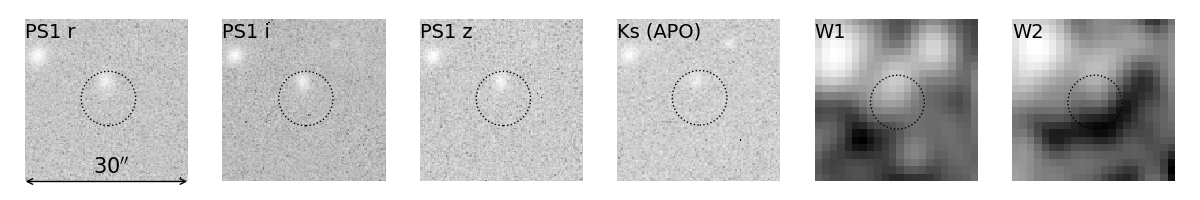}{032104$+$012934}{ has no associated radio emission detected in our 6-cm VLA imaging. There is an extended optically-bright galaxy within 6$''$ of the ACT location.}

\dsfgentry{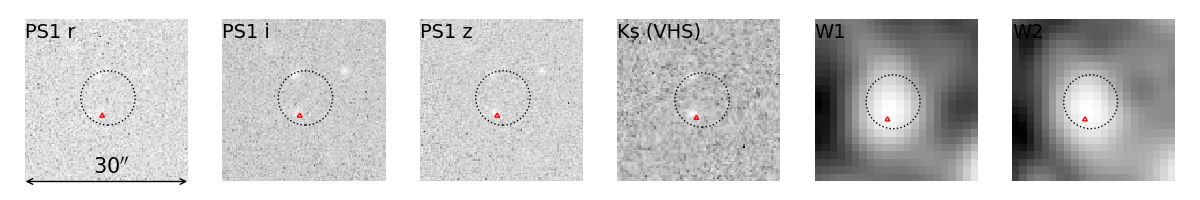}{0321221$-$000221}{ has an accurate position from the SMA. Near this position, there was no associated radio emission in FIRST or our 6-cm VLA follow-up. There is a nearby infrared bright galaxy, a lens candidate, with a photometric redshift of $z=0.8$ from SDSS.}

\dsfgentry{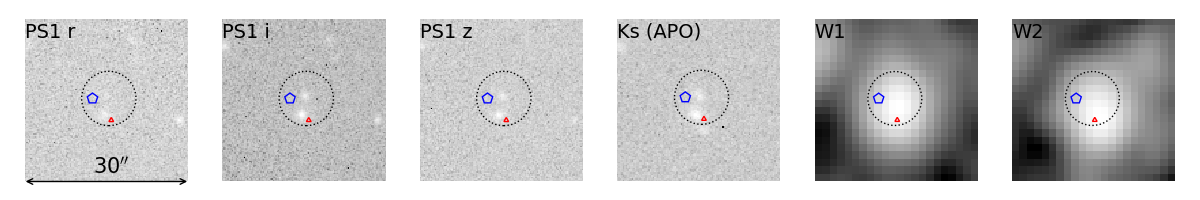}{032351$+$012801}{ was detected with the SMA. There is 6-cm radio emission ($160\pm32$~$\mu$Jy) from our VLA imaging displaced by 5$''$ from the SMA position. There are also multiple lens candidates with associated optical and infrared emission.}

\dsfgentry{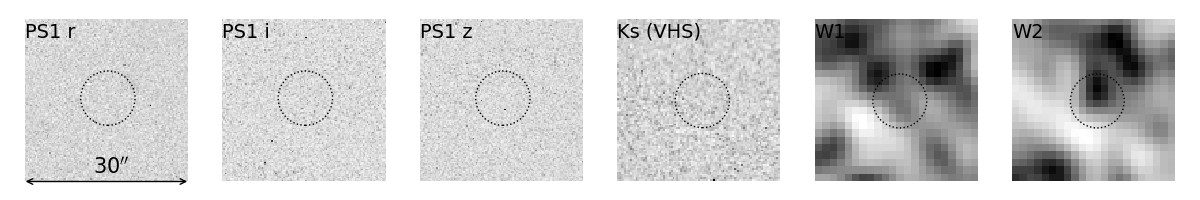}{034003$+$001627}{ shows no radio emission  in FIRST or our 6-cm VLA imaging. There are also no nearby optical or infrared bright sources.}

\dsfgentry{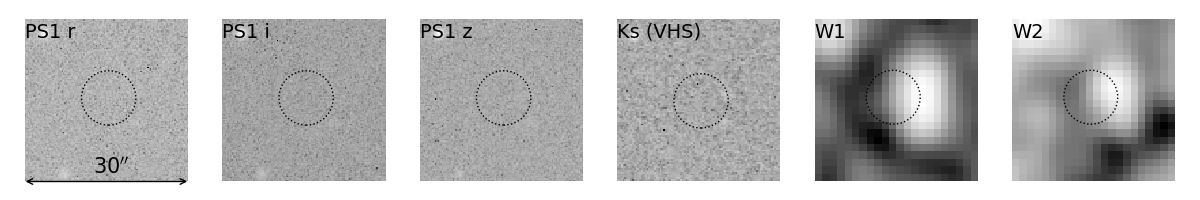}{034228$-$005644}{ has no official FIRST detection, but there is excess 21-cm brightness in the FIRST image coinciding with   3.5~$\mu$m and 4.5~$\mu$m emission west of the ACT detection.}

\dsfgentry{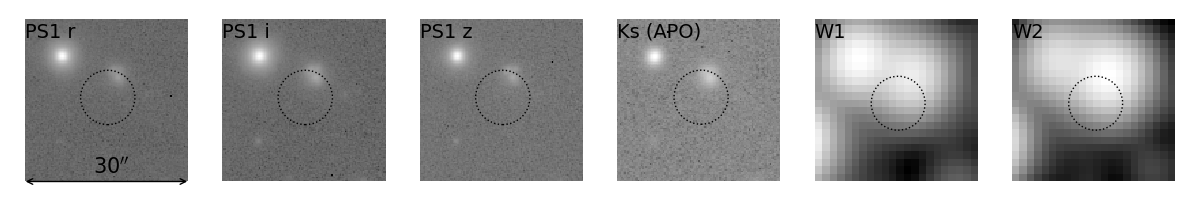}{202955$+$012054}{ is studied extensively in \cite{Roberts-Borsani2017}. A galaxy 4$''$ from the ACT location has optical and near infrared emission with the suggestion of a northern lensing arc. This complex was identified with J2029 in Combined Array for Millimeter Astronomy (CARMA) 3~mm imaging. Spectroscopy from CARMA, the Large Millimeter Telescope, and the IRAM~30m telescope establish the redshift of J2029 as $z=2.64$. A SALT spectrum of the putative lens galaxy gives $z=0.3242$. The CO spectral line energy distribution was found to be characteristic of a ULIRG/AGN, with significant excitation of the higher CO(7-6) and CO(8-7) lines. A broadened and enhanced C\textsc{I} line  suggests differential lensing of a galaxy-scale outflow.  }

\dsfgentry{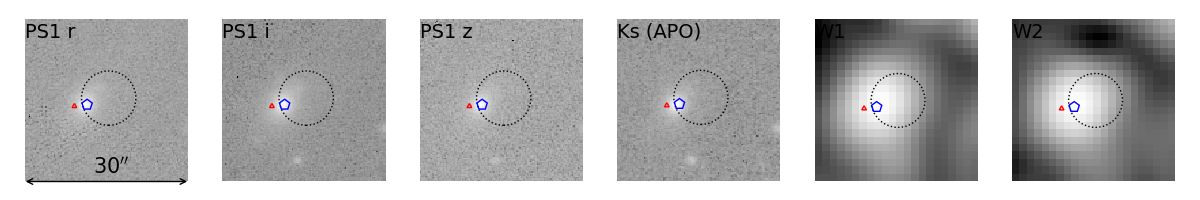}{212740$+$010921}{ has nearby (4$''$) radio emission in FIRST  (1.8 mJy) and also our 6-cm VLA follow-up (1.1 mJy). The source of the radio emission also shows extended optical and infrared emission. The precise SMA position of the source lies just to the east of the optical and infrared emission, consistent with a lensing morphology.}

\dsfgentry{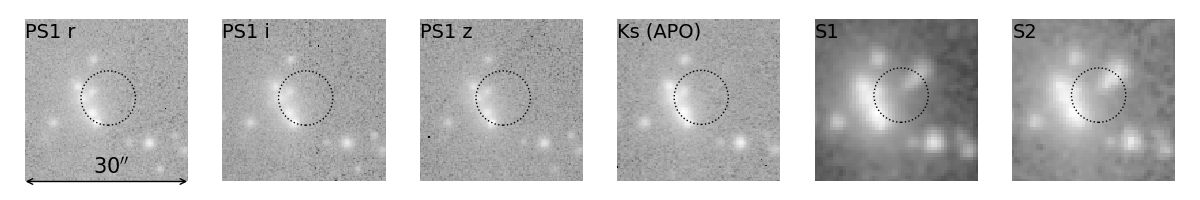}{213511$-$010255}{ is the ``Cosmic Eyelash'' \citep{Swinbank2010}, a galaxy at $z = 2.3259$ lensed by a factor of $\mu=32$ by the massive cluster MACSJ2135-010217 at $z=0.325$. As shown in Table \ref{tab:dsfg_catalog}, the 148~GHz flux density is measured to be negative for this source. The most likely interpretation of this result is that we are measuring the SZ decrement associated with MACSJ2135-010217.}

\dsfgentry{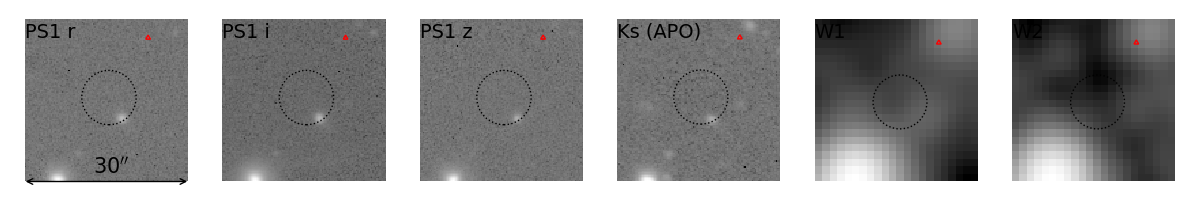}{213713$+$011156}{ does not have associated radio emission in FIRST or our 6-cm VLA follow-up. The SMA position places the source $12''$ from the ACT location near one or more galaxies detected in the infrared.}

\dsfgentry{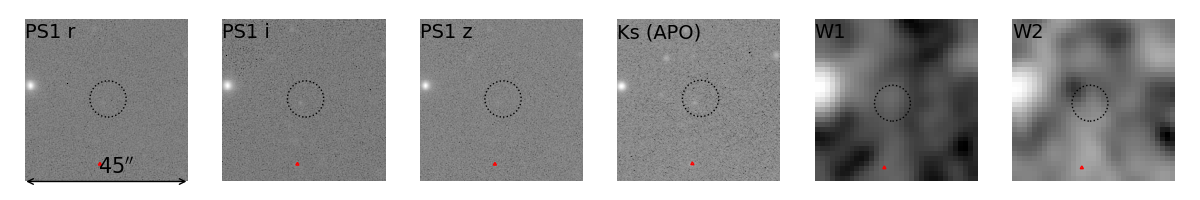}{230239$-$013333}{ has a SMA position located approximately 15$''$ south of the ACT location. At the SMA position there is no emission detected in the radio (FIRST), optical, or infrared data.}

\dsfgentry{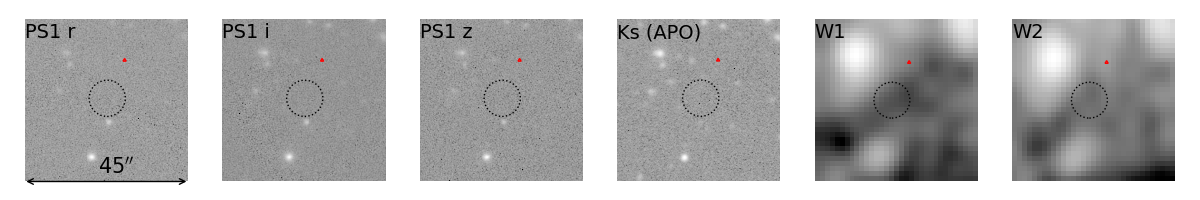}{231252$-$001524}{ has no detected radio emission in either the FIRST or our 6-cm VLA imaging. A precise SMA position puts the source  west of a cluster of galaxies, a lens candidate with a redshift of $z=0.59$ from SDSS. As shown in Table \ref{tab:dsfg_catalog}, the 148~GHz flux density is measured to be negative for this source. The most likely interpretation of this result is that we are measuring the SZ decrement associated with the $z=0.59$ galaxy cluster.}

\dsfgentry{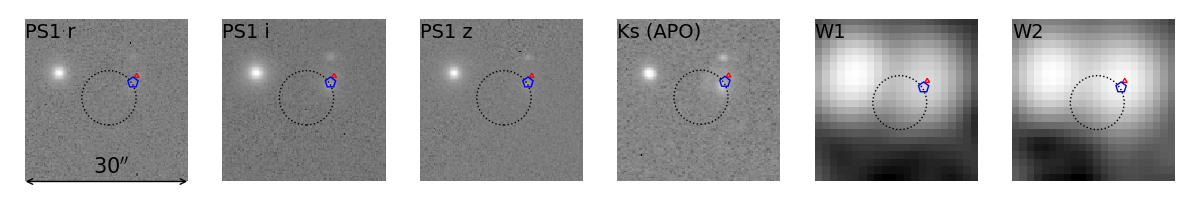}{231356$+$010910}{ has a 3.8 mJy source in FIRST near the SMA position. The SMA and FIRST detections lie between two optical and infrared-bright galaxies to the north and south. The near infrared imaging shows an arc  coincident with the SMA position.}

\dsfgentry{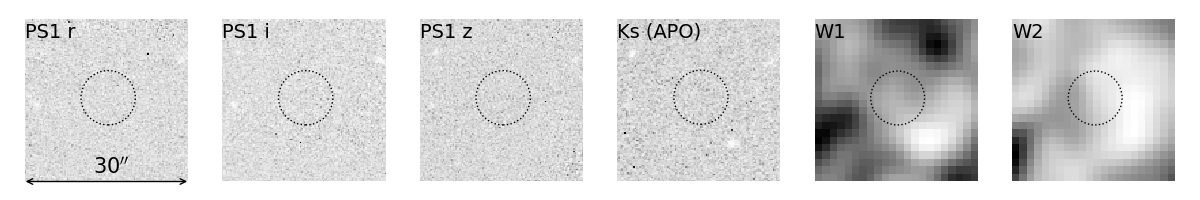}{231643$-$011325}{ has no associated radio emission in FIRST. There is an infrared bright galaxy 8$''$ away from the ACT position, but more precise astrometry for the ACT source is needed to confirm an association.}

\section{Millimeter spectral indices: Special cases} 
\label{app:alphaoutliers}

The sources discussed in this appendix have unusual millimeter spectral indices given their ACT measurements. 
We investigated these sources using optical data from SDSS and archival data from NED, and we visually inspected them in the ACT maps. Beyond curiosity, there were two main motivations for these investigations. First, sources with 148-218~GHz spectral indices that lie in between the AGN and DSFG populations could be misidentified, so the extra information could potentially clarify source categorization. Second, the debiasing methods we adopt when determining flux densities for the secondary bands assume that sources are drawn from either AGN or DSFG populations that have spectral index distributions approximated by Gaussians. For sources that have very atypical spectra, these assumptions are unlikely to hold, and thus their raw secondary band flux densities would be more reliable than what is reported as debiased. On a population level, if there is a substantial number of outliers that are misidentified or otherwise skewing the properties of the measured spectral index distributions for either the AGN or the DSFGs, this would make our debiasing less optimal.

We first restrict attention to the area of the map that has reliable and comparatively low-noise measurements in all three ACT bands, which is within $-1.2\degree<$ dec $<1.2\degree$. 
We then identify four groups of special cases: sources that have $0.5<\alpha_{148-218} < 1.5$, AGN with $\alpha_{218-277} < -5$, DSFGs with $\alpha_{218-277} > 5$ and DSFGs with $\alpha_{218-277} < 0$. 

Sources that have $0.5<\alpha_{148-218}<1.5$ are near the boundary between AGN and DSFGs, which we draw at $\alpha_{148-218}=1.0$. In this clean area of the maps, there are 13 such sources. All appear to be well-measured in the ACT data. An inspection of optical images from SDSS reveals that two coincide with large nearby elliptical galaxies, one at $z=0.07$ with a known 4C radio source, and the other at $z=0.017$. These have $\alpha_{148-218} < 1.0$ and so are (correctly) classified as AGN. Another, ACT-S J000910$-$003652, has a pair of galaxies within the ACT beam. The $\alpha_{148-218}$ for this source is 1.2, so it is classified as a DSFG. Both galaxies in this pair are at $z=0.07$, and one is a spiral galaxy and the other contains a known AGN. We note that the CO $J(2-1)$ line at 230.5~GHz can fall within the ACT 218~GHz band. The
ACT 218 GHz band is 17.0~GHz wide, with central frequency
at 219.7 GHz \citep{swetz2011}. If we approximate
the band transmission as a step function, the CO line will fall within the ACT band for sources in the range $0.01 < z < 0.09$. Thus, if present, some CO line emission could increase their 218~GHz flux density relative to their 148~GHz flux density and thus move a more typical AGN spectral index toward the dusty galaxy spectral index. In addition to these nearby galaxies, one of the other source fields contains a QSO at $z=1.77$, and the corresponding ACT AGN has $\alpha_{148-218} = 0.5$. The remaining 9 fields do not contain unambiguous matches. 

There are 7 AGN with very steep $\alpha_{218-277} < -5$. All are undetected and basically have no flux density at 277~GHz (S/N $< 1\sigma$), and all but one have 218~GHz S/N $<5$ as well. All have known radio counterparts in NED, with measurements in the 4C, NVSS, PMN and/or GBT surveys. One of these is a radio-bright star, V0711 Tau. 

There are 7 DSFGs with $\alpha_{218-277} > 5$. As might be anticipated given their spectra, these all have 218 S/N $< 3.5$, with more significant S/N $>5$ detections at 277~GHz. None are in compromised or dusty locations in the ACT maps. Two are nearby spiral galaxies, ACT-S J235106$+$010318 and ACT-S J214129$+$005340. The SDSS imaging indicates a potential galaxy cluster near a third candidate, with elliptical galaxies located at $z=0.28$. The incidence of matches with known dusty galaxies and a potential lens candidate lends confidence to the selection of these DSFG candidates, even in the absence of strong 218~GHz detections. 

There are 8 DSFGs with $\alpha_{218-277} < 0$. Of these, six are nearby dusty galaxies. One, ACT-S J210551$-$001242 is a star that is a known radio source. As discussed above, CO contamination would enhance the flux densities in the 218~GHz band relative to the 277~GHz band. Also, if the galaxies are partially resolved such that some of the 277~GHz emission falls outside the beam, that could suppress the 277~GHz flux density relative to the 218~GHz flux density. In any case, all but one of these are confirmed to be dusty galaxies.

\end{document}